\documentclass[11pt,a4paper]{article} 
\pdfoutput=1
\usepackage{jheppub, hyperref}
\usepackage{amsfonts, color, float,  amsmath}
\usepackage{amssymb}
\usepackage{subfigure}
\usepackage{url}
\usepackage{appendix}
\usepackage{array}
\usepackage{cancel}
\usepackage{slashed,xspace}
\usepackage[svgnames]{xcolor}  
\usepackage{tikz}
\usepackage{ulem} 

\usepackage[small,bf]{caption}   
\usepackage{microtype}
\usepackage[bottom,para]{footmisc}

\usepackage{booktabs}

\newcommand{\MSb}{{\overline{\rm MS}}}

\newcommand{\sand}[3]{\langle#1|#2|#3 \rangle}

\renewcommand{\Re}{{\rm Re}}

\newcommand{\CT}{\ensuremath{C_T}\xspace}
\newcommand{\CTf}{\ensuremath{C_{T5}}\xspace}
\newcommand{\CP}{\ensuremath{C_P}\xspace}
\newcommand{\CS}{\ensuremath{C_S}\xspace}

\newcommand{\Cten}{\ensuremath{C_{10}}\xspace}

\newcommand{\FH}{\ensuremath{F_{H}}\xspace}
\newcommand{\dGh}{{\ensuremath{d\hat{\Gamma}(s)}}\xspace}

\newcommand{\cost}{\ensuremath{\cos{\theta}}\xspace}

\newcommand{\SM}{\ensuremath{{\rm SM}}\xspace}

\newcommand{\Cn}{\ensuremath{C_9^{(q)}}\xspace}
\newcommand{\Cnnaive}{\ensuremath{\Cn\bigr\rvert_{\rm Naive}}\xspace}
\newcommand{\CnAnn}{\ensuremath{\Cn\bigr\rvert_{\rm Ann}}\xspace}
\newcommand{\CndAnn}{\ensuremath{C_9^{(d)}\bigr\rvert_{\rm Ann}}\xspace}
\newcommand{\CnSS}{\ensuremath{\Cn\bigr\rvert_{\rm SS}}\xspace}
\newcommand{\CnFF}{\ensuremath{\Cn\bigr\rvert_{\rm FF}}\xspace}

\newcommand{\eff}{\ensuremath{\text{eff}}\xspace}
\newcommand{\als}{\ensuremath{\alpha_s}\xspace}
\newcommand{\as}{\ensuremath{a_s}\xspace}

\def\MSb {\ensuremath{\overline{\mathrm{MS}}\xspace}\xspace}

\def\beq {\begin{equation}}
\def\eeq {\end{equation}}
\def\bal {\begin{align}}
\def\eal {\end{align}}
\def\bea {\begin{eqnarray}}
\def\eea {\end{eqnarray}}

\newcommand{\ee}{\ensuremath{e^+e^-}\xspace}
\newcommand{\Dpill}{\ensuremath{D^+\rightarrow\pi^+\tl}\xspace}

\newcommand{\Dpimumu}{\ensuremath{D^+\to\pi^+\mu^+\mu^-}\xspace}
\newcommand{\Dpiee}{\ensuremath{D^+\to\pi^+\ee}\xspace}

\newcommand{\Dll}{\ensuremath{D^0\to\tl}\xspace}
\newcommand{\Dee}{\ensuremath{D^0\to e^+ e^-}\xspace}
\newcommand{\Dmumu}{\ensuremath{D^0\to \mu^+ \mu^-}\xspace}

\newcommand{\subdecay}[2]{\ensuremath{#1(\!\to #2)}\xspace}

\newcommand{\BRof}[1]{\ensuremath{{\cal B}(#1)}\xspace}

\def\mump   {{\ensuremath{\muon^\mp}}\xspace}
\def\taupm   {{\ensuremath{\tauon^\pm}}\xspace}

\def\tauppppi0{\decay{\taupm}{\pion^\pm\pion^\mp\pion^\pm\piz\neu}}
\def\tauppporpi0{\decay{\taupm}{\pion^\pm\pion^\mp\pion^\pm(\piz)\neu}}

\def\Bdstaumupi0{\decay{\B^0_{\left( s\right) }}{\subdecay{\taupm}{\pipm\pimp\pipm\piz\neu}\mump}}
\def\Bstaumupi0{\decay{\Bs}{\subdecay{\taupm}{\pipm\pimp\pipm\piz\neu}\mump}}
\def\Bdtaumupi0{\decay{\Bd}{\subdecay{\taupm}{\pipm\pimp\pipm\piz\neu}\mump}}

\def\ux85 {\mbox{UX85}\xspace}

{

 \def\Pmu         {\ensuremath{\upmu}\xspace}                 
 \def\Pnu         {\ensuremath{\upnu}\xspace}                 
                  
 \def\Ppi         {\ensuremath{\uppi}\xspace}

 \def\Ptau        {\ensuremath{\uptau}\xspace}

\def\PDelta      {\ensuremath{\Delta}\xspace}                 
 \def\PXi      {\ensuremath{\Xi}\xspace}                 
 \def\PLambda      {\ensuremath{\Lambda}\xspace}                 
 \def\PSigma      {\ensuremath{\Sigma}\xspace}                 
 \def\POmega      {\ensuremath{\Omega}\xspace}                 
 \def\PUpsilon      {\ensuremath{\Upsilon}\xspace}                 
                  
  \def\PB      {\ensuremath{B}\xspace}                 
                  
 \def\PD      {\ensuremath{\mathrm{D}}\xspace}

 \def\PK      {\ensuremath{\mathrm{K}}\xspace}

 \def\Ps      {\ensuremath{\mathrm{s}}\xspace}

}
{

 \def\Pmu         {\ensuremath{\mu}\xspace}                 
 \def\Pnu         {\ensuremath{\nu}\xspace}                 
                  
 \def\Ppi         {\ensuremath{\pi}\xspace}

 \def\Ptau        {\ensuremath{\tau}\xspace}

 \mathchardef\PDelta="7101
 \mathchardef\PXi="7104
 \mathchardef\PLambda="7103
 \mathchardef\PSigma="7106
 \mathchardef\POmega="710A
 \mathchardef\PUpsilon="7107
                  
 \def\PB      {\ensuremath{B}\xspace}                 
                  
 \def\PD      {\ensuremath{D}\xspace}

 \def\PK      {\ensuremath{K}\xspace}

 \def\Ps      {\ensuremath{s}\xspace}

}

\def\muon       {\ensuremath{\Pmu}\xspace}

\def\tauon      {\ensuremath{\Ptau}\xspace}

\def\neu        {\ensuremath{\Pnu}\xspace}

\def\squark    {\ensuremath{\Ps}\xspace}

\def\pion  {\ensuremath{\Ppi}\xspace}
\def\piz   {\ensuremath{\pion^0}\xspace}

\def\pipm  {\ensuremath{\pion^\pm}\xspace}
\def\pimp  {\ensuremath{\pion^\mp}\xspace}

  \def\Kbar  {\kern 0.2em\overline{\kern -0.2em \PK}{}\xspace}

  \def\Dbar    {\kern 0.2em\overline{\kern -0.2em \PD}{}\xspace}

\def\B       {\ensuremath{\PB}\xspace}
\def\Bbar    {\ensuremath{\kern 0.18em\overline{\kern -0.18em \PB}{}}\xspace}

\def\Bd      {\ensuremath{B^0}\xspace}
\def\Bs      {\ensuremath{B^0_\squark}\xspace}

  \def\Y#1S{\ensuremath{\PUpsilon{(#1S)}}\xspace}

\def\Lbar {\ensuremath{\kern 0.1em\overline{\kern -0.1em\PLambda}}\xspace}

\newcommand{\decay}[2]{\ensuremath{#1\!\to #2}\xspace}          

\def\to{\ensuremath{\rightarrow}\xspace}

\newcommand{\ACP}{\ensuremath{A_{\rm CP}}\xspace}

\def\AFB      {\ensuremath{A_{\mathrm{FB}}}\xspace}

\def\AT#1     {\ensuremath{A_{\mathrm{T}}^{#1}}\xspace}

\def\C#1      {\ensuremath{\mathcal{C}_{#1}}\xspace}                        
\def\Cp#1     {\ensuremath{\mathcal{C}_{#1}^{'}}\xspace}                     
\def\Ceff#1   {\ensuremath{\mathcal{C}_{#1}^{\mathrm{(eff)}}}\xspace}         
\def\Cpeff#1  {\ensuremath{\mathcal{C}_{#1}^{'\mathrm{(eff)}}}\xspace}        
\def\Ope#1    {\ensuremath{\mathcal{O}_{#1}}\xspace}                        
\def\Opep#1   {\ensuremath{\mathcal{O}_{#1}^{'}}\xspace}

\newcommand{\tev}{\ifthenelse{\boolean{inbibliography}}{\ensuremath{~T\kern -0.05em eV}\xspace}{\ensuremath{\mathrm{\,Te\kern -0.1em V}}\xspace}}
\newcommand{\gev}{\ensuremath{\mathrm{\,Ge\kern -0.1em V}}\xspace}
\newcommand{\mev}{\ensuremath{\mathrm{\,Me\kern -0.1em V}}\xspace}
\newcommand{\kev}{\ensuremath{\mathrm{\,ke\kern -0.1em V}}\xspace}
\newcommand{\ev}{\ensuremath{\mathrm{\,e\kern -0.1em V}}\xspace}
\newcommand{\gevc}{\ensuremath{{\mathrm{\,Ge\kern -0.1em V\!/}c}}\xspace}
\newcommand{\mevc}{\ensuremath{{\mathrm{\,Me\kern -0.1em V\!/}c}}\xspace}
\newcommand{\gevcc}{\ensuremath{{\mathrm{\,Ge\kern -0.1em V\!/}c^2}}\xspace}
\newcommand{\gevgevcccc}{\ensuremath{{\mathrm{\,Ge\kern -0.1em V^2\!/}c^4}}\xspace}
\newcommand{\mevcc}{\ensuremath{{\mathrm{\,Me\kern -0.1em V\!/}c^2}}\xspace}

\def\gsim{{~\raise.15em\hbox{$>$}\kern-.85em
          \lower.35em\hbox{$\sim$}~}\xspace}
\def\lsim{{~\raise.15em\hbox{$<$}\kern-.85em
          \lower.35em\hbox{$\sim$}~}\xspace}

\def\tell1  {TELL1\xspace}
\def\ukl1   {UKL1\xspace}

\def\Bbar{\overline{B}}
\def\Kbar{\overline{K}}

\def\Lbar{\overline{L}}

\def\B{\mathcal{B}}
\def\C{\mathcal{C}}
\def\Re{\mathcal{R}e}

\def\SM{{\rm SM}}

\def\GeV{{\rm GeV}}
\def\MeV{{\rm MeV}}

\def\tl{{\ell^+\ell^-}}
\def\ee{{e^+e^-}}

\newcommand{\Da}{\ensuremath{\mathcal{T}}\xspace}

\newcommand{\res}{\ensuremath{\mathcal{O}}\xspace}

\newcommand{\hameff}{\ensuremath{\mathcal{H}_{\rm eff}}\xspace}

\newcommand{\cull}{\ensuremath{c\to u \tl}\xspace}

\title{Disentangling QCD and New Physics in \boldmath $D\to\pi\ell^+\ell^-$}

\author[1]{Aoife Bharucha,}
\author[2]{Diogo Boito,}
\author[3]{C\'{e}dric M\'{e}aux}
\affiliation[1]{Aix Marseille Univ, CNRS, CPT, Marseille, France, 13288 Marseille, France}
\affiliation[2]{Instituto de F\'{i}sica de S\~{a}o Carlos, Universidade de S\~{a}o Paulo, CP 369, 13560-970, S\~{a}o Carlos, SP, Brazil}
\affiliation[3]{CPPM Aix-Marseille Universit\'{e}, CNRS/IN2P3, Marseille, France}
\abstract{
In this paper we consider the decay \Dpill, addressing in particular the resonance contributions as well as the relatively large contributions from the weak annihilation diagrams. For the weak annihilation diagrams we include known results from QCD factorisation at low $q^2$ and at high $q^2$, adapting the existing calculation for $B$ decays in the Operator Product Expansion. The hadronic resonance contributions are obtained through a dispersion relation, modelling the spectral functions as towers of Regge-like resonances in each channel, as suggested by Shifman, imposing the partonic behaviour in the deep Euclidean. The parameters of the model are extracted using $e^+e^-\to{\rm (hadrons)}$ and $\tau\to {\rm (hadrons)}+\nu_\tau$ data as well as the branching ratios for the resonant decays $D^+\to\pi^+\,R\,(R\to\ell^+\ell^-)$, with $R=\rho$, $\omega$, and $\phi$. We perform a thorough error analysis, and present our results for the Standard Model differential branching ratio as a function of $q^2$. Focusing then on the observables \FH and \AFB, we consider the sensitivity of this channel to effects of physics beyond the Standard Model, both in a model independent way and for the case of leptoquarks.}

\emailAdd{aoife.bharucha@cpt.univ-mrs.fr}
\emailAdd{boito@ifsc.usp.br}

\date{\today}

\preprint{INT-PUB-20-048}
\begin{document}
\maketitle
\flushbottom
\section{Introduction}\label{sec:intro}
The anomalies in measurements of $b\to s$ transitions remain unresolved, and whether these can be interpreted via new particles, e.g.~leptoquarks, or an underestimation of uncertainties, i.e.~experimental errors or in the treatment of charm loops, remains uncertain.
If these are indeed a sign of physics beyond the Standard Model (BSM),
it is plausible that such physics could affect other flavour changing neutral current (FCNC) processes, amongst which the $c\to u$ transition is the least constrained (for a recent study of $K\to\pi\ell\ell^{(\prime)}$ see Ref.~\cite{Crivellin:2016vjc}).
At the same time LHCb is currently producing unprecedented numbers of $D$ mesons, the charm production cross-section, $\sigma(c\bar{c})=1200\,\,\mu$b~\cite{Aaij:2013mga}, exceeds the bottom production cross section of $\sigma(b\bar{b})=75\,\,\mu$b~\cite{Aaij:2010gn} by far.

The lack of constraints on the $c\to u$ transition is due to the relative difficulty with respect to $b\to s$ decays in making the necessary predictions. This is in part due to the reduced hierarchy between the charm mass and $\Lambda_{\rm QCD}$ compared to the corresponding hierarchy for $B$ decays, which makes expansions less effective. In addition, it is due to the fact that for $D$ decays the resonances affect a larger portion of the phase space. In the case of $B$ meson observables, the contributions due to the light resonances ($\rho^0, \omega^0$ and $\phi$) can be circumvented  
as the resulting effects in binned observable are negligible since the typical bin size is large compared to the width of the states. Nonetheless, polluting resonant effects due to the $c$-quark loop, are much larger and the kinematic regime where charmonium resonances are produced is ignored and often vetoed in the experimental analyses. However, in the case of $D$ decays the resonances affect a larger fraction of the available phase space, such that predictions are required in regions not sufficiently far from the resonance tails and modelling the structure of the hadronic resonances becomes crucial.

In a first approach to this problem in \Dpill decays, the resonances were added ``by hand", for example see Ref.~\cite{dBH15,FK2015}, by means of Breit-Wigner functions, on top of a non-resonant background described by the partonic result for the quark vacuum polarisation. In Ref.~\cite{FMS2017}, an alternative approach is advocated, where a subtracted dispersion relation is used to reconstruct the $s$- and $d$-quark vacuum polarisations from the respective imaginary parts, which must be modelled. In this approach, the partonic result is recovered asymptotically and the resonances are described following Regge trajectories with a number of simplifications, the main one being that the isospin 1 and isospin 0 channels are not treated separately, but in terms of a single tower of resonances with ``effective'' parameters.

We aim to improve upon the model of~\cite{FMS2017} by adapting the strategy applied to $B\to K\ell\ell$ in Ref.~\cite{LZ14}, where the authors investigate the charm resonance contribution, extracting the charm vacuum polarisation from  $e^+e^- \to {\rm (hadrons)}$ data by means of a dispersion relation and the optical theorem, an approach pioneered in the context of   $b\to s$ transitions in Ref.~\cite{Kruger:1996cv}. Here we extract the $u$, $d$ and $s$ vacuum polarisations from $e^+e^- \to {\rm (hadrons)}$ and  $\tau \to {\rm hadrons} +\nu_\tau$ data modelling the resonances present in the different channels with a Regge-based description similar to that of
~Ref. \cite{FMS2017}. The fact that in the low-energy region the three quark flavors are active makes the interplay between the different resonances more intricate and disentangling the contributions from the different quark flavours more challenging. The uncertainties on our final vacuum polarisations are sizeable; this stems from the uncertainties on the data sets and from the inherent limitations in modelling hadronic resonances. Given the challenges arising in obtaining a theoretical description of rare FCNC $D$ decays, it is necessary to remain conservative and seek observables which evade these limitations while being sensitive to beyond the Standard Model (BSM) physics.

The decay \Dpill has already been the subject of several studies, i.e. ~Refs.~\cite{Fajfer:2001sa,Fajfer:2005ke,Fajfer:2007dy, dBH15,FK2015,FMS2017,Bause:2019vpr,Bause:2020obd}, which focused on defining observables and determining the possible effects of BSM physics both model-independent and for specific BSM physics scenarios.
Here we propose to go beyond the existing work using the above-described novel treatment of resonances in combination with the treatment of the perturbative contribution including QCD factorisation (QCDf) corrections derived in Ref.~\cite{FMS2017} and the operator product expansion (OPE) weak annihilation corrections adapted from Ref.~\cite{Beylich:2011aq}. Note that this is the first phenomenological study of \Dpill where these QCDf and OPE corrections are implemented which, due to the large contribution of weak annihilation in this channel, should not be neglected. Note however that weak annihilation effects for $D\to V\gamma$ and three-body hadronic $D$ decays  have  been studied in the QCDf framework in Refs.~\cite{Adolph:2020ema,deBoer:2017que}.
Our results are further ameliorated by the recent advances on the form factors from the Lattice~\cite{Lubicz:2017syv,Lubicz:2018rfs} and Wilson coefficients from Ref.~\cite{Boer_WC} which we incorporate.
Our aim is to, using the theoretical framework just described, provide accurate and realistic predictions of the observables sensitive to BSM physics with little dependence on the non-perturbative physics, along with conservative error estimates. 

In this paper, we first lay out the theoretical framework necessary for the perturbative part of the calculation in Sec.~\ref{sec:theo_bkg}, most importantly introducing the operator basis and the calculation of the matrix element.
This is followed by a detailed description of our formalism for the resonant contribution in Sec.~\ref{sec:Resonances}, providing details of the model and the fit to the experimental data.
In Sec.~\ref{sec:pheno} we define a set of observables which are both sensitive to BSM physics and minimally affected by the hadronic uncertainties, and explore the related phenomenology both in the SM and for model-independent effects, keeping in mind the existing experimental constraints.
Our conclusions can be found in Sec.~\ref{sec:Conclusions}.

\section{Naive amplitude and annihilation contributions}\label{sec:theo_bkg}

In this section we present the framework for the calculation of the amplitude for \Dpill, ignoring the effect of resonances, which can be divided into a naively factorisable part and a part containing non-factorisable corrections, where by non-factorisable we mean that the diagram cannot be factorised into a perturbative part and a form factor.

\subsection{Theoretical framework}

	For $c\to u$ transitions, between the scale where the $W$ boson is integrated out ($\mu_W\sim M_W$) and the scale $\mu_b\sim m_b$, only operators $\res_1$ and $\res_2$ are present in the SM effective Hamiltonian. However, integrating out the $b$ quark the penguin operators $C_{3-9}$ are generated. We adopt the basis of operators and effective Hamiltonian used in Ref.~\cite{FK2015} and write:
	\beq
		\hameff^{\rm SM}(m_b>\mu>m_c) =  -\frac{4\,G_F} {\sqrt{2}} \sum_{q=d,b} \lambda_q \hameff^{(q)},
		\label{Eq:Heff_db}
	\eeq
	where $G_F$ is the Fermi constant, the combination of CKM matrix elements, $\lambda_q$, is given by $\lambda_q=V_{cq}^* V_{uq}$, and
	\begin{align}
    	\hameff^{(b)} = C_1 \res_1^s + C_2 \res_2^s + \sum_{i=3}^9 C_i \res_i,\qquad
		\hameff^{(d)} = C_1 (\res_1^s -\res_1^d) + C_2 (\res_2^s -\res_2^d).
	\end{align}
	However, note that since $\lambda_b \ll \lambda_d$
	, all contributions entering $\hameff^{(b)}$ are heavily CKM suppressed.  The complete set of operators used in this paper is  then composed of the current-current operators
  \begin{eqnarray}
		        \res_1^p =& (\bar{u}_{L}\gamma_\mu T^a p_L)(\bar{p}_L\gamma_\mu T^a c_{L}), \nonumber\hspace{2cm}
	        \res_2^p &= (\bar{u}_{L}\gamma_\mu p_L)(\bar{p}_L\gamma_\mu c_{L}),  \nonumber\\
	  \nonumber &&\\
	  \nonumber \mbox{where}& \,\, p=d,\mbox{ or }s,\mbox{ and the operators  }\, \mathcal{O}_{3-10},&\mbox{ which read:}  \\
	  	  \nonumber &&\\
            \nonumber   \res_3 =&(\bar{u}_{L}\gamma_\mu c_{L}) \sum_{p:m_p\leq\mu} (\bar{p}\gamma^\mu p),\hspace{2cm}	\res_4 &=(\bar{u}_{L}\gamma_\mu T^a c_{L}) \sum_{p:m_p\leq\mu}  (\bar{p}\gamma^\mu T^a p),\\
             \nonumber  \res_5 =&(\bar{u}_{L} \gamma_\mu \gamma_\nu \gamma_\rho c_{L})\sum_{p:m_p\leq\mu} (\bar{p}\gamma^\mu \gamma_\nu \gamma_\rho p),\quad\res_6 &=(\bar{u}_{L}\gamma_\mu \gamma_\nu \gamma_\rho T^a c_{L})\sum_{p:m_p\leq\mu}  (\bar{p}\gamma^\mu \gamma_\nu \gamma_\rho T^a p),\\
            \nonumber  \res_7 =&-\frac{g_{\rm e}}{16 \pi^2} m_{u}(\bar{u}_{L} \sigma^{\mu \nu} c_{R}) F_{\mu\nu},\hspace{2.2cm}\res_8 &=-\frac{g_s}{16 \pi^2} m_{u}(\bar{u}_{L} \sigma^{\mu \nu} T^a c_{R}) G_{\mu\nu}^a,\\
            \nonumber \res_9 =&-\frac{\alpha_e}{4 \pi} (\bar{u}_{L} \gamma^\mu c_{L}) (\bar{\ell} \gamma^\mu \ell),\hspace{2.7cm}\res_{10}&= -\frac{\alpha_e}{4 \pi}(\bar{u}_{L} \gamma^\mu c_{L}) (\bar{\ell} \gamma^\mu \gamma_5 \ell),
	\end{eqnarray}
	where $T^a$ are the $SU(3)_c$ generators, $q_{L/R}=(1 \mp \gamma_5)q/2$ denote the left/right-handed quark fields, $m_{u}$ is given in the \MSb ~scheme at the scale $\mu_c\sim m_c$ ,  $g_e=\sqrt{4\pi\alpha_e}$ is the electromagnetic coupling,  $\alpha_e$ the fine structure constant, and $F_{\mu\nu}$ and $ G_{\mu\nu}^a$ are, respectively, the electromagnetic and chromomagnetic field strength tensor. In the semileptonic operators, $\ell$ represents the lepton field.
	 
	To be exhaustive, the SM basis also contains the chirality-flipped operators $\res_i^\prime$ identical to the $\res_i$ up to the transformation $q_{L/R} \to q_{R/L}$. As the Wilson coefficients of these (and also that of $\res_{10}$) are negligible in the SM, these will only be included amongst the BSM contributions to the Hamiltonian. We calculate the Wilson coefficients in the SM at next-to-next-to-leading logarithmic order (NNLL) following Ref.~\cite{Boer_WC}, details can be found in appendix~\ref{App:WC}.

		This SM Hamiltonian for \Dpill can be extended to receive BSM dimension-six contributions, which give rise to additional Lorentz structures
		 
		\beq
		     \hameff = \hameff^{\rm SM} + \frac{4 G_F}{\sqrt{2}} \left( \sum \limits_{i=10,S,P,T,T5} C_i \res_i +  \sum \limits_{i=7,9,10,S,P} C_i^\prime \res_i^\prime \right),
		     \label{Eq:BSM_ham}
		\eeq
		 
    where the following operators are defined:
    \begin{align}
        \nonumber & \res_S^{(\prime)} = \frac{g_e^2}{16\pi^2}(\bar{c} u_{R(L)})(\bar{\ell} \ell), \qquad  \nonumber \res_P^{(\prime)} = \frac{g_e^2}{16\pi^2}(\bar{c} u_{R(L)})(\bar{\ell} \gamma_5 \ell), \\
        \nonumber & \res_T = \frac{g_e^2}{16\pi^2}(\bar{c}  \sigma_{\mu\nu} u)(\bar{\ell} \sigma^{\mu\nu} \ell), \qquad \res_{T5} = \frac{g_e^2}{16\pi^2}(\bar{c}  \sigma_{\mu\nu} u)(\bar{\ell} \sigma^{\mu\nu} \gamma_5 \ell).
        \end{align}
		Note that on comparing Eqs.~(\ref{Eq:BSM_ham}) and~(\ref{Eq:Heff_db}) we see that the CKM factors $\lambda_b$ and $\lambda_d$ have been absorbed into		 
		 the BSM Wilson coefficients, such that there is no assumption made about the flavour structure of the BSM physics.

		\begin{figure}[!t]

	  \begin{center}
	  \subfigure[]{
		\begin{tikzpicture}[scale=.8]
			\draw [line width = 0.7pt] (-2,0) -- (0,0);
            \draw [line width = 0.7pt] (0,0) -- (2,0);
            \node [scale=1.] {$\bigotimes$};
            \draw [fill=black] (-0.09,-0.09) rectangle (0.09,0.09);
            \node[scale=1.2, below=3mm] {$\mathcal{O}_7$};
		\end{tikzpicture}
		}
	  \qquad\qquad
	  \subfigure[]{
		\begin{tikzpicture}[scale=.8]
			\draw [ line width = 0.7pt] (-2,0) -- (0,0);
            \draw [ line width = 0.7pt] (0,0) -- (2,0);
            \node at (0,1.5) [scale=1.]  {$\bigotimes$};
            \draw [ line width = 0.7pt] (0,0) arc (-90:90:0.75cm) ;
            \draw [line width = 0.7pt] (0,0) arc (-90:-270:0.75cm) ;
            \draw [fill=black] (-0.09,-0.09) rectangle (0.09,0.09);
            \node[scale=1.2, below=3mm] {$\mathcal{O}_{1-6}$};
		\end{tikzpicture}
			}
		\qquad\qquad
	  \subfigure[]{
			\begin{tikzpicture}[scale=.8]
                	\draw [ line width = 0.7pt] (-1.5,-.8) -- (0,.2);
                	\draw [ line width = 0.7pt] (-1.5,1.2) -- (0,.2);
                	\draw [line width = 0.7pt] (1.5,-.8) -- (0,.2);
                	\draw [ line width = 0.7pt] (1.5,1.2) -- (0,.2);
                	\draw [fill=black] (-0.09,0.11) rectangle (0.09,0.29);
                    \node[scale=1.2, below=4mm] {$\mathcal{O}_{1-6}$};
                    \node at (-1,.9) [scale=1.]  {$\bigotimes$};
                    \node at (-1,-.5) [scale=1.]  {$\bigotimes$};
                    \node at (1,-.5) [scale=1.]  {$\bigotimes$};
                    \node at (1,.9) [scale=1.]  {$\bigotimes$};
			\end{tikzpicture}
	     }
	  \end{center}
			\caption{Leading contribution to $\langle \pi \tl|\hameff^{(q)}|D\rangle$ in an expansion in the strong coupling. The circled cross marks the  possible insertion of a virtual photon.}
			\label{Fig:BFS_LO}
	\end{figure}
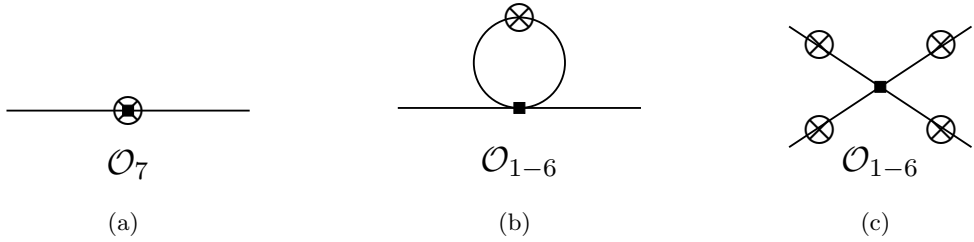

	In the SM, the leading contributions in an expansion in the strong coupling \als and $1/m_c$ for the \Dpill decay arise from:
		\begin{itemize}
		\itemsep0em
			\item interactions via the semileptonic operator $\res_{9}$,
			\item interactions where the charged lepton pair originates from a virtual photon $\gamma^*$ emitted via the EM dipole operator $\res_7$ (as shown in Fig.~\ref{Fig:BFS_LO}(a)) or via the 4-quark operators (as in Fig.~\ref{Fig:BFS_LO}(b)),
			\item weak annihilation as seen in Fig.~\ref{Fig:BFS_LO}(c).
		\end{itemize}
		In naive factorisation, only the first two are taken into account. Annihilation topologies fall under the set of so-called non-factorisable contributions. However, in contrast to the corresponding $B$ decays, the annihilation topology is not CKM suppressed, and therefore, as observed first in \cite{FMS2017}, the decay amplitude turns out to be dominated by non-factorisable dynamics.

		The closed fermion loop in Fig.~\ref{Fig:BFS_LO}(b) is calculable perturbatively as long as the $q\bar{q}$ pairs remain off-shell and for invariant dilepton masses away from hadronic resonances.

		In the region where the invariant dilepton mass squared $s$ corresponds to that of hadronic resonances, non-perturbative methods are required.

\subsection{The naive amplitude}\label{Sec:naive_amp}

	Based on the  effective Hamiltonian described above, we can write the amplitude $\mathcal{M}(D^+(p)\to\pi^+\ell^-\ell^+)$ as in Ref.~\cite{FK2015}:
	 
	\begin{align}
		\mathcal{M}(D^+(p)\to\pi^+\ell^-\ell^+) =& i\frac{G_F \alpha_e}{\sqrt{2}\pi} \big[ F_V p^\mu (\bar{\ell}\gamma_\mu\ell) + F_A p^\mu (\bar{\ell} \gamma_\mu \gamma_5 \ell)   \nonumber \\
		& +(F_S+ \cos{\theta} ~ F_T )(\bar{\ell}\ell) +(F_P+\cos{\theta}~F_{T5} )(\bar{\ell}\gamma_5\ell)  \big],
	\end{align}
	where the coefficients $F_A, F_S, F_P, F_T$, and $F_{T5}$ are functions of the dilepton invariant mass squared $s$, and are expressed in terms of the $D$ to $\pi$ transition form factors $f_0$, $f_+$ and $f_T$, defined in appendix~\ref{App:FF}, and the Wilson coefficients as
	\begin{align}
	\label{Eq:Fs}
        \nonumber F_V(s)=&  \left[ \lambda_b C_9^{(b)}(s) + \lambda_d C_9^{(d)}(s) \right]f_+(s) + \frac{8 m_l}{m_D +m_\pi}C_T\,f_T(s),\\
       \nonumber F_A(s)=& (C_{10} +C_{10}^\prime)~f_+(s),\\
        \nonumber F_S(s)=& \frac{m_D^2-m_\pi^2}{2m_c}(C_S+C_S^\prime)~f_0(s),\\
        \nonumber F_P(s)=& \frac{m_D^2-m_\pi^2}{2m_c}(C_P+C_P^\prime)~f_0(s) \\
        \nonumber &-m_\ell (C_{10}+C_{10}^\prime)\left( f_+(s)-\frac{m_D^2-m_\pi^2}{s} \left[ f_0(s)-f_+(s) \right] \right),\\
        \nonumber F_T(s)=& \frac{2\beta(s)\lambda(s)^{1/2}}{m_D+m_\pi}C_T\,f_T(s),\\
         F_{T5}(s)=&\frac{2\beta(s)\lambda(s)^{1/2}}{m_D+m_\pi}C_{T5}\,f_T(s).
	\end{align}
 The function $\beta(s)$ and the K\"{a}llen function $\lambda(s)$ entering the expression of the tensor functions $F_T(s)$ and $F_{T5}(s)$ are  given by
    	\begin{align}
    		\lambda(s)=(m_D^2 + m_\pi^2 + s)^2 -4(m_D^2 m_\pi^2 + m_D^2 s+ m_\pi^2 s)\quad\mbox{and}\quad\,
    		\beta(s) =\sqrt{1-4m_\ell^2/s}\,.
    		\label{Eq:Kinematic_func}
    	\end{align}

In the SM, the only non-vanishing structure is the contribution of $f_+$ to $F_V(s)$,
	 
	\beq
		F_V^{\rm SM}(s)=  \left[ \lambda_b C_9^{(b)}(s) + \lambda_d C_9^{(d)}(s) \right] f_+(s).
		\label{Eq:FV_SM}
	\eeq
where the functions  $C_9^{(b)}(s)$ and $C_9^{(d)}(s)$ can be expressed as~\cite{FMS2017}
	\beq
		C_9^{(q)}(s) = \delta^{qb} C_9 + \frac{2 m_c}{m_D}\frac{\Da^{(q)}(s)}{f_+(s)},
		\label{Eq:C9q}
	\eeq
in terms of the  ``hard kernels" $\Da^{(q)}$.
In naive factorization these hard kernels are given by
	\beq
		\Da^{(q)}(s)\bigr\rvert_{\rm Naive} = -f_+ (s)C^{(0,q)}(s),
		\label{Eq:GeneFF_naive}
	\eeq
	where the coefficient $C^{(0,q)}$ contains only factorizable and leading contributions:
	\beq
		C^{(0, q)}(s) = -\delta^{qb} C_7 -\frac{m_D}{2 m_c} Y^{(q)}(s).
		\label{Eq:Cq}
	\eeq

	The 1-loop functions $Y^{(q)}$ combine the contribution from the four-quark operators $\res_{1-6}$,
	\begin{align}
		Y^{(d)}(s) =& -  \frac{4}{9} \left( \frac{2}{3} C_1 +  \frac{1}{2} C_2 \right) \left[ \tilde h_s(s) - \tilde h_d(s) \right], \label{Eq:Yd} \\
		Y^{(b)}(s) =&  \frac{4}{9} \biggr[ \left( 7 C_3+ \frac{4}{3} C_4 + 76 C_5 + \frac{64}{3} C_6 \right) \left[ \tilde h_c(s) + \tilde h_u(s) \right]\nonumber \\
		&-\left( \frac{2}{3}C_1 + \frac{1}{2}C_2 + 3 C_3 +30 C_5 \right) \tilde h_s(s)\nonumber \\
		&-\left(3 C_3 + 30 C_5 \right)   \tilde h_d(s) + 2\left(3 C_3 + 16 C_5 + \frac{16}{3} C_6 \right) \biggr] ,  \label{Eq:Yb}
	\end{align}
	where $\tilde h_q(s)$ is the closed-quark loop function, discussed  in detail the next section.
	Note that our closed fermion loop $\tilde h_q(s)$ is related to that defined in \cite{FMS2017}, denoted $h(s,m_q)$, by $\tilde h_q(s) = \frac{9}{4}h(s,m_q)$.
 For completeness, we give the explicit expression of $\tilde h_q(s)$ in perturbation theory, $\tilde h^{(\rm pt)}(s)$, at leading order
	\beq
\tilde h_q^{(\rm pt)}(s)=-\ln \frac{m_q^2}{\mu^2}+\frac{2}{3}+\zeta-(2+\zeta)\sqrt{1-\zeta}\ln\left[\frac{1+\sqrt{1-\zeta}}{\sqrt{-\zeta}}\right] ,\label{hpt}
\eeq
where $m_q$ is the mass of the quark $q$ and $\zeta=4 m_q^2/(s+i\epsilon)$. The quark-loop functions $\tilde h(s,m_q)$ are needed in the resonance region. The perturbative description of these functions does not include any long-distance hadronic effects. In the next section, we will include these by reconstructing the functions $\tilde h(s,m_q)$ from their imaginary part by means of a dispersion relation, where the imaginary part is obtained by constraining the parameters of a hadronic model for the spectral functions~\cite{FMS2017}.
	Finally, we remark that there is a partial cancellation in the combination of Wilson coefficients that participate in Eq.~(\ref{Eq:Yd}) which leads to an accidental suppression of the LO contribution.

 	Combining Eqs.~(\ref{Eq:C9q}) and (\ref{Eq:GeneFF_naive}) and (\ref{Eq:Cq}), the functions \Cn in naive factorisation, denoted \Cnnaive, can be summarised by
 	\begin{align}
 		C_9^{(d)}(s)\bigr\rvert_{\rm Naive} &= Y^{(d)}(s), \\
 		C_9^{(b)}(s)\bigr\rvert_{\rm Naive} &=  C_9 +Y^{(b)}(s)+ \frac{2m_c}{m_D+m_\pi}C_7 \frac{f_T(s)}{f_+(s)}.
 		\label{Eq:C9naive}
 	\end{align}
	As discussed previously, the naive result is not sufficient, primarily due to the weak annihilation diagrams. We will now consider corrections, calculated within  
	QCD factorisation in the low $q^2$ region and in the OPE framework in the high $q^2$ region (above the $\phi$ pole).

\subsection{Annihilation contributions} \label{Sec:Anni}

The expressions for the QCDf corrections were calculated in \cite{FMS2017}, where the authors adapted expressions for the $B\to K^{*}\ell^+\ell^-$ from Ref.~\cite{BFS2001} to the case of $D\to \rho(\pi)\ell^+\ell^-$.
		Following \cite{BFS2001}, in QCDf the non-factorisable contributions can be classified under two categories:
	\begin{itemize}
		\item A first category where the spectator quark participates in the hard scattering; annihilation topologies enter in this category. Here the spectator quark participates in the FCNC process via hard gluon exchange. The calculation of these processes leads to the so-called \emph{hard spectator scattering} corrections.
		\item A second category where the spectator quark is connected to the hard process only through soft interactions and the hadronic transition can be described by the form factors. The calculation of which leads to the so-called \emph{form-factor} corrections.
	\end{itemize}
		 QCDf is applicable in the combined heavy-quark and large-energy (recoil) limit, where energy $E$ refers to that of the final state meson, related to the dilepton invariant mass $s$ via:
	\beq
		E = \frac{m_{D}^2 + m_{\pi}^2 - s}{2 m_{D}} \to ~ \sim\frac{m_D}{2}.
		\label{Eq:Def_E}
	\eeq
 In QCDf the decay amplitude is schematically expressed as
	\beq
		\langle \pi \tl|\hameff^{(q)}|D\rangle \sim C^{(q)}f(s) + \phi_D^\pm \otimes T^{(q)} \otimes \phi_\pi + \res(\frac{1}{m_c}).
		\label{Eq:Schem_QCDF}
	\eeq
	The non-factorisable corrections enter the coefficients $C^{(q)}$ as well as the second term made of a convolution product between the so-called hard kernel $T^{(q)}$ where $\phi_D^\pm$ is the LCDA of the $D$ meson. More explicitly, the naive factorisation results in Eq.(\ref{Eq:GeneFF_naive}) can be extended by
	\beq
		\Da^{(q)}(s) = -C^{(q)} f_+(s) + \frac{\pi^2}{N_c}\frac{f_{D} f_{\pi}}{m_{D}} \sum_\pm \int \frac{d\omega}{\omega} ~\phi_{D, \pm} (\omega) \int_0^1 du~ \phi_{\pi}(u) ~T^{(q)}_\pm(u, \omega).
		\label{Eq:GeneFF_QCDF}
	\eeq
	where the $\pm$ subscript refers to the projection of the amplitude on the $D$ meson LCDA. The perturbative quantities $C^{(q)}$ and $T_\pm^{(q)}$ are given by
	\begin{align}
		C^{(q)} &= C^{(0, q)} + \as C_F C^{(1, q)}_\pm, \nonumber \\
		T_\pm^{(q)} &= T^{(0, q)}_\pm + \as C_F T^{(1, q)}_\pm,
		\label{Eq:Spec_bd}
	\end{align}
	where we remind the reader that $\as\equiv\als/(4\pi)$ and $C^{(0, q)}$ is defined in Eq.~(\ref{Eq:Cq}). The form-factor corrections are contained in $C^{(1, q)}$, the annihilation corrections in $T^{(0, q)}$ and the hard spectator scattering corrections in $T^{(1, q)}$, expressions for all these quantities can be found in Ref.~\cite{FMS2017}.
	Finally, in the QCDf framework the function \Cn defined in Eq.~(\ref{Eq:C9q}) can be written as:
	\beq
	    \Cn = \Cnnaive + \CnAnn + \CnSS +  \CnFF,
	    \label{Eq:C9QCDf}
	\eeq
	where \Cnnaive has been defined in Eq.~(\ref{Eq:C9naive}) and \CnAnn, \CnSS, \CnFF are the annihilation, the spectator scattering, and the form-factor corrections to the \Cn functions, respectively. Note that unlike Ref.~\cite{FMS2017} we have chosen to use the pole mass for the charm quark, and have therefore set the quantity $\Delta M$ accordingly the definition of which can be found in Ref.~\cite{Beneke:2004dp}.  We have studied these sets of QCDf corrections numerically in Sec.~\ref{Sec:results_SM} (see Tab.~\ref{Tab:Num_contrib}, and find that, in agreement with \cite{FMS2017}, the only one numerically relevant to our analysis, given the size of the theoretical errors, is \CnAnn. We will therefore discuss these annihilation corrections in more detail.

	The four annihilation diagrams  shown in Fig.~\ref{Fig:BFS_LO}(c) contribute at different powers in the $1/m_c$ expansion.
	With the convention that the $\pi^+$ meson momentum is nearly light-like in the minus light-cone direction, the amplitude for the surviving contributions depend only on the minus component and $T^{(0)}_+ = 0$.
	The result depends on the charge factor of the spectator quark, in our case $e_d=-1/3$,
		\begin{align}
			T^{(0,b)}_-(\omega) &= \frac{e_d}{m_c} \frac{4 m_D \omega}{\omega - s/m_D - i\epsilon} \left[-C_3 - \frac{4}{3}(C_4+12 C_5+16 C_6) \right],\\
			T^{(0,d)}_-(\omega) &= \frac{e_d}{m_c} \frac{4 m_D \omega}{\omega - s/m_D - i\epsilon} ~ 3C_2.
		\end{align}
	$T^{(0,d)}_-$ being $u$-independent, corrections to \Cn are then given by
		\begin{align}
		C_9^{(d)}(s)\bigr\rvert_{\rm Ann} &= 8  e_d \frac{\pi^2}{N_c}\frac{f_D f_\pi}{m_D}\frac{1}{f_+(s)}\frac{1}{\lambda_D^-(s)} ~3 C_2, \label{Eq:CndAnn}\\
			C_9^{(b)}(s)\bigr\rvert_{\rm Ann} &= 8 e_d \frac{\pi^2}{N_c}\frac{f_D f_\pi}{m_D}\frac{1}{f_+(s)}\frac{1}{\lambda_D^-(s)} \left[-C_3-\frac{4}{3}(C_4+12 C_5+16 C_6) \right],
		\end{align}
	with the $s$-dependent moment $\lambda_D^-(s)$  given by
		\beq
			\frac{1}{\lambda_D^-(s)} = \int_0^\infty d\omega \frac{\phi_D^-(\omega)}{\omega-s/m_D - i \epsilon}.\label{Eq:lDpartonic}
		\eeq
	We note that \CnAnn leads to a sizeable contribution as $C_2$ appears without any cancellation from other Wilson coefficients and, as a result, it turns out that annihilation gives very large contributions to the decay rate. It should also be mentioned that Eq.~(\ref{Eq:lDpartonic}) is purely partonic and does not include any effect due to the resonance spectrum. In our final results we will modify $\lambda_D^{-}(s)$ to include those effects employing the ansatz of Ref.~\cite{FMS2017}. We remark that the calculation of the annihilation contribution could also be performed in light-cone sum rules as applied to the $B$-decay case in Ref.~\cite{Lyon:2012fk,Lyon:2013gba}. We have verified that the two results are approximately compatible. Given the dominance of and the large uncertainties on the contribution of the hadronic resonances, we don't expect significant changes in the results should the LCSR framework be adopted for this contribution.

    The QCD factorisation framework is valid at small $q^2$, as here the pion is energetic, and in the heavy-quark limit. To be precise, the energy of the pion $E_\pi$ should be large compared to $\Lambda_{\rm QCD}$.  We would also like to be able to make predictions at large $q^2$, away from the dominant hadronic resonances, where the theoretical uncertainties are  under better control. Since the QCDf weak annihilation provides such a large contribution in the low-$q^2$ region one wonders whether this is  also true at high $q^2$ and how to rigorously estimate these contributions in this regime.
         Several papers exist which tackle this issue for $B\to K^{(*)} \ell\ell$ transitions~\cite{Grinstein:2004vb, Beylich:2011aq}. Here we follow Ref.~\cite{Beylich:2011aq} where one can perform an operator product expansion (OPE) exploiting the large size of $q^2$ compared to the pion energy ($\sqrt{q^2}\gg E_\pi,\Lambda_{\rm QCD}$) to expand the amplitude in powers of $E_\pi/\sqrt{q^2}$. The amplitude can then be easily factorised into the form factor and a coefficient which can be calculated perturbatively to order $\alpha_s$. The results obtained in Ref.~\cite{Beylich:2011aq} for the weak annihilation contribution (which arises at leading order in  $\alpha_s$)  to $C_9$ can easily be adapted to the case of \Dpill, and we find
    \beq
        C_9^{(d)}(s)\bigr\rvert_{\rm Ann}^{\rm OPE}= -1/3 \frac{8 \pi^2  C_2  f_D f_\pi}{s f_+(s)}.\label{eq:WAOPE}
    \eeq
   Being proportional to $C_2$, as opposed to $C_4+C_3/3$ for the case of $B\to K\ell\ell$, annihilation also dominates in the high-$q^2$ regime (we neglect the other contributions which are Cabibbo suppressed).

    In our implementation, the QCDf and OPE corrections must be adopted only in the appropriate range in $q^2$, i.e.~ for QCDf $E_\pi\gg\Lambda_{\rm QCD}$ and for the OPE $\sqrt{q^2}\gg E_\pi,\Lambda_{\rm QCD}$. These conditions can be satisfied if we use the QCDf corrections only up to the $\phi$ pole, and for higher energies  the corrections in the OPE framework. Note that the latter are not valid up to the lowest recoil point, and therefore in our phenomenological analysis we only consider $q^2$ up to~2.3 GeV$^2$.

\section{Hadronic resonances}\label{sec:Resonances}
As discussed in the previous section, in \Dpill decays, contributions with a closed quark-loop arise at leading
order,  where the $d$- and
$s$-quark vacuum polarisations intervene. These contributions are encoded in  the $\tilde h(s,m_q)$ functions of Eqs.~(\ref{Eq:Yd}) and~(\ref{Eq:Yb}). Here we discuss in detail how we obtain a description for these functions beyond perturbation theory, including effects of the hadronic resonances.

  Due to phase space restrictions, any realistic phenomenological study of  \Dpill decays will require a description of resonance effects. The resonances must be included, ideally, without
losing contact with the  first-principles calculation.
In
Ref.~\cite{dBH15}, the resonances were added by means of
Breit-Wigner functions, on top of a non-resonant background described
by the partonic result for the quark vacuum polarisation (at leading order).
A somewhat more sophisticated approach was advanced in Ref.~\cite{FMS2017}
in which the vacuum polarisations are modelled following
Ref.~\cite{Shifman:2000jv}. In this approach, the partonic result
is recovered asymptotically, away from the resonant region, and the resonances are modelled following
Regge trajectories with a number of simplifications, the main one
being that the isospin 0 and isospin 1 channels are not treated
separately, but in terms a single tower of resonances with
``effective'' parameters~\cite{FMS2017}.

Here, we improve on the description of Ref.~\cite{FMS2017} by implementing
a strategy similar to that of Ref.~\cite{LZ14}, where the charm
vacuum polarisation was extracted from  $e^+e^- \to {\rm (hadrons)}$ data by means of a dispersion relation.
The general framework is similar to Ref.~\cite{LZ14}, but the fact that we are dealing with
light quarks leads to a number of complications. Below the charm threshold, quarks $u$, $d$ and $s$
contribute to the vacuum polarisation probed in $e^+e^- \to {\rm (hadrons)}$, with several resonances that
interfere.  In the case of charm, however, one can safely subtract the perturbative background arising from the light-quark
contributions and work with data that are ``pure" $c\bar c$. Here, in order to be able to disentangle the different channels better
we will  make use of $\tau \to ({\rm hadrons})+\nu_\tau$ data as well,  which are pure isospin 1, as opposed to the $e^+e^-\to \gamma^* \to ({\rm hadrons})$ data which are a mixture of isospin 0 and 1.

However, no  treatment of the resonances in the vacuum polarisations needed for the description of $D \to \pi \ell^+
\ell^-$ is completely sound from a theoretical
perspective. 
The reconstruction of the vacuum polarisation functions using a dispersion relation requires knowledge about their imaginary parts from threshold to infinity. This description cannot be done within a unique framework from the low-$q^2$ region, where chiral symmetry plays a role, up to 
the high-$q^2$ regime, where perturbative QCD is the main 
contribution, passing by resonance peaks and residual 
resonance oscillations.
In the end, non-factorisable corrections will 
also play a role and the quark-loop functions that appear in the  \Dpill amplitude are expected to be 
modified with respect to the description of the quark vacuum polarisations. The different treatments of the resonant 
contribution should serve,
in the end, to obtain a reasonable estimate of their effect, with a conservative error band, in order to 
identify regions of the spectrum where room for BSM physics still exist and where theory is under control.
With these caveats in mind, we discuss now how the framework
of Ref.~\cite{LZ14}, together with the model from Refs.~\cite{FMS2017,Shifman:2000jv}, can be adapted to our case.

In the calculation of \Dpill one needs the $d$ and $s$ quark vacuum polarisation
functions, the imaginary parts of which intervene in the observable
\beq
R(q^2) =\frac{3s}{4\pi\alpha_e^2}\sigma(e^+e^- \to {\rm hadrons})\simeq\frac{\sigma(e^+e^- \to {\rm hadrons})}{\sigma(e^+e^- \to \mu^+\mu^-)},
\eeq
where the second (approximate) equality is valid
at LO and for values of $s$ for which the muon mass can safely be neglected. 
Below the charm threshold, experimental data for $R(q^2)$ contains the entangled contributions from  $u$, $d$ and $s$ quarks.

On the theory side, $R(q^2)$ can be written in terms of the  imaginary part of the $\bar q q$ contributions to the photon vacuum polarisation as
\beq
R(q^2) =  12\pi\,{\rm Im}\Pi(q^2),\label{RPi}
\eeq
where the vacuum polarisation is defined as
\beq
\Pi_{\mu\nu}(q^2)= i\! \int d^4x\, e^{i x\cdot q}\sand{0}{T\{J^{\rm EM}_\mu(x) J^{\rm EM}_\nu(0)\}}{0} = (q_\mu q_\nu - g_{\mu\nu}q^2)\Pi(q^2),\label{eq:PiEM}
\eeq
with $s=q^2$ and the electromagnetic current $J_\mu^{\rm EM}$ given by
\beq
J^{\rm EM} _\mu=Q_u(\bar u\gamma_\mu u)+Q_d(\bar d\gamma_\mu d)+Q_s(\bar s\gamma_\mu s)\label{EMcurrent}
\eeq
(where we omitted the sum over colour indices). Henceforth we will
disregard the mixed quark flavour contribution which can occur through disconnected diagrams that are $1/N_c$ suppressed~\cite{Boito:2018yvl}. Below charm threshold, we  then write the $R$ ratio as
the sum of the three quark-flavour contributions $R = \sum_q R_q =R_u+R_d+R_s$.

We then define the correlators $\Pi^{(q)}(q^2)$ in terms of the flavour-singlet currents $j^{(q)}_\mu=\bar q\gamma_\mu q$,
in complete analogy with Eq.~(\ref{eq:PiEM}).
To make contact with Ref.~\cite{FMS2017,LZ14} it is convenient to define
the  function $\tilde h(q^2)$ as
\beq
\tilde h_q(q^2) = \frac{12\pi^2}{N_c}\Pi^{(q)}(q^2).
\eeq
At sufficiently high momenta in the deep Euclidean region,  $q^2\to -\infty$,  we have, after normalisation,
\beq
\tilde h_q(q^2) \underset{q^2\to -\infty}{\longrightarrow} - \log\left(-\frac{q^2}{\mu^2}\right)   +  ({\rm constant}).
\eeq
Upon analytic continuation one obtains, for $q^2>0$, in the chiral limit, ${\rm Im}\, \tilde h_q(q^2) =\pi$.

For the purpose of modelling the resonances it is appropriate to
consider the different isospin channels separately. The
electromagnetic current of Eq.~(\ref{EMcurrent}) contains an $I=1$ and
 $I=0$ light-quark part, as well as the strange-quark contribution,
the respective dominant vector resonances being the $\rho(770)$, the
$\omega(782)$, and the $\phi(1020)$. Taking into account the quark-charge
factors, the electromagnetic current can be written as
\beq
J^{\rm EM} _\mu= \frac{1}{2}\left( \bar u \gamma^\mu u -\bar d \gamma^\mu d \right) +\frac{1}{6}\left( \bar u \gamma^\mu u+\bar d \gamma^\mu d \right) -\frac{1}{3}\bar s \gamma_\mu s,
\eeq
where we made the $I=1$ and $I=0$ light-quark contributions  explicit in the
first and second terms in between parenthesis on the right-hand side, respectively, where $J^\mu_{1/0}=1/\sqrt{2}(\bar u\gamma^\mu u\mp \bar d\gamma^\mu d)$.
For the $R$ ratio one has then\footnote{For related discussions see Ref.~\cite{Boito:2018yvl,Daub:2015xja}}
\begin{align}
  R_{uds} &=  N_c\left(\frac{1}{2}\frac{{\rm Im} \tilde h_{1}(q^2)}{\pi} + \frac{1}{18}\frac{{\rm Im} \tilde h_{0}(q^2)}{\pi} + \frac{1}{9}\frac{{\rm Im} \tilde h_{s}(q^2)}{\pi}  \right),\label{Ruds}
\end{align}
where $\tilde h_{1/0}(q^2)$ represent the light-quark $I=1$ and $I=0$
contributions and $\tilde h_s(q^2)$ is the contribution from the
strange quark. The imaginary part of these functions are proportional to the
respective spectral functions.  In the case of the    $I=1$ current,
additional experimental data for the spectral function exists from   $\tau \to ({\rm hadrons})+\nu_\tau$ decays~\cite{Davier:2013sfa,Ackerstaff:1998yj,Boito:2012cr}.
We will use these data sets as a way to help disentangling the different contributions in $R(q^2)$.
    Ultimately, in the application to \Dpill, within our assumptions,  we need
the functions $\tilde h_d(q^2)$ and $\tilde h_s(q^2)$. The former can be
obtained from  $\tilde h_{1/0}(q^2)$,  which contribute
equally to the $d$-quark vacuum polarisation.

We then need  a concrete model for the imaginary part of the different
$\tilde h_I(q^2)$ functions of Eq.~(\ref{Ruds}), with $I=1,0$ or $s$. Our description is based on
the model suggested in App.~B of Ref.~\cite{FMS2017} which, in turn, is
based on a proposal by Shifman~\cite{Shifman:2000jv}. The model can be
summarised as follows. The imaginary part of each channel is modelled
with a dominant vector resonance plus the sum of an infinite tower of
resonances, starting from the first excitation, with masses following
Regge trajectories. The dominant resonance in each channel  is modelled by a Breit-Wigner  function, $f_{\rm BW}^{(R)}$, while
the  sum over the infinite tower of resonances
is performed analytically within Shifman's model~\cite{Shifman:2000jv}. Concretely, after summing over  all the excited states,  we have
\beq
{\rm Im}h_I(q^2) =  {\rm Im}f_{\rm BW}^{(R)}(q^2) - {\rm Im}\left[\frac{\Psi(z_I+a_I)}{1-b_I/\pi}\right]. \label{Imhmodel}
\eeq
The first term on the r.h.s. corresponds to the dominant vector resonance and is discussed below,
the second term represents the sum
over  the infinite tower of equally spaced resonances with masses
\beq
M_I^2(n)= (n+a_{I})\sigma_I^2 \label{MassRegge}
\eeq and widths
\beq
\Gamma_I(n)= b_I\,M_I(n). \label{WidthRegge}
\eeq  In Eq.~(\ref{Imhmodel}), $\Psi(z)$ is the
digamma function and
\beq
z_I=\left(\frac{-q^2-i\epsilon}{\sigma_I^2}\right)^{1-b_I/\pi},
\eeq
for $I=1,0$ while, for the strange quark,
\beq
z_s=\left(\frac{4\,m_K^2-q^2-i\epsilon}{\sigma_s^2}\right)^{1-b_s/\pi}.
\eeq
(We are treating the light quarks and the pions as massless.)
The term with the $\Psi$ function corresponds to the  description of Ref.~\cite{Shifman:2000jv}, reviewed in \cite{FMS2017}, to which we refer for more  details about the model. The description of the tower of resonances adds three parameters per channel to the model ($\sigma_i$, $a_i$, and $b_i$), therefore nine in total.

The dominant resonances are the $\rho(770)$ in the $I=1$ channel,  the $\omega(782)$ in $I=0$, and the $\phi(1020)$ in the $s\bar  s$ channel. For the description of the leading resonances we use the following Breit-Wigner function
\beq
f_{\rm BW}^{(R)} = n_{R}e^{i\alpha_R} \frac{M_{R}^2}{M_{R}^2-q^2 -i \sqrt{q^2}\,\Gamma_t},\label{BW}
\eeq
where $M_R$ and $\Gamma_t$ are the Breit-Wigner parameters related to the mass and  the total width. The phase $\alpha_{\rho(770)}$ is taken to be zero and is used as a reference for the phases of the $\omega(782)$ and $\phi(1020)$.
 The imaginary part in the denominator of the $f_{\rm BW}^{\phi(1020)}$ is multiplied by $\Theta(q^2-4m_K^2)$ 
 to account for the non-zero kaon mass.

  We fix the parameters of our model from a comparison to $e^+e^- \to
 {\rm (hadrons)}$ and $\tau \to {\rm (hadrons)} + \nu_\tau$ data --- a strategy similar to the one of Ref.~\cite{LZ14}. It is known
 that models related to the one we are employing here agree well with the data in
 an asymptotic region~\cite{Boito:2012cr,Boito:2014sta,Boito:2018yvl}, where $q^2$ is large enough (in practice this means $q^2 \gtrsim 1.5$~GeV$^2$, but this value is channel dependent). In the case of the $\tau$ data, which is pure $I=1$, this type of description has also been extended to lower energies with the inclusion of  a
 Breit-Wigner for  the $\rho(770)$~\cite{Shifman:2000jv}.

Here we use the publicly available Particle Data Group
compilation of $R(q^2)$ data~\cite{PDG2018} supplemented with  $R(q^2)$
measurements from the BES and KEDR collaborations published recently in
Refs.~\cite{Ablikim:2009ad,Anashin:2015woa,Anashin:2016hmv}.  No correlations are publicly available for these data, and
the data are therefore treated as uncorrelated.  These data are a mixture  of the three channels $I=1$, $I=0$, and $s\bar s$ as described in Eq.~(\ref{Ruds}).
To better disentangle the three channels, we also use the ALEPH~\cite{Davier:2013sfa} and OPAL~\cite{Ackerstaff:1998yj} data for the vector-isovector spectral function
from $\tau\to {(\rm hadrons) +\nu_\tau}$. In the case of OPAL data we use the  updated version of Ref.~\cite{Boito:2012cr}.
These data sets are also treated as uncorrelated, to be fully coherent with our treatment of the data for $R(q^2)$.

\begin{figure}[!t]
\begin{center}
\includegraphics[width=.9\columnwidth,angle=0]{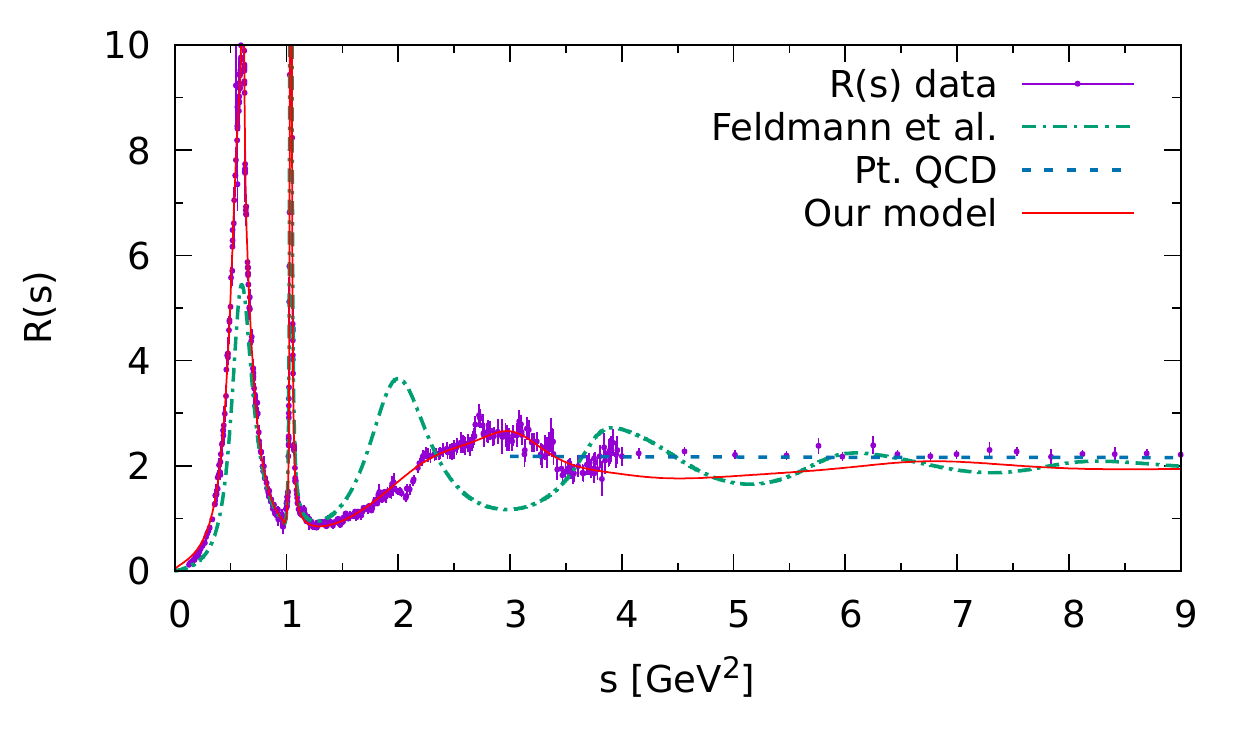}
\caption{$R(q^2)$ described by Eq.~(\ref{Ruds}) with parameters given in Tab.~\ref{Tab:fitparms} (solid line). Results from the model of Ref.~\cite{FMS2017} are also shown for comparison (dash-dotted line). The perturbative QCD result at four loops with $N_f=3$ in the chiral limit is also shown (dashed line). }
\label{RudsFig}
\end{center}
\end{figure}

\begin{figure}[!t]
\begin{center}
\includegraphics[width=.9\columnwidth,angle=0]{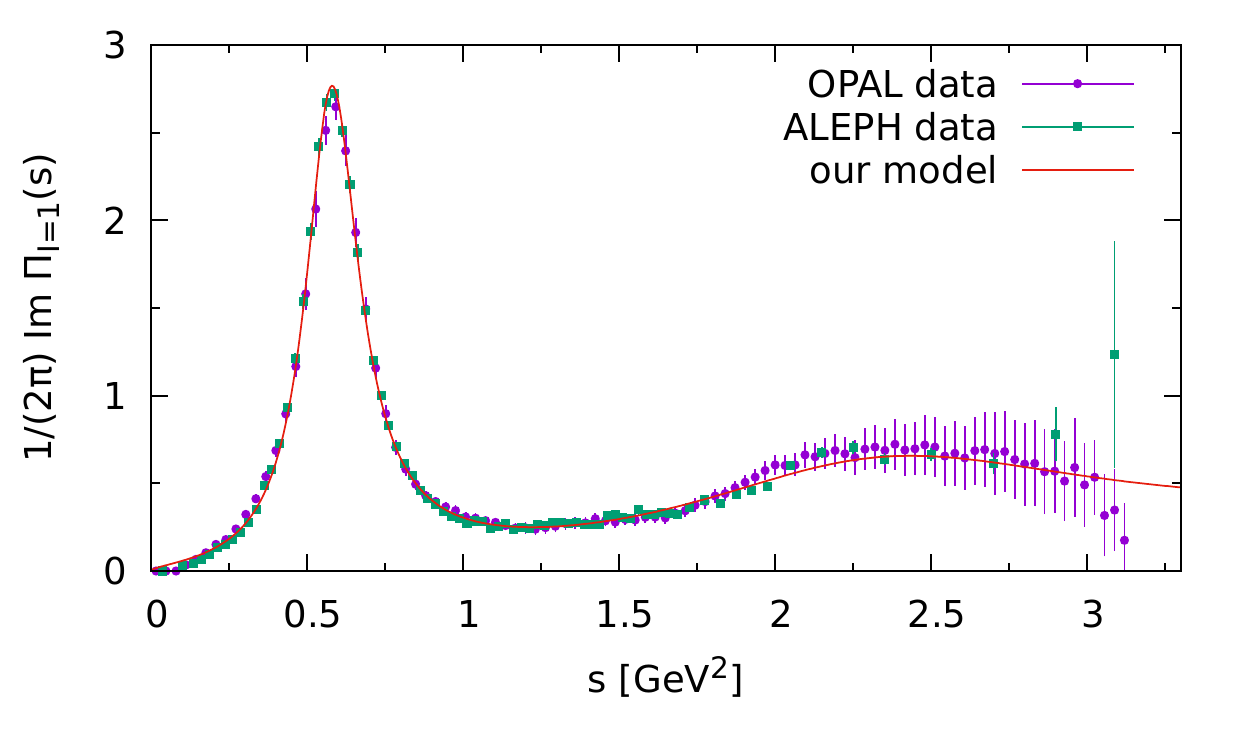}
\caption{Vector isovector spectral function from hadronic tau decay data~\cite{Ackerstaff:1998yj,Davier:2013sfa,Boito:2012cr} compared with our model with parameters given in Tab.~\ref{Tab:fitparms}. }
\label{TauFig}
\end{center}
\end{figure}

Since the number of free parameters in the model is quite large\footnote{In total there are 20 parameters: three masses, three widths, and three normalisations in each channel, plus nine parameters (3 per channel) to describe the Regge towers, as well as  two phases. } and the interplay between the different contributions to $R(q^2)$ is intricate, we need to make a few assumptions in order to fix all the parameters in our description. The isospin 1 parameters $a_1$ and $\sigma_1$  from Eq.~(\ref{MassRegge}) are fixed, from a fit to the $\tau$-decay data, to be $\sigma^2_1= 2.476$~GeV$^2$ and $a_1=0.974$, which is in the ballpark of values expected from Regge behaviour  (2~GeV$^2$ and 1.0, respectively~\cite{FMS2017}).   We also fix the parameter $b_{0}$, related to the widths of the resonances in the $I=0$ channel, to be $b_0 = 0.2 \approx \frac{\Gamma_{\omega(1640)}}{M_{\omega(1640)}}$. Additionally, we  assume that $\sigma_1^2 =\sigma_0^2$ --- which amounts to the assumption that the masses of higher resonances with $I=1$ and $I=0$ follow a similar pattern.
We then build an uncorrelated  $\chi^2$ function from the $R(q^2)$ and $\tau$ data combined. We use all the $\tau$-decay data available, which gives 174 data points. From the $R(q^2)$ data we exclude the  low-energy data below 0.3~GeV$^2$ since the use of our Breit-Wigner functions is not fully reliable at such low energies. We also exclude 8 data points around 2.1~GeV$^2$ where the data shows a fluctuation downwards, clearly visible in Fig.~\ref{RudsFig}, which cannot be described within our model. The inclusion of these data points do not change the values of the parameters significantly, but leads to a worse $\chi^2$ value. In total we use 637 points from the $R(q^2)$ data set, with the highest energy bin being at 4.0~GeV$^2$. Little information is added if one includes data beyond this point (mainly from BES and KEDR), since data points become scarce.  The fit gives then the value $\chi^2_{\rm min}=828.1$ for
795 degrees of freedom. The resulting parameters from this minimisation are shown in Tab.~\ref{Tab:fitparms}. In Figs.~\ref{RudsFig} and~\ref{TauFig} we compare the results of the fit with the $R(q^2)$ and $\tau$-decay data.
The errors given in our Tab.~\ref{Tab:fitparms} are quite small
for some of the parameters. However, to accommodate further
non-factorisable effects, in our practical use of these
results we will allow for a variation of the resonance phases, $\alpha_R$ of Eq.~(\ref{BW}). In the end, this variation will be one of the main sources of error in the description of  resonances  in \Dpill decays.

Our model is sufficient to achieve a reasonably good representation of the data. One should note that, at higher energies, the model is systematically below the data due to the lack of perturbative corrections, which are of the order of~15\%.
With this description, we  obtain the imaginary parts of the three  non-perturbative functions $\tilde h_{I}(q^2)$.

 From the imaginary part of $\tilde h_I(q^2)$ we can
reconstruct the full function using a dispersion relation.  We follow the suggestion of Ref.~\cite{FMS2017} and use a once-subtracted dispersion relation with the
subtraction constant fixed from the perturbative
description. Accordingly, the subtraction point is chosen in the deep
Euclidean at $q^2=-s_0$ and the functions $\tilde h_I(q^2)$ are given by
\beq
\tilde h_I(q^2) = \tilde h_I^{(\rm pt)}(-s_0) + \frac{1}{\pi}\int_0^\infty ds' \frac{s_0 + q^2}{s_0 + s'} \frac{{\rm Im}\, \tilde h_I(s')}{s'-q^2-i \epsilon}.\label{DispRel}
\eeq
The subtraction constant is calculated from the perturbative description  (without $\alpha_s$ corrections)
which is given by Eq.~(\ref{hpt}).
(For the light-quark contributions we  take Eq.~(\ref{hpt}) in the chiral limit.) When reconstructing
the functions $\tilde h_q(q^2)$ we have checked that  using a dispersion relation with more subtractions leads to very similar results. We have also  checked that the results are stable upon variation of the subtraction point in Eq.~(\ref{DispRel}). For our final results we use $s_0 = 10$~GeV$^2$ and $\mu^2=(1.5~{\rm GeV})^2$.

\begin{table}[!ht]
\begin{center}{
\begin{tabular}{c|c|c}
\toprule
Parameter & Central value &  Relative error  \\
\midrule
\quad $n_\rho$ & 3.070 & 0.24\%  \quad\\
\quad $m_\rho$ (GeV) & 0.7653&0.034\%  \quad\\
\quad $\Gamma_\rho$ (GeV)  & 0.1374& 0.40\%  \quad\\
\quad $b_{I=1}$ & 0.323&1.2\% \quad\\
\quad $\sigma^2_{I=1}$  ( GeV$^2$) & 2.476& fixed  \quad\\
\quad $a_{I=1}$ & 0.974 & fixed  \quad\\
\quad $n_\omega$ &  2.51 &1.2\%  \quad\\
\quad $m_\omega$ (GeV) &  0.78234&0.0072\%  \quad\\
\quad $\Gamma_\omega$ (GeV) & 0.0088&1.4\%  \quad\\
\quad $b_{I=0}$ & 0.2&fixed  \quad\\
\quad $\sigma^2_{I=0,1}$ ( GeV$^2$)& 2.476& fixed  \quad\\
\quad $a_{I=0}$ & 1.5 & 22\%  \quad\\
\quad $n_\phi$  &  1.9&0.3\%  \quad\\
\quad $m_\phi$ (GeV) &  1.01921 &0.0010\%  \quad\\
\quad $\Gamma_\phi$ (GeV) &  0.00421&0.54\%  \quad\\
\quad $\sigma_s^2$ ( GeV$^2$) & 3.6 & 24\%  \quad\\
\quad $a_s$ &   0.60&20\%  \quad\\
\quad $b_s$ & 0.20&12\%  \quad\\
\bottomrule
\end{tabular}
}\end{center}
\caption{Parameters for the spectral functions}
\label{Tab:fitparms}
\end{table}

\subsection{Branching ratios and the \boldmath \texorpdfstring{$\phi$}{phi} normalisation}
\label{sec:BRs}
The direct application of the results from the above fit to $e^+e^-$ and $\tau$ data assumes that the quark loop can in fact be factorised from the mesonic $D\to\pi$ transition. While in Ref.~\cite{LZ14} it was argued that this is a good approximation  for the analogous charm loop in $B\to K\ell\ell$,  it is prudent to verify the extent to which it can be trusted here. We cannot follow the strategy of Ref.~\cite{LZ14} for want of direct measurements of the resonance spectrum in the \Dpill decays. We do this by comparing our predictions for the branching ratio $D^+ \to R\pi^+\to \pi^+ \mu^+\mu^-$, with $R=\phi$, $\rho$ or $\omega$ to the limited experimental results available.\footnote{Note that we ignore the  contributions from the narrow pseudoscalar mesons, e.g. $\eta$, $\eta^\prime$ in our \Dpill description.} In the case of $D^+ \to \phi\pi^+ \to \pi^+ \mu^+\mu^-$,    there is only one  measurement  \cite{Abazov:2007aj} which gives $\B(D^+ \to \phi\pi^+\to \pi^+ \mu^+\mu^-) = (1.8\pm 0.8)\times 10^{-6}$,
and is consistent with the estimate obtained from the product of the individual
branching ratios of $D^+ \to \phi\pi^+$ and $\phi\to \mu^+\mu^-$.  The calculation of this  branching ratio within our description (in the SM), and using the parameters of Tab.~\ref{Tab:fitparms} for the functions $\tilde h_I(s)$, gives $1.3\times 10^{-8}$, which is still compatible within a little more than 2$\sigma$ with the experimental result, given the large uncertainties.\footnote{The values for all the parameters that enter this calculation can be found in Tab.~\ref{Tab:Num_inputs}.} We decide, nevertheless, to re-scale the parameter $n_\phi$ by a factor of 11.8, in order to account for this mismatch, presumably a result of the combination of an experimental fluctuation and non-factorisable effects. After this 
re-scaling, the central value for $\B(D^+ \to \pi^+\phi\to \pi^+ \mu^+\mu^-)$ calculated with our description reproduces the central value of the  experimental measurement.
Integrating our decay distribution around the $\rho-\omega$ region we find a branching ratio of $\sim 10^{-7}$. The experimental counterpart to this value can be estimated from the sum of $\B(D^+ \to R\pi^+) \times \B (R\to \mu^+\mu^-)$ with $R=\rho, \omega$ \cite{Zyla:2020zbs}, which gives $(6\pm 1)\times 10^{-8}$,  not too far from the value we obtain. Therefore, we do not perform any further rescaling of the normalisation of the resonances.
\begin{table}[!t]
			\center
			\begin{tabular}{c|c|c|c}
					\toprule
					Region & $s$-bin/GeV$^2$ & 90\% C.L. Limit & 95\% C.L. Limit\\
					\midrule
					 
					Reg.~I & $[0.250^2, 0.525^2]$& $2.0 \times 10^{-8}$ &  $2.5 \times 10^{-8}$\\
					Reg.~II & $>1.25^2$ & $2.6 \times 10^{-8}$ & $2.9 \times 10^{-8}$\\
					\bottomrule
				\end{tabular}
				\caption{Current best-world limits on \BRof{\Dpimumu} in two bins of the dilepton invariant mass $s$ \cite{Aaij:2013sua}. In the experimental search, the  branching fraction excluding the resonant contributions is extrapolated assuming a phase space model. \label{Tab:Meas_Dpimumu1}}
			\end{table}

Upper bounds on the non-resonant branching fractions exist from LHCb \cite{Aaij:2013sua}.
The limits are shown in the Tab.~\ref{Tab:Meas_Dpimumu1} in two regions, Region I (low $q^2$) and Region II (high $q^2$). 
After the re-scaling of the $\phi$ contribution (which affects the branching ratio in the non-resonant region through the tale of the resonance) we find, in the SM,\footnote{The calculation of the branching ratio in the high-$q^2$ region requires integration up to the kinematic end point of the spectrum. This is beyond the expected validity range of the OPE approach to the weak annihilation ($q^2\leq 2.3$ GeV$^2$) contribution. This uncontrolled systematic error should be relatively small since the main contributions to the integral arise from the region where we expect the OPE to be valid. } 
\begin{align}
&\mathcal{B}(\Dpimumu)\Big|_{{\rm low}\, q^2}^{\rm SM}=(8.1_{-6.1}^{+5.9})\times 10^{-9} ,\\
&\mathcal{B}(\Dpimumu)\Big|_{{\rm high}\, q^2}^{\rm SM} = (2.7^{+4.0}_{-2.6})\times 10^{-9}, 
\end{align}
which fulfil the experimental bounds. (Our errors are 68\% C.L. We postpone to Sec.~\ref{Sec:MCUB}  a detailed discussion about the estimate of theoretical uncertainties.) These results appear to be in line with the findings of Refs.~\cite{dBH15,FK2015} although there they were not given in the same form. Note that the purely short-distance contribution to the functions $\tilde h(s)$ would lead to much smaller branching fractions, of the order of $10^{-12}$, again in agreement with the results of Refs.~\cite{dBH15,FK2015}. 
Branching fractions of the order of $10^{-9}$ can be at LHCb reach once the full Run2 data is analysed.

We conclude that  the re-scaling the $\phi$ normalisation  is compatible with all known experimental bounds and,  from now on, we will use the value of $n_\phi$ given in Tab.~\ref{Tab:fitparms} multiplied by 11.8.

\section{Numerical and phenomenological analysis} \label{sec:pheno}

We will now perform a thorough phenomenological analysis of the decay \Dpill: we first define the observables of interest; we present our results in the SM, where the set of observables reduces to the differential Branching Ratio; allowing BSM contributions to the Wilson coefficients, we then assess the existing experimental bounds on such contributions and determine how the aforementioned observables are affected.

\subsection{Definition of the observables}
\label{Sec:DecayWidth}

Following \cite{BHP2007, FK2015}, the double differential \Dpill decay width with respect to $s$ the dilepton invariant mass and $\theta$, the angle between the three-momenta of $D^+$ and $\ell^-$ in the rest frame of the lepton pair, is given by
		\beq 
		    \frac{d^2\Gamma(\Dpill)}{ds~ d\cost} = N \lambda^{1/2} \beta\, \left[\,a_\ell +b_\ell\cost +c_\ell\cos^2{\theta}\,\right]
		    \label{Eq:DecayWidth}
		\eeq
		where the kinematic function $\beta$  and $\lambda$ are given in Eq.~(\ref{Eq:Kinematic_func}), and the dependence on $s$ is assumed to be understood. The normalisation factor $N$ is given by
		\beq
		     N=\frac{G_F^2 \alpha_e^2}{(4\pi)^5 m_D^3}
		\eeq
		and the three angular coefficients $a_\ell$, $b_\ell$  and $c_\ell$ by
		\begin{align}
		\nonumber  a_\ell&=\frac{\lambda}{2}(|F_V|^2+|F_A|^2) + 8 m_\ell^2 m_D^2 |F_A|^2 + 2s (\beta^2|F_S|^2+|F_P|^2),\\
		\nonumber 	b_\ell&= 4\,\mathrm{Re} \left[ s(\beta^2 F_S F_T^* + F_P F_{T5}^*) + m_\ell(\sqrt{\lambda} \beta F_V F_S^* + (m_D^2-m_\pi^2 +s)F_A F_{T5}^*)\right], \\
	\label{Eq:abc}		c_\ell&=-\frac{\lambda \beta^2}{2}(|F_V|^2+|F_A|^2)+2s(\beta^2|F_T|^2+|F_{T_5}|^2) + 4m_\ell\beta\sqrt{\lambda}\,\mathrm{Re}[F_V F_T^*].
		\end{align}
		Note that these three angular coefficients form the basis of the observables we will construct, and given that there are three independent coefficients (in the absence of CP violation) there will clearly only be three independent observables: we choose the differential decay width, the flat term, and the forward-backward asymmetry. On the other hand, in the SM the three angular coefficients are not independent, one finds the expressions:
			\begin{align}
			b_\ell&= 0, \nonumber\\
			a_\ell &= -\frac{c_\ell}{\beta^2} =\frac{\lambda}{2}|F_V^\SM|^2, \label{Eq:abcSM}
			\end{align}
			where $F_V^\SM$ is the SM contribution to $F_V$ defined in Eq.~\eqref{Eq:Fs}, depending on $C_9^q(s)$, $f_+(s)$ and CKM factors. There is therefore only one independent observable in the SM, the decay rate, and all others will vanish. 

	    Coming back to the general case, in order to obtain the first of these observables, the differential decay width, we integrate Eq.~\eqref{Eq:DecayWidth} with respect to \cost, 
		\beq
		    \frac{d\Gamma(\Dpill)}{ds} = 2 N \lambda^{1/2} \beta  \left[ a_\ell + \frac{c_\ell}{3} \right] \equiv N \lambda^{1/2} \beta\, d\hat{\Gamma}(s).
		    \label{Eq:decayratedist}
		\eeq
		Here it is convenient to introduce \dGh as it arises in the definition of several observables:
		\begin{align}
			d\hat{\Gamma}(s) =&\frac{2}{3} \lambda ( |F_V^\SM|^2+|C_{10}|^2 f_+^2) + (m_D^2-m_\pi^2 ) \frac{s }{m_c^2}f_0^2(|C_P|^2+|C_S|^2) \nonumber\\
			&+\frac{16}{3}\lambda s\frac{ f_T^2(|C_T|^2+|C_{T5}|^2)}{(m_D+m_\pi)^2},
			\label{Eq:dHat}
		\end{align}
 and depends on all the considered Wilson coefficients and on the $D$ to $\pi$ transition form factors, $f_+(s)$, $f_0(s)$ and $f_T(s)$.
\vspace{1cm}\\
Following~\cite{FK2015}, the next observable, the flat term, \FH, can be defined as
		     \begin{align}
		      \hspace{-1cm}  F_H(s) \equiv &\,\frac{a_\ell+c_\ell}{a_\ell+c_\ell/3} \nonumber \\
		        = &\,\frac{s}{d\hat{\Gamma}(s)}\left( \frac{m_D^2-m_\pi^2 }{m_c^2}f_0^2(|C_P|^2+|C_S|^2)+16\lambda\frac{ f_T^2\,(|C_T|^2+|C_{T5}|^2)}{(m_D+m_\pi)^2}\right), \label{Eq:FH}
		    \end{align}
		     
		  	where the second line in Eq.~(\ref{Eq:FH}) is valid only in the limit $m_\ell\to 0$ or in the high-$s$ region where $\beta\sim 1$. In the SM, up to $\res(m_\ell)$ corrections the flat term is only a function of $\beta$:
		    \beq
		    	F_H^{\rm SM}(s) = \frac{1-\beta^2}{1-\beta^2/3} + \res(m_\ell),
		    \eeq

		    as the Wilson coefficients and form factors cancel, clearly resulting in a small uncertainty on the helicity suppressed SM distribution. We note that \FH is numerically very small for $q^2\gg m_\ell^2$. Effects of physics beyond the Standard Model would therefore stand out for this observable, which is primarily sensitive to $C_P$, $C_S$, $C_T$ and $C_{T_5}$~\cite{FK2015,dBH15}.
		    
		   Noticing that $b_\ell=0$ might no longer hold in BSM scenarios, it is interesting to build observables sensitive to the angular coefficient $b_\ell$, one example of which is the forward-backward asymmetry $A_{\rm FB}$. Following~\cite{FK2015}, $A_{\rm FB}$ can be defined as
			\begin{align}
				\label{Eq:AFB_FK}
				A_{\rm FB}(s)\equiv &  \frac{ \left(\int_0^1 -\int_{-1}^0  \right)d\cost \frac{d^2\Gamma}{ds~d\cost} }{\int_{-1}^1 d\cost  \frac{d^2\Gamma}{ds~d\cost}}  \nonumber \\
				= &\,\,\frac{ b_\ell}{2\left[ a_\ell+c_\ell/3 \right]} \nonumber \\
				= &\,\, \frac{2 s \sqrt{\lambda}}{m_c\,d\hat{\Gamma}(s)}f_0 f_T \left[ \Re(C_{T5} C_P^*) + \Re(C_T C_S^*) \right] + \res(m_\ell).
			\end{align}
		    As visible in Eq.~(\ref{Eq:AFB_FK}), \AFB depends, up to $\res(m_\ell)$ corrections, only on BSM Wilson coefficients and $\AFB \simeq 0$ in the SM.
		    Moreover, no BSM Wilson coefficient can alone give rise to a non-vanishing $\AFB$, this requires BSM contributions to a combination of Wilson coefficients, either to both \CTf and \CP or \CT and \CS. 

            Finally, in order to explore the sensitivity to possible BSM CP phases, we consider the $CP$-asymmetry (\ACP) defined by \cite{dBH15}: 
		    \beq\label{Eq:ACP}
		        \ACP(s)=\frac{d\Gamma/ds - d\bar{\Gamma}/ds}{\int_{s_\text{min}}^{s_\text{max}} ds (d\Gamma/ds + d\bar{\Gamma}/ds)},    
		    \eeq
		    where $d\bar{\Gamma}/ds$ is the differential decay rate of the CP-conjugate mode, $D^-\to\pi^-\ell^+\ell^-$.
		    \ACP is zero in the SM (up to $\res(m_\ell)$ terms) and clearly also in BSM scenarios where Wilson coefficients are real.
		    For all the above observables $\mathcal{O}=$ \AFB, \FH and \ACP, we define the integrated observables $ \langle\, \mathcal{O}\, \rangle$ by
		    \begin{equation}
		    \label{eq:IntObs}
		        \langle\, \mathcal{O}\, \rangle =\int_{q^2_{\rm min}}^{q^2_{\rm max}}\, dq^2\,\, \mathcal{O}(q^2).
		    \end{equation}
            where the range if integration in the dilepton mass squared is between $q^2_{\rm min}$ and $q^2_{\rm max}$.

		\subsection{Results in the Standard Model}\label{Sec:results_SM}

			In the SM, making use of Eq.~\eqref{Eq:abcSM}, the double differential decay rate can be written
			\begin{align}
	 			\frac{d^2\Gamma(\Dpill)}{ds ~d \cost} =& N \lambda^{1/2} \beta \,a_\ell \,\left(1-\beta^2\right)\cos^2\theta
	 			= N \frac{\lambda}{2} \left(1-\beta^2\cos^2\theta \right)|F_V^\SM(s)|^2,
			\end{align}
				 such that in the SM, beyond the kinematic functions ($\lambda$ and $\beta$), the differential decay width only depends on the function $F_V^\SM(s)$.
			Integrating over $\cos{\theta}$, and again using Eq.~\eqref{Eq:abcSM},  we find
			\beq
				\frac{d\Gamma(\Dpill)}{ds} = N \lambda^{3/2} \beta \left( 1-\frac{\beta^2}{3}\right)|F_V^\SM(s)|^2.
				\label{Eq:DecayRate}
			\eeq
 It is therefore of interest to examine the different contributions to $F_V^\SM(s)$ in more detail. 
 
	Before doing this we first need to discuss the numerical values of the parameters used in our work, summarised in Tab.~\ref{Tab:Num_inputs} along with the appropriate references. The first set of parameters are the masses of the quarks, where we adopt the $\overline{\mbox{MS}}$ values of the strange, bottom and top masses and the pole mass of the charm quark as explained in Sec.~\ref{Sec:Anni}, and the on-shell mass of the $W$ boson, taken from Ref.~\cite{PDG2018}. For the scales, $\mu_c$ is set to the pole mass of the $c$ quark, for the scale $\mu_b$ we use the $\overline{\mbox{MS}}$ mass $m_b(m_b)$, and for the scale $\mu_W$ the on-shell mass of the $W$ boson. These scales $\mu_c$, $\mu_b$, $\mu_W$ are varied between $\mu/\sqrt{2}$ and $\sqrt{2}\,\mu$, to account for residual uncertainties related to renormalisation-scale variation.  
	
	Next we come to the hadronic parameters characterising the decay constants of the charged pion and $D$ meson, which we take from Ref.~\cite{PDG2018}, and the pion and $D$ meson light-cone distribution amplitudes (LCDAs); definitions of $\omega_0$ for the $D$ meson LCDA and $a_2$ and $a_4$ for the pion LCDA can be found in App.~\ref{App:LCDA}.
   Unlike the LCDA for of the $B$ meson, very little is known about that of the $D$ meson. Therefore, following Ref.~\cite{FMS2017}, we adopt an ad-hoc range for $\omega_0$, keeping in mind the naive expectation from heavy-quark symmetry and choosing a sufficiently large uncertainty to remain conservative.
    As can be seen in Eq.~\eqref{Eq:LCDA_pi}, the only numerical input required for the twist-2 $\pi$ LCDA are the Gegenbauer moments. These can only be calculated via non-perturbative methods. While there has been a great deal of progress from the Lattice in calculations of $a_2$~\cite{Braun:2015axa}, the most recent calculation being that of the RBC collaboration with $N_f = 2 + 1$ flavours of dynamical Wilson-clover fermions~\cite{Bali:2019dqc}, $a_4$ has not so far been calculated. We therefore adopt $a_{2,4}$(1 GeV) from Ref.~\cite{Khodjamirian:2011ub}, where the light-cone sum rules (LCSR) result for the pion electromagnetic form factor~\cite{Bijnens:2002mg} is fitted to experimental data~\cite{Huber:2008id}. The extracted values, $a_2$(1 GeV) = $0.17 \pm 0.08$ and $a_4$(1 GeV) = $0.06 \pm 0.10$, where the errors reflect both experimental and theoretical uncertainties, are consistent with previous results from sum rules and Lattice QCD.\footnote{Very recently, a new result for $a_{2,4}$ appeared in Ref.~\cite{Cheng:2020vwr}. The values are in agreement with ours within uncertainties, and would produce negligible changes in our results.}
    
    For the form factors, we adopt the recent calculation on the Lattice by the ETM collaboration~\cite{Lubicz:2017syv, Lubicz:2018rfs} (again the details of the parametrisation can be found in App.~\ref{App:FF}). The parameters given in Tab.~\ref{Tab:Num_inputs} are taken from Refs.~\cite{Lubicz:2017syv, Lubicz:2018rfs}. Note that in the low-$q^2$ region, these form factors were also calculated in LCSR in Ref.~\cite{Khodjamirian:1998vk}. 
		
			\begin{table}[!t]
				\center
				\begin{tabular}{c|c|c}
					\toprule
					Parameters & Value &\,\, Reference \,\,\\
					\midrule
					$m_s(m_s)$ [\MeV] & $95\pm3$ & \cite{PDG2018} \\
					$m_c$ [\GeV] & $1.67_{-0.07}^{+0.07}$ & \cite{PDG2018} \\
					$m_b(m_b)$ [\GeV] & $4.18_{-0.03}^{+0.04}$ & \cite{PDG2018} \\
					$m_t(m_t)$ [\GeV]& $163.3\pm 2.7$ & \cite{Alekhin:2012py} \\
					$M_W$ [\GeV] & $80.385\pm0.015$ & \cite{PDG2018} \\
					\midrule
					$\omega_0 [\MeV]$ & $450\pm300$ & \cite{FMS2017}\\
					$f_{\pi^+}$ [\MeV] & $130.5\pm0.16$ & \cite{PDG2018}\\
					$f_{D^+}$ [\MeV]& $212.15\pm1.45$  & \cite{PDG2018}\\
					$a_2 (1\,\mbox{GeV})$ & $0.17\pm0.08$ & \cite{Khodjamirian:2011ub}\\
					$a_4 (1\,\mbox{GeV})$ & $0.06\pm0.1$   & \cite{Khodjamirian:2011ub}\\
					\midrule
					$f(0)$  & $0.6117\pm0.0354$ &\cite{Lubicz:2017syv}\\
						$f_T(0)$& $0.5063\pm0.0786$ &\cite{Lubicz:2018rfs}\\
					$c_+$ & $-1.985\pm0.347$ &\cite{Lubicz:2017syv}\\
					$c_0$  & $-1.188\pm0.256$ &\cite{Lubicz:2017syv}\\
					$c_T$ & $-1.10\pm1.03$& \cite{Lubicz:2018rfs}\\ 
					$P_V$& $0.1314\pm0.0127$ & \cite{Lubicz:2017syv}\\
					$P_S$& $0.0342\pm0.0122$ & \cite{Lubicz:2017syv}\\
					$P_T$ & $0.1461\pm0.0681$ &  \cite{Lubicz:2018rfs}\\
					\midrule
					$\tau_{D^+}$ [ps] & $1040\pm7$ &  \cite{PDG2018} \\
					$|V_{ud}|$ & $0.97420\pm 0.0002$  &  \cite{PDG2018}\\
					$|V_{cd}|$ & $ 0.218 \pm 0.004$  &  \cite{PDG2018}\\
					$|V_{ub}|$ & $ (4.09 \pm 0.39) 10^{-3}$  &  \cite{PDG2016}\\
					$|V_{cb}|$ & $(40.5 \pm 1.5) 10^{-3}$  &  \cite{PDG2016}\\
					$\gamma$ & $(73.2^{+6.3}_{-7.0})^{^\circ}$ &  \cite{PDG2016}\\
					\bottomrule
				\end{tabular}
				\caption{Summary of the numerical input used in the study of \Dpill.\label{Tab:Num_inputs}}
			\end{table}

			However,  Tab.~\ref{Tab:Num_inputs} does not provide the values of the Wilson coefficients $C_{1-9}$ defined in Sec.~\ref{App:WC}.
			The numerical results for the Wilson coefficients, calculated as described in App.~\ref{App:WC} for the central value of these scale (as given in Tab.~\ref{Tab:Num_inputs}), are summarised in Tab.~\ref{Tab:WC_values}. We note that only $C_1$, $C_2$ and $C_9$ have sizeable values, where however the signs of $C_1$ and $C_2$ are opposite, the Wilson coefficients related to the strong penguin and the dipole operators are numerically small. Further, we have checked that for the same set of input parameters, our results agree with those given in Ref.~\cite{deBoer:2016dcg}.
			\begin{table}[!t]
			\small
			\hspace{-.25cm}
			     \centering
			     \begin{tabular}{c|c|c|c|c|c|c|c|c|c}
			            \toprule
			            & $C_1$ & $C_2$ & $C_3$ & $C_4$ & $C_5$ & $C_6$ & $C_7^\eff$ & $C_8^\eff$ & $C_9$ \\
			            \midrule
			            LL  &  -0.890 & 1.072 & -0.002 & -0.041 & 0.000 & 0.000 & 0.057 & -0.042 & -0.095 \\
			            NLL & -0.603 & 1.029 & -0.003 & -0.065 & 0.000 & 0.000 &  0.035 & -0.045 & -0.270 \\
			            NNLL & -0.529 & 1.026 & -0.004 & -0.063 &0.000 & 0.000 & 0.036 & -0.048 & -0.413 \\
			            \bottomrule
			     \end{tabular}
			     \caption{Value of the Wilson coefficients at the NNLL approximation given at scale $\mu_c=1.67~\GeV$, for $\mu_b=4.18~\GeV$ and $\mu_W=80.4~\GeV$.}
			     \label{Tab:WC_values}
			 \end{table}

		Having provided the numerical values of all parameters that enter our analysis, we are now ready  to examine the contributions to $F_V^\SM(s)$, 
			\beq \label{Eq:FVSM}
				F_V^{\rm SM}(s)=  \left[ \lambda_b C_9^{(b)}(s) + \lambda_d C_9^{(d)}(s) \right] f_+(s),
			\eeq
			where 
			\beq 
    			\Cn = \Cnnaive+ \CnAnn + \CnSS +  \CnFF. 
			\eeq
			In naive factorization, 
			\begin{align}
			\nonumber	C_9^{(d)}(s)\bigr\rvert_{\rm Naive} &= Y^{(d)}(s), \\
				C_9^{(b)}(s)\bigr\rvert_{\rm Naive} &= Y^{(b)}(s) + C_9 + \frac{2m_c}{m_D+m_\pi}C_7\frac{f_T(s)}{f_+(s)}, \label{Eq:C9bnaive}
			\end{align}
			where the QCDf corrections $\CnAnn$, $\CnSS$ and $\CnFF$ are provided in Sec.~\ref{sec:theo_bkg}. The numerical results for these corrections at $s=0.5~\GeV^2$  are presented in Tab.~\ref{Tab:Num_contrib}. We note that $Y^{(q)}$ (containing the resonance contribution) and \CnAnn are the only sizeable contributions. We further remind the reader that the ($q=b$) contributions are CKM suppressed since $\lambda_b \ll \lambda_d$. Hence, $F_V^{\rm SM}(s)$ can be approximated by
			\beq
				F_V^{\rm SM}(s) \sim  \lambda_d \left( Y^{(d)} + \CndAnn \right) f_+(s).\label{eq:FVSM2}
			\eeq
			As was first noticed in Ref.~\cite{dBH15}, $Y^{(d)}$ is further suppressed due to the cancellation between $C_1$ and $C_2$ occurring at the scale $m_c$ in the factor $(2/3 C_1 + 1/2 C_2) \sim 0.34$.
			
			In our final results, we further modify the $q^2$-dependent moment, $\lambda_D^-(s)$, of Eq.~(\ref{Eq:lDpartonic}) to take into account effects due to the hadronic resonances as
\begin{equation}
			\frac{1}{\lambda_D^-(s)} = \int_0^\infty d\omega \frac{\phi_D^-(\omega)n_d j_d(s)}{\omega-s/m_D - i \epsilon},\label{Eq:lDhadronic}
\end{equation}
where 
\beq
j_d(s)=\frac{1}{\pi}{\rm Im}\,\tilde h_d(s).
\eeq
The constant $n_d$ is introduced to ensure that the perturbative result is recovered for $q^2\ll -m_D\Lambda_{QCD}$, as discussed in App. B of Ref.~\cite{FMS2017}.

			\begin{table}[!t]
				\center
				\begin{tabular}{c|c|c}
					\toprule 
					Contribution & $\propto \lambda_b$  & $\propto \lambda_d$ \\ 
					\midrule
					$C_9$          & -0.413 & 0 \\
					$Y^{(q)}$ & $-1.303 + 0.034 i$ & $1.345+ 0.981 i$\\
					\CnFF           & $-0.287 - 0.457 i$ & $-0.028 - 0.002 i$\\
					\CnAnn            & $0.013 - 0.054 i$ & $0.503 - 2.100 i$\\
					\CnSS  & 	 $0.028 - 0.033 i$ & $0.005 + 0.002 i$  \\
					\bottomrule 
				\end{tabular}
				\caption{Breakdown of individual contribution at NLO for the \Dpill decay at $s=0.5 ~\GeV^2$.}
				\label{Tab:Num_contrib}
			\end{table}

     \begin{figure}[!t]
			    \centering
			    \includegraphics[width=.8\textwidth]{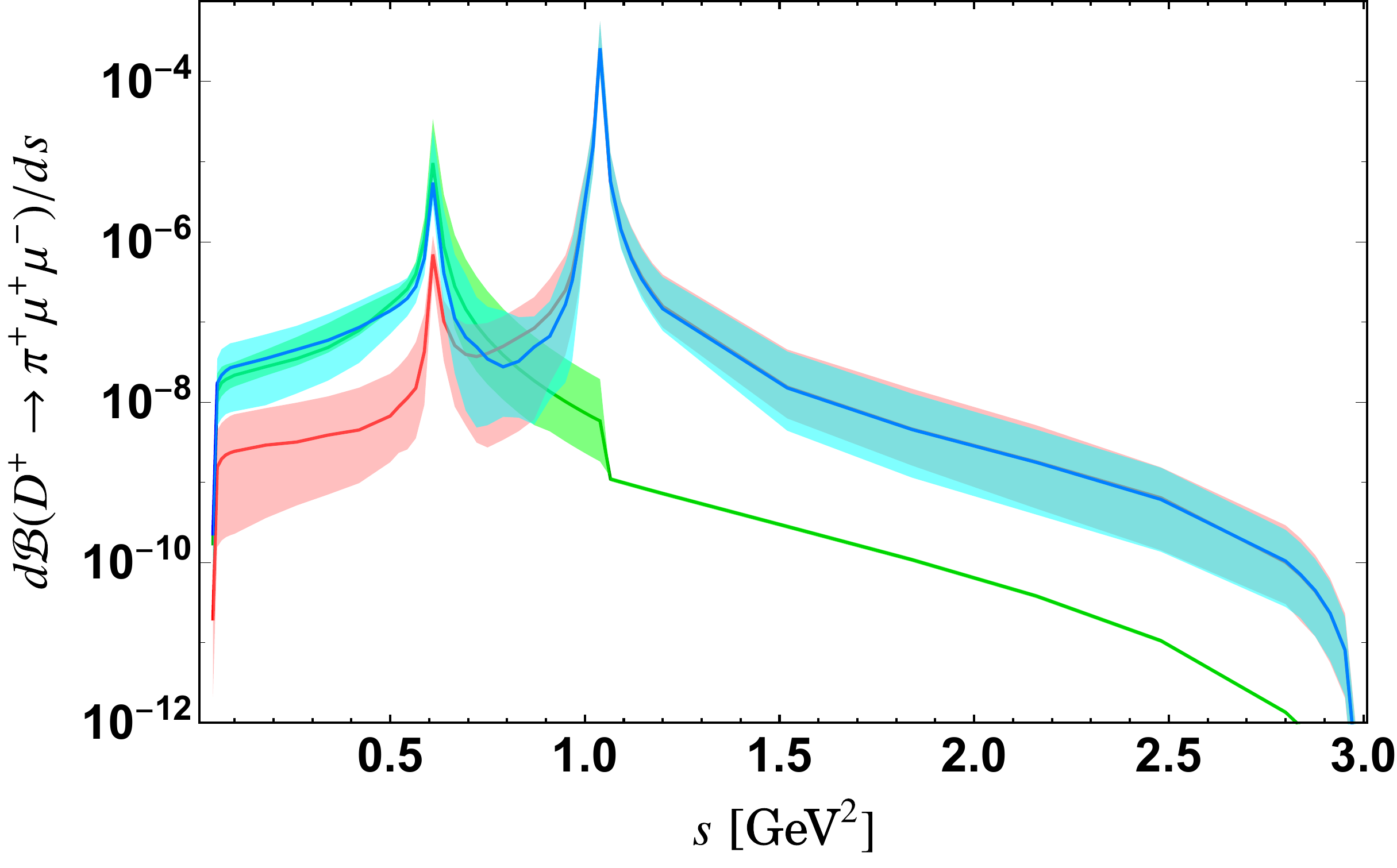}
			    \caption{Distribution of the \BRof{\Dpimumu} in the SM. In blue the total branching ratio, in red the contribution from $Y^{(d)}$, and in green that of $\CndAnn$ including resonance effects as per Eq.~(\ref{Eq:lDhadronic}). The error bands on the SM prediction, determined as explained in Sec.~\ref{Sec:MCUB}, are shaded}\label{Fig:BR_SM}
			\end{figure}

						The branching ratio distribution is related to the decay width distribution defined in Eq.~(\ref{Eq:decayratedist}) via
			\beq
				\frac{d\BRof{\Dpimumu}}{ds} = \frac{1}{\Gamma_D}  \frac{d\Gamma(\Dpill)}{ds}
				\label{Eq:BRdistri}
			\eeq
			where $\Gamma_D$ is the total decay width of the $D^0$ meson. 
The final result for the differential branching fraction in the SM, taking into account all uncertainties, is given in Fig.~\ref{Fig:BR_SM}. (Below, in Sec.~\ref{Sec:MCUB}, we will discuss in detail how the uncertainty bands are obtained.) In the same figure, we also show the two separate contributions in Eq.~(\ref{eq:FVSM2}): in red the component due to $Y^{(d)}$ and in green the weak annihilation contribution, $\CndAnn$. The decay width is largely  dominated by weak annihilation for $q^2 \lesssim 0.7$GeV$^2$ as well as by the contribution from the narrow $\eta$ resonance, which is not included in our description~\cite{dBH15,FK2015}. For $q^2$ above the $\phi$ peak the contribution from $Y^{(d)}$ becomes dominant. (We remind the reader that at the $\phi$ peak we start using the OPE expression for the weak annihilation contribution, given in  Eq.~(\ref{eq:WAOPE}). At the switch point, the two descriptions are compatible within one sigma.)

\begin{figure}[!t]
			    \centering
			    \includegraphics[width=.8\textwidth]{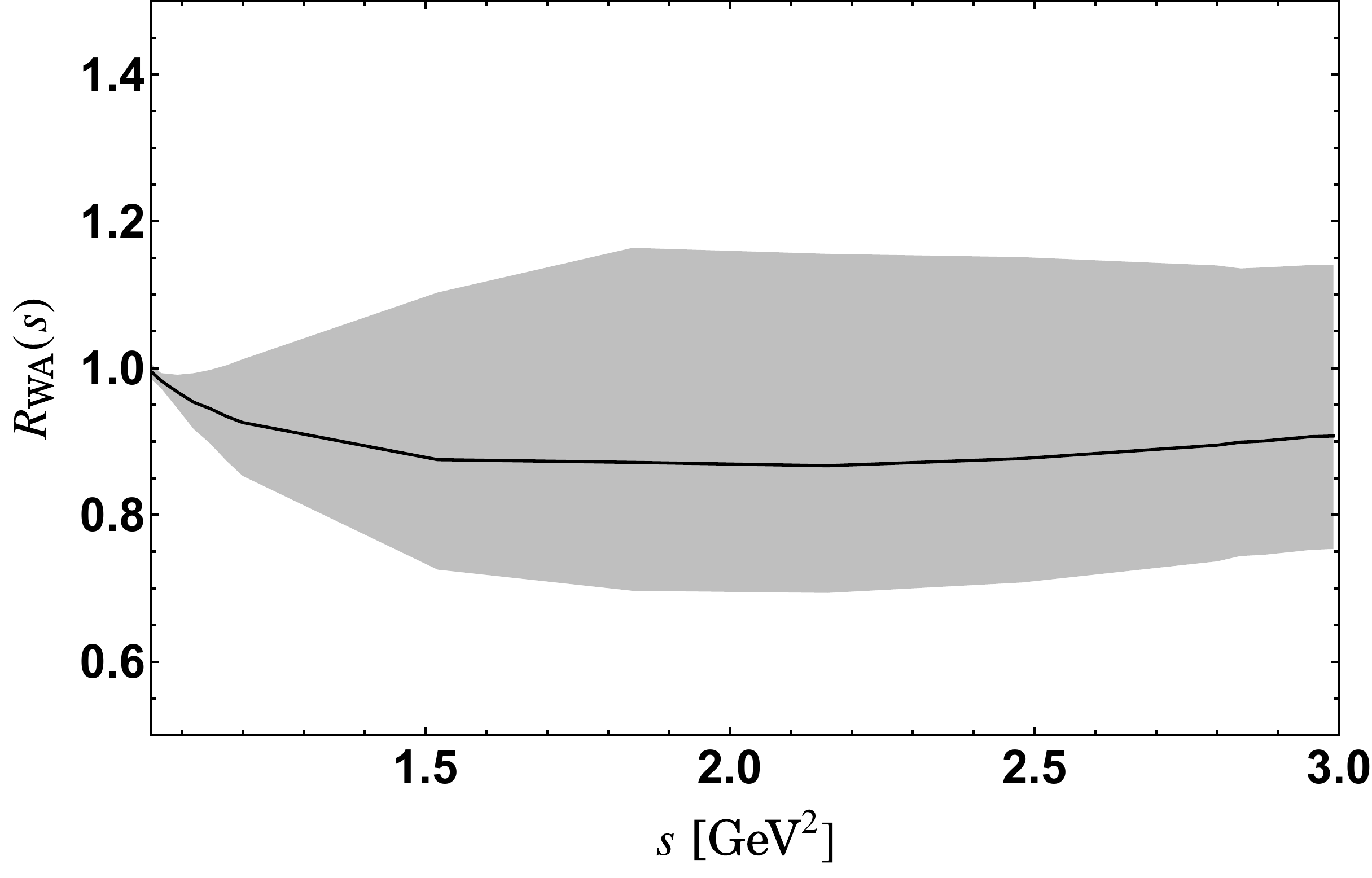}
			    \caption{Ratio of the differential branching ratios obtained with and without the weak annihilation contribution, as defined in Eq.~(\ref{Eq:RWA}). The shaded error band is determined as explained in Sec.~\ref{Sec:MCUB}}.\label{Fig:RatioWA}
			\end{figure}

The effect of weak annihilation for the higher-$q^2$ part of the spectrum is not completely negligible, though. To quantify it
we plot, in  Fig.~\ref{Fig:RatioWA}, the ratio of the branching ratio obtained with the weak annihilation contribution to the branching ratio without this contribution
\begin{equation}
R_{\rm WA}(s) = \frac{\frac{d\mathcal{B}(D \to \pi \mu \mu)}{ds}\Big |_{\rm with \,\, WA}}{\frac{d\mathcal{B}(D \to \pi \mu \mu)}{ds}\Big |_{\rm without \,\, WA} } .    \label{Eq:RWA}
\end{equation}
For our central values, including the weak annihilation results in a  decrease in the branching fraction of about  $10\%$.  But the magnitude and sign of the effect depends on interferences that are sensitive to the resonance phases. Within 68\% CL, this means that, for the differential branching ratio, either a decrease of up to $30\%$ or an increase by up to $15\%$ is possible, locally.  Since the contribution of weak annihilation can be sizeable, we include it in our final results.
			
Finally, as we mentioned in Sec.~\ref{sec:BRs}, our integrated branching ratios are compatible with other results found in the literature, although our values tend to be a bit larger. The results of Fig.~\ref{Fig:BR_SM} can also be compared locally. For example,  comparing our differential branching ratio at $q^2=0.5$~GeV$^2$ with the results of  Ref.~\cite{dBH15} we find, in our description, values ranging from $7.1\times 10^{-8}$ to  $2.8\times 10^{-7}$, while in Ref.~\cite{dBH15}  one has values ranging from $3.6\times10^{-9}$ to $1.1\times10^{-7}$, which is fully compatible, although our upper boundary is larger by a factor of almost 3. At $q^2=2.5$~GeV$^2$, we find a total branching ratio between $1.4\times 10^{-10}$  and $1.5\times 10^{-9}$  whereas in Ref.~\cite{dBH15}  the corresponding result is between $3\times 10^{-11}$  and $6.2\times 10^{-10}$. The two ranges are again compatible, but the upper bound of our result is a factor of 2.4 times larger.

			\subsection{Calculation of the uncertainty band} \label{Sec:MCUB}

    We have identified a set of seven parameters playing a major role in the uncertainties entering our predictions. These consist of the phases of the resonances ($\phi_\omega$ and $\phi_\Phi$), three parameters entering the $\phi$ resonance structure and the $s\bar s$ resonance excitations ($n_\Phi$, $\sigma^2_\Phi$ and $a_\Phi$) as well as two of the three renormalisation scales which enter the Wilson coefficients ($\mu_c$ and $\mu_W$).
   The dependence on these parameters is highly non-linear and there is no reason to assume a Gaussian distribution for the scale variation. Therefore,
    we use a Monte Carlo method in order to calculate the uncertainty band. In the case of the SM decay width, for a list of carefully chosen $q^2$ values within the kinematic range,
    we calculate the value of the observable $N=1000$ times varying the parameters listed above following a specific distribution (uniform for the phases and the scales and Gaussian for the other parameters). We then use the median value as the central value, the uncertainty band being obtained looking for the interval that contains 68\% of the values around the median. (The upper and lower uncertainty bands shown in Fig.~\ref{Fig:BR_SM} are the result of an interpolation.)
    
    Note that on calculating the observables in BSM Physics scenarios, it is again impractical to apply this MC technique
     to a large number of points in the plane of BSM Physics contributions to the Wilson coefficients, due to the extensive computing time that it  requires.  
     We therefore perform  the full Monte-Carlo error propagation ($N=1000$) on the desired observable for a small subset of 9 points spread equally across the plane and extrapolate between these points.

		\subsection{Constraints on the BSM Wilson coefficients}\label{Sec:NP_Bounds}

		Before investigating the possible size of the effect of BSM physics on our observables, we first need to verify the existing constraints on the Wilson coefficients from the \Dll and \Dpill branching ratios. Note that here we do not consider the interplay with direct searches at the LHC, studied in Ref.~\cite{Fuentes-Martin:2020lea}. We further compare these constraints with the possible sizes of these Wilson coefficients in certain BSM models.

		\subsubsection{Constraints from upper limits on \boldmath \texorpdfstring{\Dll}{D to ll}}

			The measurement of the \Dll branching ratio provides constraints on the BSM Wilson coefficients \Cten, \CS and \CP. The expression for this branching ratio is given by~\cite{dBH15}:
			\beq
				\BRof{\Dll} = \frac{1}{\Gamma_D}\frac{G_F^2\alpha_e^2}{64\pi^3}f_D^2 m_D^5 \beta \left( |P|^2 + \beta^2|S|^2 \right),
			\eeq
			where in $\beta$ we replace $s$ by $m_D^2$, and 
			\begin{align}
				P&= \frac{1}{m_c}(C_P-C_P^\prime)+ \frac{2 m_\ell}{m_D^2}(C_{10} - C_{10}^\prime),\quad\mathrm{and}\nonumber \\
				S&=\frac{1}{m_c}  (C_S - C_S^\prime). \label{Eq:P_Dll}
			\end{align}
            Note that by simply adapting the result for $B_s\to\ell^+\ell^-$ for $D\to\ell^+\ell^-$, one neglects the possibility of additional long-distance contributions which might arise at the scale $m_c$, thereby increasing the uncertainty on this constraint, an estimate of these effects can be found in Ref.~\cite{Burdman:2001tf}.

			The constraints obtained from the \Dee decay are much weaker than those obtained from \Dmumu. For \Dmumu, using the current  best upper limit at 90\%C.L., $6.2\times 10^{-9}$~\cite{Aaij:2013cza}, we find:
			\beq
			    |C_S - C^\prime_S|^2 +  |C_P - C^\prime_P + 0.1(C_{10} - C^\prime_{10})|^2 \lesssim 0.008.
			\eeq
            Barring cancellations between coefficients, this leads to
			\begin{align}
			\label{Eq:Dllconstraints}
			    &|C_{10} - C^\prime_{10}| \leq 0.86, \\
			    &|C_{P/S} - C^\prime_{P/S}| \leq 0.087.
			\end{align}
			These constraints are similar to that obtained in Ref.~\cite{dBH15}. We note that the constraint on \Cten is ten times weaker than those on \CP and \CS. This is due to the fact that the axial-vector contribution to the branching ratio is helicity suppressed.

		\subsubsection{Constraints from upper limits on \boldmath \texorpdfstring{\Dpimumu}{D to pi mu mu}}

			As for \Dll, the constraints obtained from the $c\to u e^+e^-$ transition are much weaker than those obtained from the $c\to u\mu^+\mu^-$ transition. Compared to the leptonic \Dmumu decay, which is only sensitive to effects in $C_{10}^{(\prime)}$, $C_{S}^{(\prime)}$ and $C_{P}^{(\prime)}$, the \Dpimumu decay is sensitive to a more diverse range of BSM Physics effects, namely $C_{10}^{(\prime)}$, $C_{S}^{(\prime)}$, $C_{P}^{(\prime)}$, \CT, \CTf but also to possible BSM contributions to $C_7$ and $C_9$. In the SM, these Wilson coefficients are multiplied by $\lambda_b$ as visible in Eq.~(\ref{Eq:C9bnaive}) and (\ref{Eq:FV_SM}). We follow Ref.~\cite{FK2015}, defining possible BSM contributions to $C_7$ and $C_9$ as
			\begin{align}
				C_7 =C_7^{\rm SM} + \frac{C_7^{\rm BSM}}{\lambda_b},\\
				C_9 =C_9^{\rm SM} + \frac{C_9^{\rm BSM}}{\lambda_b},
			\end{align}
			where we assume that  $C_7^{\rm BSM}$ and $C_9^{\rm BSM}$ are not necessarily subjected to the same CKM suppression as the SM contributions are.

			The constraints on the BSM Wilson coefficients from the upper limits on the binned branching ratios of \Dpimumu are determined using the results at 90\% C.L. given in Tab.~\ref{Tab:Constraints_Dpimumu}. We use the central value of the parameters given in Tab.~\ref{Tab:Num_inputs} and our expression for the branching ratio distribution given in Eq.~(\ref{Eq:BRdistri}), integrated over two bins at low and high $q^2$ to compute the constraints on the Wilson coefficients $C_7^{\rm BSM}$, $C_9^{\rm BSM}$, \Cten, $C_{S/P/T/T5}$ in the two regions Reg.~I and Reg.~II respectively. These constraints can be found in Tab.~\ref{Tab:Constraints_Dpimumu}.
			
			\begin{table}[!t]
				\center
				\begin{tabular}{c|c|c}
					\toprule
					& Reg.~I & Reg.~II \\
					\midrule
					$|C_7^{\rm BSM}|$ & $\leq$\,1.58 & $\leq$\,0.67 \\
					$|C_9^{\rm BSM}|$ & $\leq$\,2.17 & $\leq$\,0.84 \\
					$|C_{10}+C_{10}^\prime|$ & $\leq$\,0.938 & $\leq$\,1.1 \\
					$|C_S +C_S^\prime|$ & $\leq$\,3.81 & $\leq$\,0.60 \\ 
					$|C_P+C_P^\prime|$ & $\leq$\,3.28 & $\leq$\,0.60 \\ 
					$|C_T|$ & $\leq$\,3.50 & $\leq$\,0.68 \\ 
					$|C_{T5}|$ & $\leq$\,2.48 & $\leq$\,0.77 \\ 
					\bottomrule
				\end{tabular}
				\caption{Constraints on the maximal values of the Wilson coefficients from the 90\% C.L. limit on the low-$q^2$  (Reg.~I) and high-$q^2$ (Reg.~II) bins of \Dpimumu branching ratio~\cite{Aaij:2013sua}. \label{Tab:Constraints_Dpimumu}}
			\end{table}
			It is relevant to note that while the upper limits lie very close to the SM prediction in Reg.~I compared to Reg.~II; bounds on BSM coefficients are not stronger in this region. This can be understood by the fact that this region is dominated by weak annihilation, depending on $C_1$ and $C_2$ for which we do not consider BSM contributions.

		\subsubsection{Implications of the constraints for BSM models}

			If we assume that the BSM Wilson coefficients are real and that the chirality-flipped Wilson coefficients are all equal to zero, the maximal values allowed by the current experimental limits on \BRof{\Dll} and \BRof{\Dpill} are summarised in Tab.~\ref{Tab:BSM_bounds}. 
			\begin{table}[!t]
				\center
				\begin{tabular}{c|c}
					\toprule
					BSM Wilson coeff.  & Maximal values \\
					\midrule
					$C_7^{\rm BSM}$ & 1.03 \\
					$C_9^{\rm BSM}$ & 1.3 \\
					\Cten & 0.86 \\
					\CS & 0.087 \\ 
					\CP & 0.087 \\ 
					\CT &  0.84 \\ 
					\CTf & 0.90 \\ 
					\bottomrule
				\end{tabular}
				\caption{Summary of the maximal value allowed by the current experimental limits on \Dll and \Dpill at 90\% C.L. if we assume the BSM Wilson coefficients are real and that the chirality-flipped Wilson coefficients are all equal to zero. \label{Tab:BSM_bounds}}
			\end{table}

Several models generating \cull transition have been studied in the past, e.g.~Minimal Supersymmetric Standard Models \cite{Burdman:2001tf, Wang:2014uiz,Fajfer:2001sa, Fajfer:2007dy}, two Higgs doublet models \cite{Fajfer:2001sa}, Little Higgs model \cite{Paul:2011ar, Fajfer:2005ke}, vector-like quark singlet \cite{Fajfer:2007dy} and leptoquarks \cite{FK2015,dBH15}. It is of interest to compare the size of the Wilson coefficients allowed by the current experimental constraints on \BRof{\Dmumu} and \BRof{\Dpimumu} summarised in Tab.~\ref{Tab:BSM_bounds} to the values obtained in concrete BSM models. 
						
We will focus on scalar and vector leptoquarks (LQs), due both to the significance of these models in light of the $B$ anomalies~\cite{Crivellin:2017zlb,Becirevic:2018afm,Angelescu:2018tyl,Crivellin:2019dwb} and the interesting effects that they can generate in $c\to u\ell\ell$ transitions~\cite{dBH15,FK2015}. Following Ref.~\cite{dBH15}, the contributions of the LQs to the Wilson coefficients $C_9^{(\prime)}$, $C_{10}^{(\prime)}$, $C_S^{(\prime)}$, $C_P^{(\prime)}$, $C_T$ and $C_{T_5}$ in terms of the couplings $\lambda_{L/R}^{u\ell/c\ell}$, only considering the case $\ell=\mu$, can be expressed as
\begin{align}\label{Eq:WCsLQ}
 \nonumber&C^{(\prime)}_{9,10}=\frac{\sqrt2\pi}{G_F\alpha_e}k^{(\prime)}_{9,10}\frac{\lambda_{i(j)}^I\left(\lambda_{i(j)}^J\right)^*}{M^2}\,,&C_T=\frac{\sqrt2\pi}{G_F\alpha_e}k_T\left(\frac{\lambda_i^I\left(\lambda_j^J\right)^*}{M^2}+\frac{\lambda_j^I\left(\lambda_i^J\right)^*}{M^2}\right),\\
 &C_{S,P}^{(\prime)}=\frac{\sqrt2\pi}{G_F\alpha_e}k_{S,P}^{(\prime)}\frac{\lambda_{j(i)}^I\left(\lambda_{i(j)}^J\right)^*}{M^2}\,\,\,,&C_{T_5}=\frac{\sqrt2\pi}{G_F\alpha_e}k_{T_5}\left(\frac{\lambda_i^I\left(\lambda_j^J\right)^*}{M^2}-\frac{\lambda_j^I\left(\lambda_i^J\right)^*}{M^2}\right) \, ,
\end{align}
where  $i,j=L,R$, and $M$ is a generic scale of the leptoquarks.
We observe that, beyond (semi)leptonic $D$ decays, the only observables constraining the couplings $\lambda_{L/R}^{u\ell/c\ell}$ are the branching ratios of $K^+\to\pi^+\nu\bar{\nu}$ for $S_{1L}$ and $K^0_L\to\mu\mu$ for $S_{2R}$, $V_2$ and $V_3$, as explained in Ref.~\cite{dBH15}. As these constraints are very strong, the Wilson coefficients $C_9$, $C_{10}$, $C_S$ and $C_P$ affected can be neglected. In Tab.~\ref{Tab:LQscenarios} we provide details of the remaining scalar and vector leptoquarks after taking into account these constraints.

The Scalar LQs $S_1$ with quantum numbers $(3, 1, -1/3)$ and $S_2$ with $(3, 2, -7/6)$ are interesting as they contribute to all the WCs we consider.
For $S_1$, we assume that the left-handed up-quark-muon coupling can be neglected, i.e.~$\lambda_{L}^{u\mu}\sim 0$, such that the combination $\lambda_{R}^{u\mu}\lambda_{R}^{c\mu}$ controls  $C'_ 9 = C'_ {10} $, and $\lambda_{R}^{u\mu}\lambda_{L}^{c\mu}$ controls $C' _S = C' _P = C_T/2 = C_ {T_ 5}/2 $.
Analogously for $S_2$, we assume that the right-handed up-quark-muon coupling can be neglected, i.e.~$\lambda_{R}^{u\mu}\sim 0$, such that the combination $\lambda_{L}^{u\mu}\lambda_{L}^{c\mu}$ controls $C'_9 = - C'_{10}$, and $\lambda_{L}^{u\mu}\lambda_{R}^{c\mu}$ controls $C'_S = -C'_P = C_T/2 = -C_{T_5}/2 $. 
As for the Vector LQs, these only contribute to $C^{(\prime)}_9$ and $C^{(\prime)}_{10}$. After taking into account the constraints from kaon decays, the only vector LQs which can give rise to non-negligible Wilson coefficients are $\tilde{V}_1$ with quantum numbers $(3,1,-5/3)$ with $C'_9=C'_{10}$ and $\tilde{V}_2$ with $(3,2,1/6)$ with $C'_9=-C'_{10}$. Large values of un-primed $C_9$ and $C_{10}$ cannot be generated.
In addition to our model independent analysis, we will also study these models in the following subsection.
\begin{table}[!t]
 \centering
 \begin{tabular}{c|ccccccccccc}
 \toprule
                &  $I$     &  $J$     &  $i$   &  $j$   &    $k_9'$      &  $k_{10}'$   &   $k_{S/P}'$  &  $k_T$       &  $k_{T5}$  \\
\midrule
  $S_1\,(3, 1, -1/3)$         &  $(cl)$  &  $(ul)$  &  $L$   &  $R$   &    $-\frac14$  &  $-\frac14$  &   $\mp\frac14$  &  $-\frac18$  &  $-\frac18$  \\
  $S_2\,(3, 2, -7/6)$         &  $(ul)$  &  $(cl)$  &  $R$   &  $L$   &  $-\frac14$  &  $\frac14$   &   $\mp\frac14$  &  $-\frac18$  &  $-\frac18$  \\
  $\tilde V_1\,(3,1,-5/3)$  &  $(ul)$  &  $(cl)$  &  --  &  $R$    &  $\frac12$   &  $\frac12$   &    $0$         &  $0$         &  $0$  \\
  $\tilde V_2\,\,(3,2,\,\,\,1/6)$  &  $(cl)$  &  $(ul)$  &  --  &  $L$   &  $\frac12$   &  $-\frac12$  &    $0$         &  $0$         &  $0$  \\
  \bottomrule
 \end{tabular}
 \caption{Coefficient matrix for the LQ Wilson coefficients defined in Eq.~(\ref{Eq:WCsLQ}), following Ref.~\cite{dBH15}.}
 \label{Tab:LQscenarios}
\end{table}

			\subsection{Results for physics beyond the SM}
						In Fig.~\ref{Fig:BR_SM} we see that the resonances have a larger effect on the differential Branching Ratio in the lower half of the $q^2$ region. In this section, we therefore choose to study the sensitivity of the observables defined in Eqs.~\eqref{Eq:FH}, \eqref{Eq:AFB_FK} and \eqref{Eq:ACP} in the high $q^2$ region to possible BSM effects, integrating these observables over a large bin in $q^2$ in order to reduce the impact of the OPE assumptions adopted here for the weak annihilation corrections. We first do this in a model independent way, considering generic BSM contributions to pairs of Wilson coefficients, and then study the specific case of leptoquark models.
						
			\subsubsection{The model independent case}
\begin{figure}[!t]
			    \centering
			    \includegraphics[width=.6\textwidth]{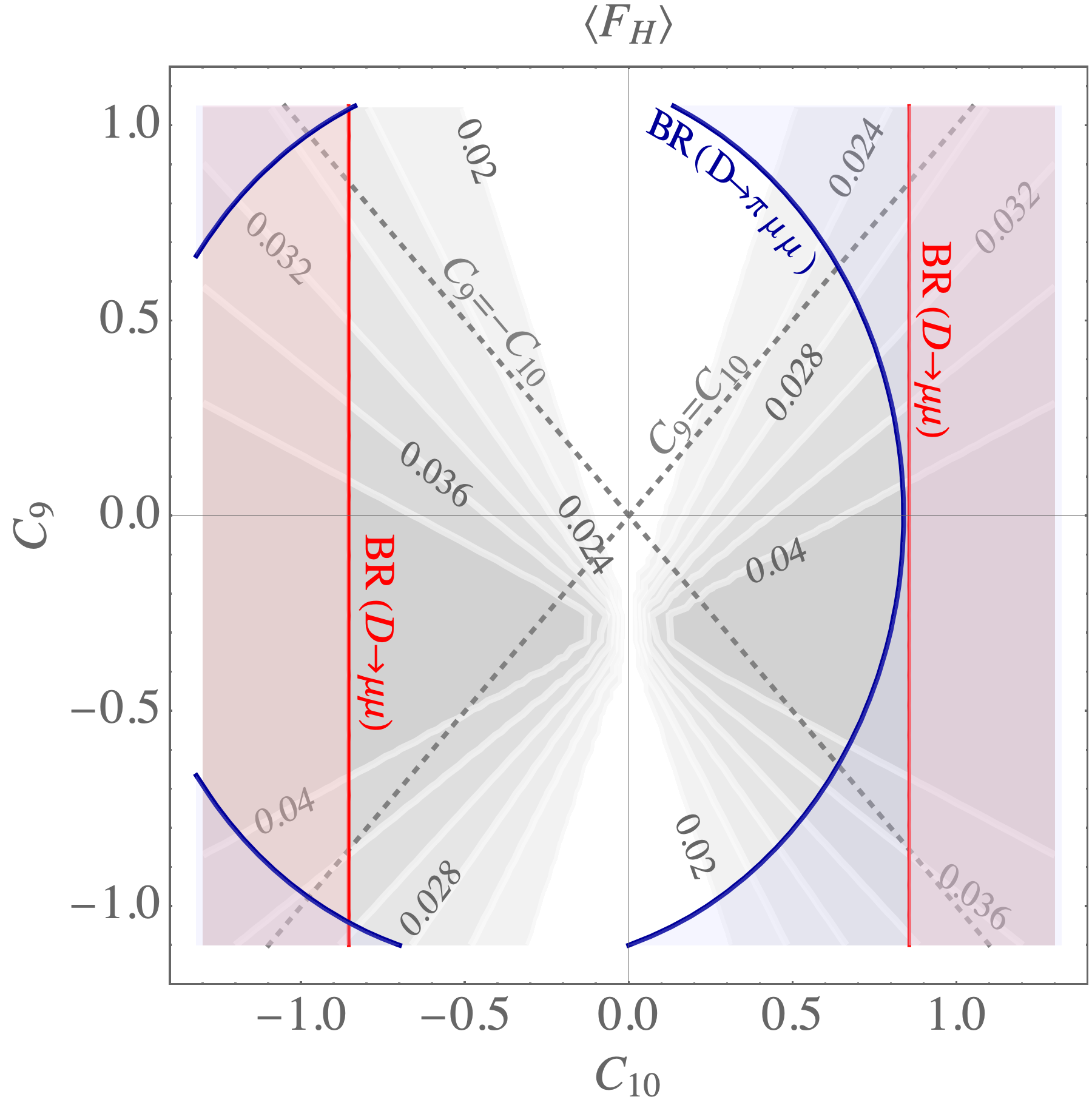}\caption{Contours (white) of the observable $\langle F_H\rangle$, as defined in the text, in the $C_{10}$ -- $C_9$ plane. The red lines and shaded areas indicate the exclusion from the LHCb bounds on the branching ratio \BRof{\Dmumu}~\cite{Aaij:2013cza}, and for illustration the red dotted line shows the exclusion on increasing the bound by a factor three. The blue lines and shaded areas indicate the exclusion from the LHCb bounds on the branching ratio \BRof{\Dpimumu}~\cite{Aaij:2013sua}.\label{Fig:FH-C10-C9}}
			\end{figure}

			In order to study the potential size of these observables in BSM physics models we start by considering $\langle F_H\rangle$, as defined in Eq.~\eqref{eq:IntObs}, integrated from  $q_{\rm min}^2$=1.8 $\mathrm{GeV}^2$ to $q_{\rm max}^2$=2.3 $\mathrm{GeV}^2$. The range is chosen for the following reasons: the OPE is valid for the case $\sqrt{q^2}\gg E_\pi,\Lambda_{\rm QCD}$; on approaching the $\phi$ resonance the uncertainties increase. Note that for those cases where the BSM contributions are real and the tensor operators vanish, this is the only observable apart from the Branching Ratio.  We start by looking at the effect of BSM contributions to $C_9$ and $C_{10}$ in Fig.~\ref{Fig:FH-C10-C9}. We include the constraints from $\mathcal{B}(D\to\mu^+\mu^-)$ and $\mathcal{B}(D^+\to\pi^+\mu^+\mu^-)$. We find that for the allowed ranges of $C_9$ and $C_{10}$, values of $\langle F_H\rangle$ almost reaching 0.04. This small value can be attributed to the fact that $C_{10}$ is restricted from \BRof{\Dmumu}. Further it turns out that the errors on $\langle F_H\rangle$ in this plane are too large to be able to differentiate between different values of the Wilson coefficients. To be more precise, given the small variation in $\langle F_H\rangle$ in the $C_9$ -- $C_{10}$ plane, the uncertainty, of order 20\% when not very close to the axis $C_{10}=0$, is too large to distinguish between different points in this plane.  Note that we have checked that our results for \FH and \AFB are compatible with those in Ref.~\cite{Bause:2019vpr} (taking into account the different definition of these observables), in spite of differences in the treatment of the decay amplitude, including the inclusion of weak annihilation and the treatment of the resonances.
			
\begin{figure}[!t]
			    \centering
			    \includegraphics[width=.6\textwidth]{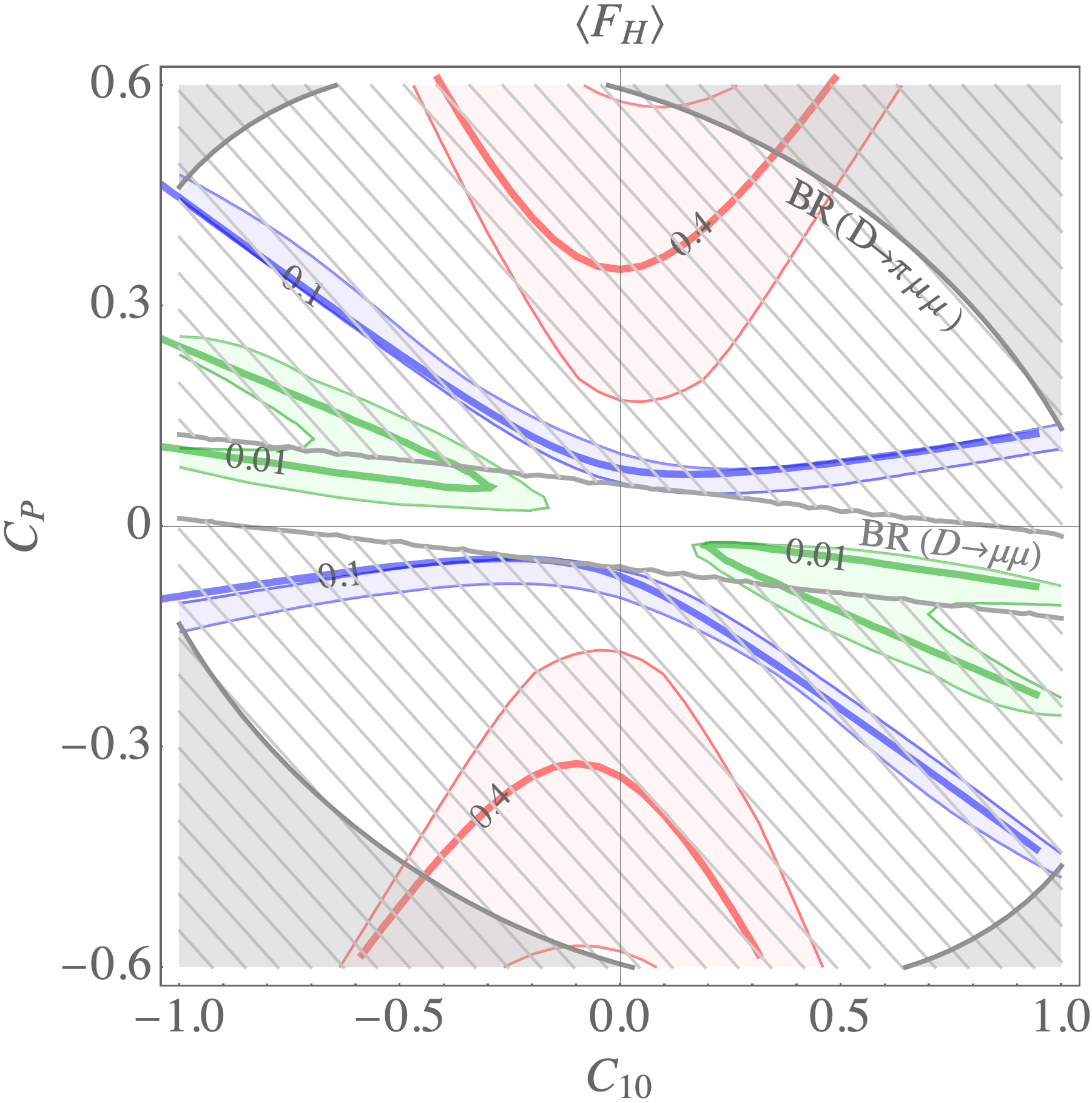}\caption{Contours of the observable $\langle F_{H}\rangle$, as defined in the text, in the $C_{10}$ -- $C_{P}$ plane.
			    The bands around the contours signify the theory uncertainty. The grey lines and shaded areas indicate the exclusion from the LHCb bounds on the branching ratios \BRof{\Dmumu}~\cite{Aaij:2013cza} and \BRof{\Dpimumu}~\cite{Aaij:2013sua} as labelled.\label{Fig:FH-C10-CP}}
			\end{figure}

	        Let us now consider a case where the uncertainties are small enough such that different values of the Wilson coefficients can be distinguished between. Since the largest contribution to the uncertainties come from the resonance structure affecting $C_9$, it turns out that observables can be predicted more accurately as a function of BSM contributions to other Wilson coefficients.
			As a first example in Fig.~\ref{Fig:FH-C10-CP} we vary $C_{10}$ and $C_{P}$, again including constraints from $\mathcal{B}(D\to\mu^+\mu^-)$ and $\mathcal{B}(D^+\to\pi^+\mu^+\mu^-)$. The uncertainty bands are shown around the contours of constant values of $\langle F_{H}\rangle$. We see that, for this case, the uncertainty bands are narrow, particularly in the upper-left and lower-right quadrants of the plane. Note that while the bounds from the di-muon $D$ decay are highly constraining, on introducing a right-handed Wilson coefficient, i.e.~making the replacements $C_{10}\to (C_{10}+C'_{10})/2$ and $C_{P}\to (C_{P}+C'_{P})/2$, the results for \Dpimumu will be unchanged and the NP contribution to the branching ratio $\mathcal{B}(D\to\mu^+\mu^-)$ will vanish. For this reason in the plots this constraint is only shown faintly. The largest value of $\langle F_{H}\rangle$ shown is 0.4, which can clearly be distinguished from the other contours shown, 0.01 and 0.1. 
			
\begin{figure}[!t]
			    \centering
			     \includegraphics[width=.6\textwidth]{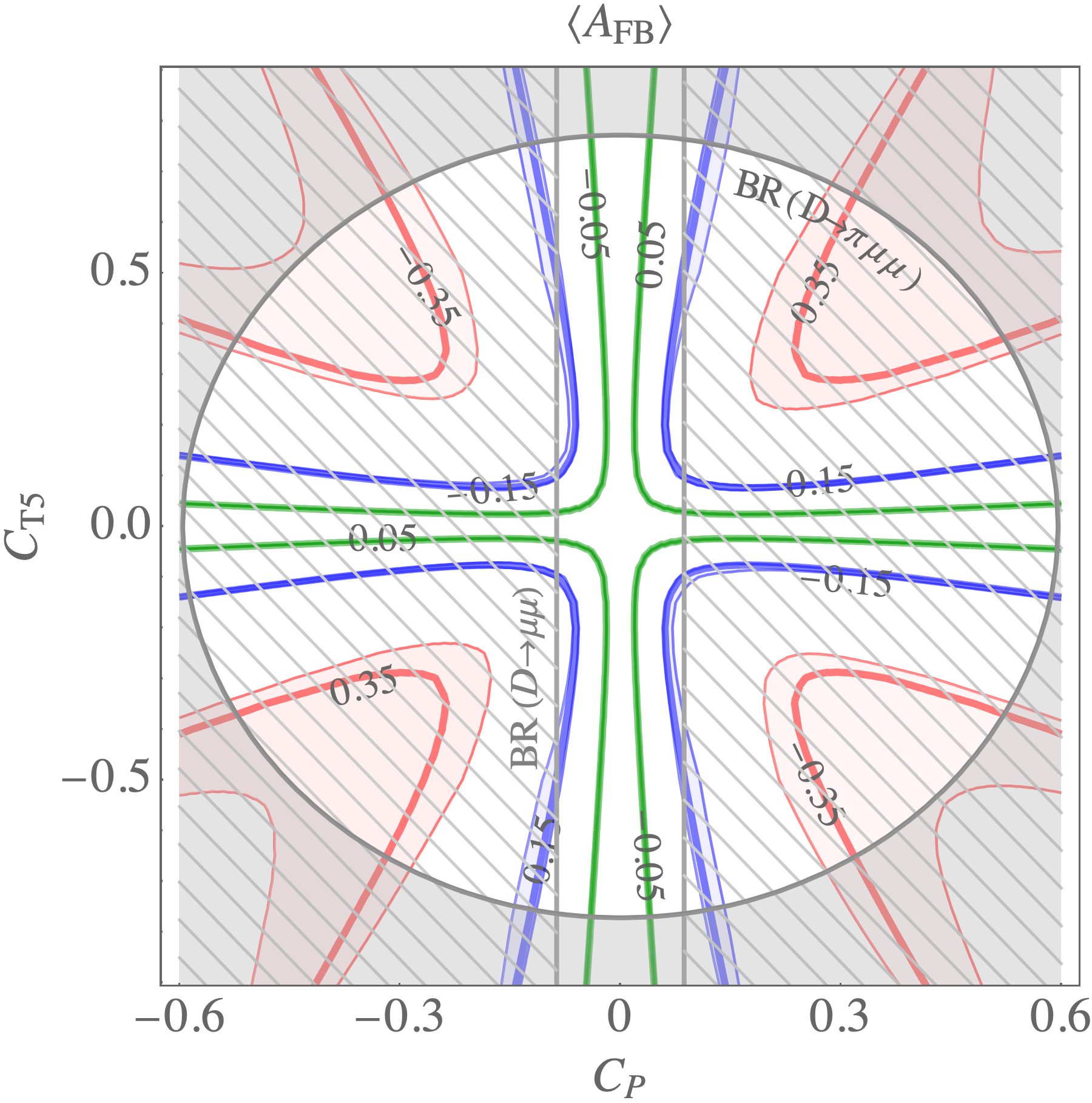}\caption{Contours (blue, red, green) of the observable $\langle A_{\rm FB}\rangle$, as defined in the text, in the $C_{P}$ -- $C_{T_5}$ plane. The bands around the contours signify the theory uncertainty. The grey lines and shaded areas indicate the exclusion from the LHCb bounds on the branching ratios \BRof{\Dmumu}~\cite{Aaij:2013cza} and \BRof{\Dpimumu}~\cite{Aaij:2013sua} as labelled.\label{Fig:AFB-CP-CT5}}
			\end{figure}

		 Now, allowing non-vanishing tensor Wilson coefficients we can  consider the observable $\langle A_{\rm FB}\rangle$, also defined in Eq.~\eqref{eq:IntObs}, which vanishes in the SM and is helicity suppressed unless the combination of Wilson coefficients $C_{S}$ and $C_{T}$ or $C_{P}$ and $C_{T_5}$ are non-zero. In Fig.~\ref{Fig:AFB-CP-CT5}, we study this observable in the $C_{P}$ and $C_{T_5}$ plane, again including constraints from $\mathcal{B}(D\to\mu^+\mu^-)$ and $\mathcal{B}(\Dpimumu)$.
		 We show the uncertainty band around the contours of constant values of $\langle A_{\rm FB}\rangle$. We find that the uncertainty on the results is small, particularly for small $C_P$ or $C_{T5}$. This seems to be a promising observable for studies of the Wilson coefficients considered, with allowed values ranging from $-0.35$ to $0.35$.
	
			  We have not yet contemplated the possibility of complex contributions to the Wilson coefficients. The ideal observable to look for CP violation is the CP asymmetry, $\langle A_{\rm CP}\rangle$ defined in Eq.~\eqref{Eq:ACP}. However, on studying this CP asymmetry we find that the theoretical uncertainties are much larger that the size of the asymmetry. Our findings slightly differ from those of Ref.~\cite{Bause:2019vpr}, this can be understood for the following reasons: unlike in Ref.~\cite{Bause:2019vpr} we choose to study the asymmetry integrated over a region in $q^2$ above the resonances, to reduce the dependence on the non-perturbative contribution, the size of the asymmetry is very much dependent on the strong phases arising due to the treatment of the resonances, which is very different in the two approaches. Note that the size of the asymmetry is larger at lower $q^2$. We therefore choose not to show plots of the CP asymmetry as it would be difficult to disentangle any BSM Physics from the hadronic physics.
            
To summarise, in Fig.~\ref{Fig:FH-C10-C9} we see the following: not only is the size of $\langle F_H\rangle$ very small when BSM contributions only affect $C_9$ and $C_{10}$, but the errors on the contours are so large that it is difficult to differentiate between them.
On the other hand in Fig.~\ref{Fig:FH-C10-CP} we see that larger values of the observable $\langle F_H\rangle$ are achievable in the $C_{10}$--$C_{P}$ plane, where the uncertainties are smaller and an experimental sensitivity below the $\mathcal{O}(10\%)$ level could already provide very interesting information. As for Fig.~\ref{Fig:AFB-CP-CT5}, we see that probing $\langle A_{\rm FB}\rangle$ down to the 10\% level would also result in important constraints on the combination of Wilson coefficients $C_P$ and $C_{T5}$, where a non-zero value of $\langle \AFB\rangle$ would be a decisive sign that either $C_P$ and $C_{T5}$ or $C_S$ and $C_{T}$ are both non-zero. On the other hand, in order to perform a precise fit to the Wilson coefficients an experimental sensitivity of $\mathcal{O}(1\%)$ would be preferable.
		
\subsubsection{The model dependent case}
	As mentioned earlier, we are interested in the possible affects of leptoquarks for the observables $\langle\FH \rangle$ and $\langle\AFB \rangle$. In Sec.~\ref{Sec:NP_Bounds} we discuss two scenarios for scalar and vector LQs which we would now like to investigate further. It turns out that the vector LQ scenarios, $\tilde{V}_1$ with quantum numbers $(3,1,-5/3)$ with $C'_9=C'_{10}$ and $\tilde{V}_2$ with $(3,2,1/6)$ with $C'_9=-C'_{10}$, were already investigated in Fig.~\ref{Fig:FH-C10-C9}, where the diagonal dashed lines represent these two scenarios as indicated. However, the uncertainties on $\langle\FH \rangle$ are too large to allow information about the Wilson Coefficients $C'_9$ and $C'_{10}$ to be extracted. We therefore turn to the scalar LQs $S_1$ and $S_2$ defined in Sec.~\ref{Sec:NP_Bounds}.
	
\begin{figure}[!t]
			    \centering
			     \includegraphics[width=.6\textwidth]{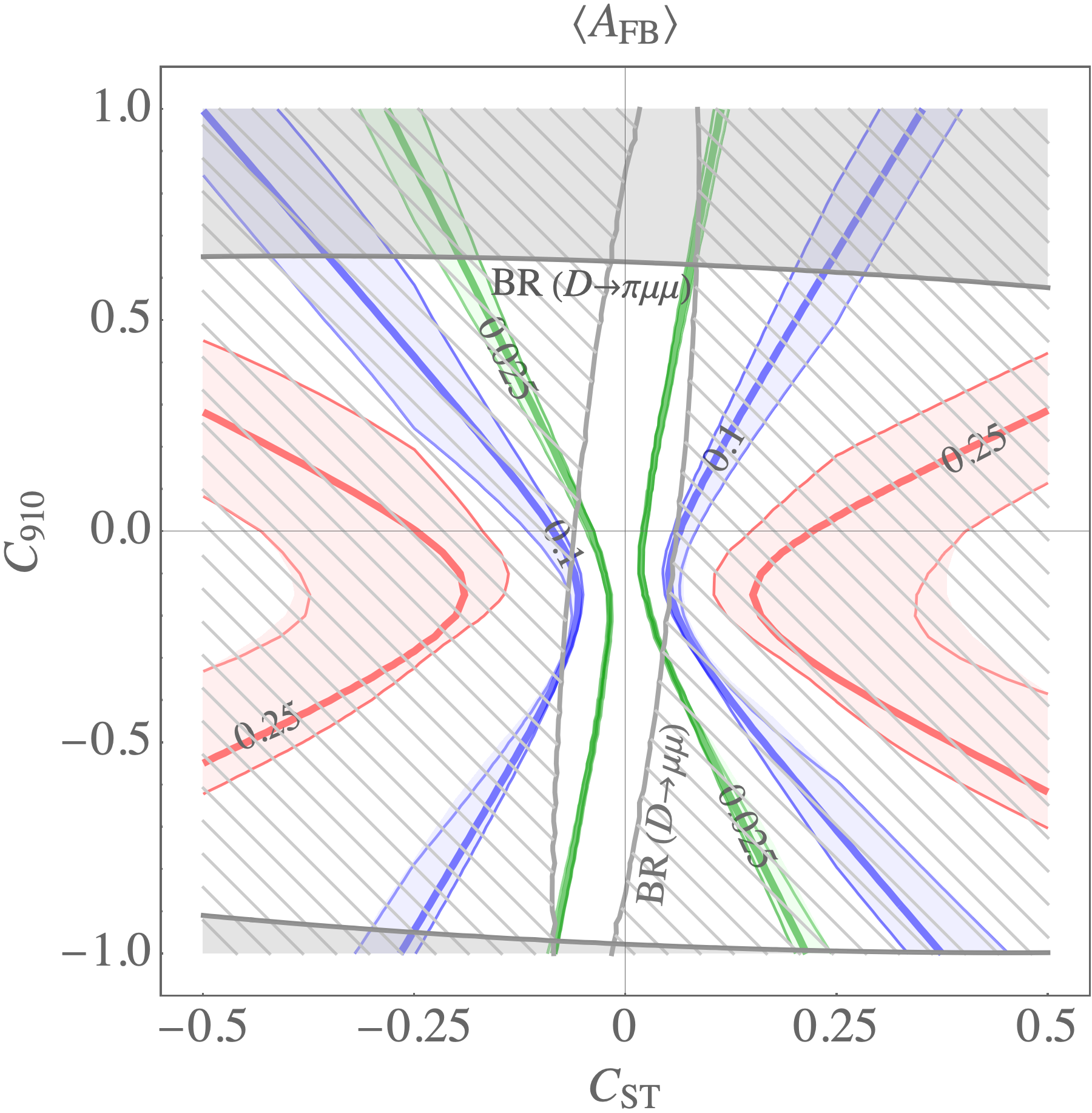}\caption{Contours (blue, red, green) of the observable $\langle \AFB\rangle$, as defined in the text, in the $C_{ST}$ -- $C_{910}$ plane. The bands around the contours signify the theory uncertainty. The grey lines and shaded areas indicate the exclusion from the LHCb bounds on the branching ratios \BRof{\Dmumu}~\cite{Aaij:2013cza} and \BRof{\Dpimumu}~\cite{Aaij:2013sua} as labelled.\label{Fig:AFB-LQ-S1}}
			\end{figure}

We begin with the case of the leptoquark $S_1$, in which there are two free parameters, $C'_ 9 = C'_ {10} \equiv C_{910}$, and $C' _S = C' _P = C_T/2 = C_ {T_ 5}/2 \equiv C_{ST}$.  In Fig.~\ref{Fig:AFB-LQ-S1} we study the observable $\langle\AFB \rangle$ in the $C_{910}$ -- $C_{ST}$ plane. We see that the uncertainty bands on the three contours shown, 0.025, 0.1 and 0.25, do not overlap, and the bands on the first two contours are particularly narrow. Note that in this scenario the primed and unprimed Wilson coefficients are fixed, such that the constraint coming from the branching ratio of $D\to\mu\mu$ cannot be avoided. This imposes strict constraints on $C_{ST}$, and only values lying approximately between $-0.1$ and 0.1 are viable, restricting $\langle\AFB \rangle$ to a maximum value of 0.1. Fortunately the uncertainties in this region are small, such that different scenarios within this region would be theoretically distinguishable.

\begin{figure}[!t]
			    \centering
			     \includegraphics[width=.6\textwidth]{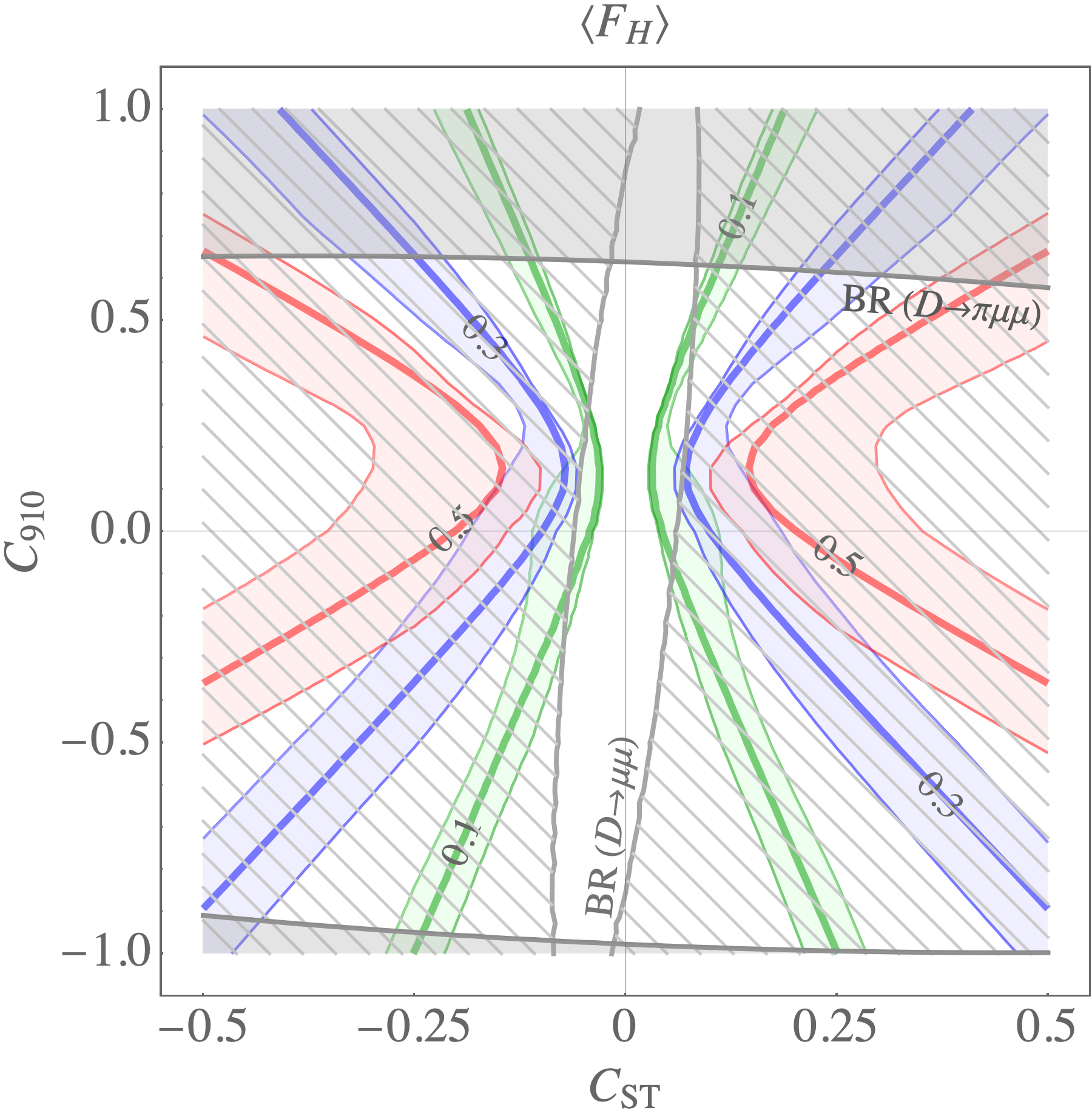}\caption{Contours (blue, red, green) of the observable $\langle \FH\rangle$, as defined in the text, in the $C_{ST}$ -- $C_{910}$ plane. The bands around the contours signify the theory uncertainty. The grey lines and shaded areas indicate the exclusion from the LHCb bounds on the branching ratios \BRof{\Dmumu}~\cite{Aaij:2013cza} and \BRof{\Dpimumu}~\cite{Aaij:2013sua} as labelled.\label{Fig:FH-LQ-S2}}
			\end{figure}

Now coming to leptoquark scenario $S_2$, in Fig.~\ref{Fig:FH-LQ-S2} we plot contours of constant $\langle \FH \rangle$ for the process $D^+\to\pi^+ e^+e^-$ in the in the $C_{910}$ -- $C_{ST}$ plane, where here  $C'_9 = - C'_{10}=C_{910}$, and $C'_S = -C'_P = C_T/2 = -C_{T_5}/2=C_{ST} $. The decision to explore the case of electrons in the final state here can be explained in terms of Eqs.~\eqref{Eq:FH} and \eqref{Eq:abc}, where we see that the dependence of the observable on the Wilson coefficients is much cleaner in the case of vanishing lepton masses. Replacing the muon mass by the electron mass not only makes it easier to interpret the results as the relationship between the observable and the Wilson coefficients is simpler, but also means that certain terms containing $F_V$ and hence the resonances are highly suppressed. This reduces the theoretical uncertainty on our predictions. From Eqs.~\eqref{Eq:abc} and \eqref{Eq:AFB_FK} we see that this reduction in uncertainty in \Dpiee is the case for the observable $\FH$ but only true for $\AFB$ when $C_T$ is present. When $C_T$ is zero (and therefore $C_{T_5}$ is non-zero in order for \AFB to be non-zero), the uncertainties cannot be reduced by choosing the electron channel. We find that again the bounds from $D\to\mu^+\mu^-$ strongly constrain the parameter space, such that the maximum value of $\langle \FH \rangle$ is 0.25 for small values of $C_{910}$ and $C_{ST}$. However in this region the theoretical uncertainties are large. Moving away from this zero point towards $C_{910}\sim 0.3$ the uncertainties decrease but only smaller values of $\langle \FH \rangle$ are attainable $\sim 0.1$.

We conclude that in order to probe these two leptoquark scenarios, uncertainties on $\langle \AFB \rangle$ (for \Dpimumu) and $\langle \FH \rangle$ (for \Dpiee) at the percent level would be required. We will comment further on the experimental feasibility of such measurements below.

			\subsubsection{Experimental prospects for \boldmath \texorpdfstring{$D^+\to\pi^+\ell^+\ell^-$}{D to pi l l}}
			The future prospects for \Dpill from LHCb, Belle-II and BES-III are very promising. Rough estimates find that the sensitivity of LHCb to the angular asymmetries and branching ratio for $D^+\to\pi^+\ell\ell$ can be found in Tab.~\ref{Tab:ExpProspects}. Note that 50 fb$^{-1}$ will only be collected with the Upgrade I at end of Run 4 in $\sim$2030. Then with the Upgrade II, by the end of Run 4 in $\sim$2038, 300 fb$^{-1}$ should be collected. Belle-II is particularly suited to the electron channel, interesting as the results can be interpreted more easily in terms of the Wilson coefficients. By scaling BaBar results~\cite{Lees:2011hb}, assuming a similar efficiency, one finds the projected upper limit of the order $10^{-8}$ on the branching ratio.\footnote{We are grateful to Vishal Bhardwaj and Guilia Casarosa from the Belle-II collaboration for providing us with this estimate} We note that this sensitivity to the branching ratio is of the same order as that of LHCb with 50 fb$^{-1}$, such that extrapolating we can say that a full angular analysis with percent-level sensitivity should also be possible at Belle-II. Finally, interesting results from BESIII in Ref.~\cite{Ablikim:2018gro} show that it will also have an important role to play for $D^+\to\pi^+\ell\ell$. 
			
			Looking back at our results in the context of these estimated future sensitivities, it seems that by the time that LHCb has 50 fb$^{-1}$ of data, the experimental errors will have far overtaken theoretical uncertainties, such that the various contours shown in our plots can easily be distinguished between. However, even before 50 fb$^{-1}$ is collected, for \FH and \AFB which are highly suppressed in the SM a sensitivity at the 10\% level could be sufficient to provide evidence for BSM physics. An experimental sensitivity at the 1\% level would be enough to perform a precise fit to Wilson coefficients, the primary hindrance being theoretical uncertainties. Note that for leptoquark scenarios, angular observables are constrained to be at most $\mathcal{O}(10\%)$, such that probing these scenarios would require sensitivities at the $\mathcal{O}(1\%)$. If Belle-II carries out the angular analysis for the electron case this will further provide important complementary information.
			\begin{table}[!t]
 \centering
 \begin{tabular}{c|cc|c}
 \toprule
  Experiment & Measurement & Sensitivity &Ref. \\
\midrule
 LHCb &Angular observables &  $\sim 0.2\%$ \,\,  with\quad 50 fb$^{-1}$,&~\cite{Prospects19,Prospects20}\\
 && $\sim 0.08\%$ \,\,with 300 fb$^{-1}$ &\\
 LHCb &Branching ratio & $\sim 10^{-8}$ \,\, with\quad 50 fb$^{-1}$,&~\cite{Prospects19}\\
 & &$\sim 3\times 10^{-9}$ with 300 fb$^{-1}$ &\\
 Belle-II &Branching ratio & $\sim 10^{-8}$ (rescaling BaBar)&~\cite{Lees:2011hb}\\
 \bottomrule
 \end{tabular}
 \caption{Summary of estimated projected experimental sensitivities from LHCb at the Upgrade I (50 fb$^{-1}$) and at the Upgrade II (300 fb$^{-1}$) for \Dpimumu and Belle-II for $D^+\to\pi^+e^+e^-$, as discussed in the text.}
 \label{Tab:ExpProspects}
\end{table}

\subsection{Consequences of updated theoretical framework on phenomenology}

In this paper we advocate a dispersive approach to describe the resonance structure, employing the Shifman model and fitting to $e^+e^-$ and $\tau$ data, as well as including weak annihilation corrections in both the low and high-$q^2$ regimes. Here we would like to address the question of how this formalism has an impact on the phenomenology. In considering this question, the following two factors are relevant: in our formalism, the residual effect of higher-order resonances are also taken into account, leading to larger uncertainties in the high-$q^2$ region; further in the high-$q^2$ region we include the weak annihilation corrections calculated in the OPE, and this constrains us to $\sqrt{q^2}\gg E_\pi,\Lambda_{\rm QCD}$. The impact on the phenomenology is as follows:
\begin{itemize}
\item  As a consequence of the range of validity of the OPE and the fact that uncertainties are still large above the $\phi$, we propose to integrate observables in the range $q^2=[1.8,2.3]~\mbox{GeV}^2$.
\item We find on allowing BSM contributions to certain Wilson coefficients (mainly $C_9$), the observables are subject to large theoretical uncertainties as a result of the resonance structure, such that, given the existing experimental bounds, determining the Wilson coefficient from the observable becomes very difficult. However the relation between the BSM contributions to certain pairs of Wilson coefficients and the integrated observables is clean, e.g.~$(C_{10},C_P)$ and $\langle \FH \rangle$ or $(C_{P},C_{T5})$ and $\langle \AFB \rangle$.
\item Within our framework we find that the uncertainty on $\ACP$ is very large (an order of magnitude large than the central value), due to the large phase uncertainty. While the strong phase of the SM contribution is unknown, differentiating between possible BSM phases would be difficult. However, a measurable $\ACP$ would probably be a sign of complex BSM couplings.
\end{itemize}
Therefore a number of differences emerge in the phenomenology as a result of the theoretical framework we adopt.

\section{Conclusions}\label{sec:Conclusions}
Charm physics is gaining increased interest, due to the large numbers of charm mesons produced at LHCb and Belle-II. 
The decay \Dpill, being a neutral current process is a particularly interesting probe of BSM physics, especially in light of potential links with the Flavour Anomalies observed in $b\to s$ transitions.

We have conducted a comprehensive analysis of this decay, tackling the increased complexity compared to its $b\to s$ counterpart resulting from the fact that the expansion in $\Lambda_{\mathrm{QCD}}/m_c$ is less effective than $\Lambda_{\mathrm{QCD}}/m_b$, and that resonances affect larger regions in phase space.
Our calculation of \Dpill includes weak annihilation within QCDf at low $q^2$ and also in the OPE at high $q^2$. To the best of our knowledge,  this is the first time both QCDf corrections have been analysed in detail in a phenomenological analysis of \Dpill, and that the OPE-approach has been applied to the weak annihilation contribution for these decays. We were motivated to include these contributions due to the fact the strong hierarchy $\lambda_b\ll\lambda_d$ means that they are not suppressed as in the $b\to s$ case, see Tab.~\ref{Tab:Num_contrib}.
Our work benefits from recent Lattice QCD results for the form factors, as well as recent calculations of the Wilson coefficients at next-to leading order.
Since the effect of resonances on the SM prediction extends over the entire kinematic regime, it is important that these are modelled carefully. We have employed a novel method, fitting the Shifman model for the resonance structure to $e^+ e^-\to{\rm (hadrons)}$ and $\tau\to{\rm (hadrons)}+\nu_\tau$ data, and further using the experimental value for the $D^+\to\pi^+\,R\,(R\to\ell^+\ell^-)$, with $R=\rho$, $\omega$, and $\phi$,  branching ratios.
We have performed a thorough and conservative error analysis involving Monte Carlo error propagation, taking into account the uncertainties from the resonance model, which dominate, as well as from the renormalisation scales.

Our prediction for the differential branching ratio as a function of $q^2$, in the entire kinematic region, along with the uncertainty band is provided in Fig.~\ref{Fig:BR_SM}. 
This can serve as a conservative prediction of the branching ratio and uncertainties throughout the phase space, which would be an important input for backgrounds in experimental searches. 
An important result is that, due to the large weak-annihilation contribution as well as the residual resonance contributions away from the resonance peaks, the integrated non-resonant branching ratios could be of the order of $10^{-9}$.
Note that the sensitivity of LHCb to the branching ratio will be $\sim 10^{-8}$ with 50 fb$^{-1}$ and $\sim 10^{-9}$ with 200 fb$^{-1}$~\cite{Prospects19}.

We studied CP conserving observables, focusing on the observables $\FH$ and $\AFB$. These are of interest as they are close to or equal to zero in the SM but can be strongly enhanced by BSM physics; to be more precise any contributions arising from the Wilson coefficients which are non-zero in the SM are helicity suppressed.
We focus on the regions at large $q^2$ where the uncertainty from the resonances is best under control, and the experimental constraints are weakest. This calls for the OPE estimate for the weak annihilation contribution, such that we integrate over a certain range in $q^2$ in order to obtain more accurate predictions.

We considered effects of both model-independent BSM and a specific model, i.e.~leptoquarks, on $\langle\FH\rangle$ and $\langle\AFB\rangle$.
We observe that BSM contributions to the vector Wilson coefficient $C_9$ are subject to large theoretical uncertainties, and do not contribute significantly to the observables $\FH$ and $\AFB$. We therefore focused on combinations $C_{10}$--$C_P$ for $\FH$ in Fig.~\ref{Fig:FH-C10-CP} and $C_P$--$C_{T5}$ for $\AFB$ in Fig.~\ref{Fig:AFB-CP-CT5}, where the uncertainties are particularly small and different values of the Wilson coefficients could be distinguished between (theoretically) in the allowed region of parameter space.
Of the leptoquark scenarios considered, we find that for vector leptoquarks $\AFB$ vanishes and $\FH$ suffers large theoretical uncertainties. On the other hand, in scalar leptoquark scenarios, while being strongly constrained by $D$ meson decays to di-muons, these observables, in particular $\AFB$, can be precisely predicted in the remaining parameter space.

We therefore look forward to the upcoming results for \Dpill from LHCb, Belle-II and BES-III (see Tab.~\ref{Tab:ExpProspects}), from which we will obtain much-improved bounds on the Wilson coefficients and the models discussed. We urge the experimental collaborations to measure the observables advocated in this paper $\langle\FH\rangle$ and $\langle\AFB\rangle$ in the range $q^2$ from $\sim$ 1.8 to 2.3 GeV$^2$, both for \Dpimumu and \Dpiee, stressing that an experimental sensitivity of $\mathcal{O}(10\%)$ to these observables would already provide evidence for BSM physics in certain scenarios, and furthermore a sensitivity of $\mathcal{O}(1\%)$ would make a precise fit to the Wilson coefficients achievable.

\section*{Acknowledgements}
We thank Vishal Bhardwaj, Guilia Casarosa, J\'{e}r\^{o}me Charles, Francesco Dettori, Danny van Dyk, Thorsten Feldmann,  Martin Hoferichter, Alexander Khodjamirian, Alberto dos Reis and Roman Zwicky for useful discussions. This work has been carried out thanks to the support of the OCEVU Labex (ANR-11-LABX-0060) and the A*MIDEX project (ANR-11- IDEX-0001-02) funded by the ”Investissements d’Avenir”. The
work of DB was supported by the S\~ao Paulo Research Foundation (FAPESP) grant
No. 2015/20689-9 and by CNPq grant No. 309847/2018-4. DB thanks the Centre de Physique Th\'eorique  (Marseille) and AB thanks the S\~ao Carlos Institute of Physics-USP (Sao Carlos)  for hospitality. 
\appendixpage
\addappheadtotoc
    \appendix
    
    \section{Form factors and light-cone distribution amplitudes}
    In this appendix we provide definitions of the form factors and light-cone distribution amplitudes, and details of the parameterisations adopted in our analysis.
     \subsection{Definition and parameterisation of the form factors}\label{App:FF}
	
        As stated previously, there are only three independent form factors needed to describe the $D$ to $\pi$ transition, the scalar $f_0$, the vector $f_+$ and the tensor $f_T$ form-factors. These are defined via
 		\begin{align}
 		\nonumber \langle \pi(k)|\bar{u}\gamma^\mu (1\pm\gamma_5) c|D (p)\rangle =&\, f_+(q^2) \left( (p+k)^\mu -q^\mu \frac{m_D^2 -m_\pi^2}{q^2} \right) + f_0(q^2) \frac{m_D^2 -m_\pi^2}{q^2}q^\mu ,\\
			\langle \pi(k)|\bar{u}\sigma^{\mu\nu}(1\pm\gamma_5) c| D(p)\rangle =&\, i\frac{f_T(q^2)}{m_D+m_\pi} \biggr( (p+k)^\mu q^\nu - (p+k)^\nu q^\mu \nonumber \\
			&\quad
			\pm i \epsilon^{\mu\nu\alpha\beta}(p+k)_{\alpha} q_\beta \biggr).
			\label{Eq:FF_T}
		\end{align}
			Note that, at zero momentum transfer, the relation $f_+(0) = f_0(0)$ holds.
			
\begin{figure}[!t]
				\center
				\includegraphics[width=.6\textwidth]{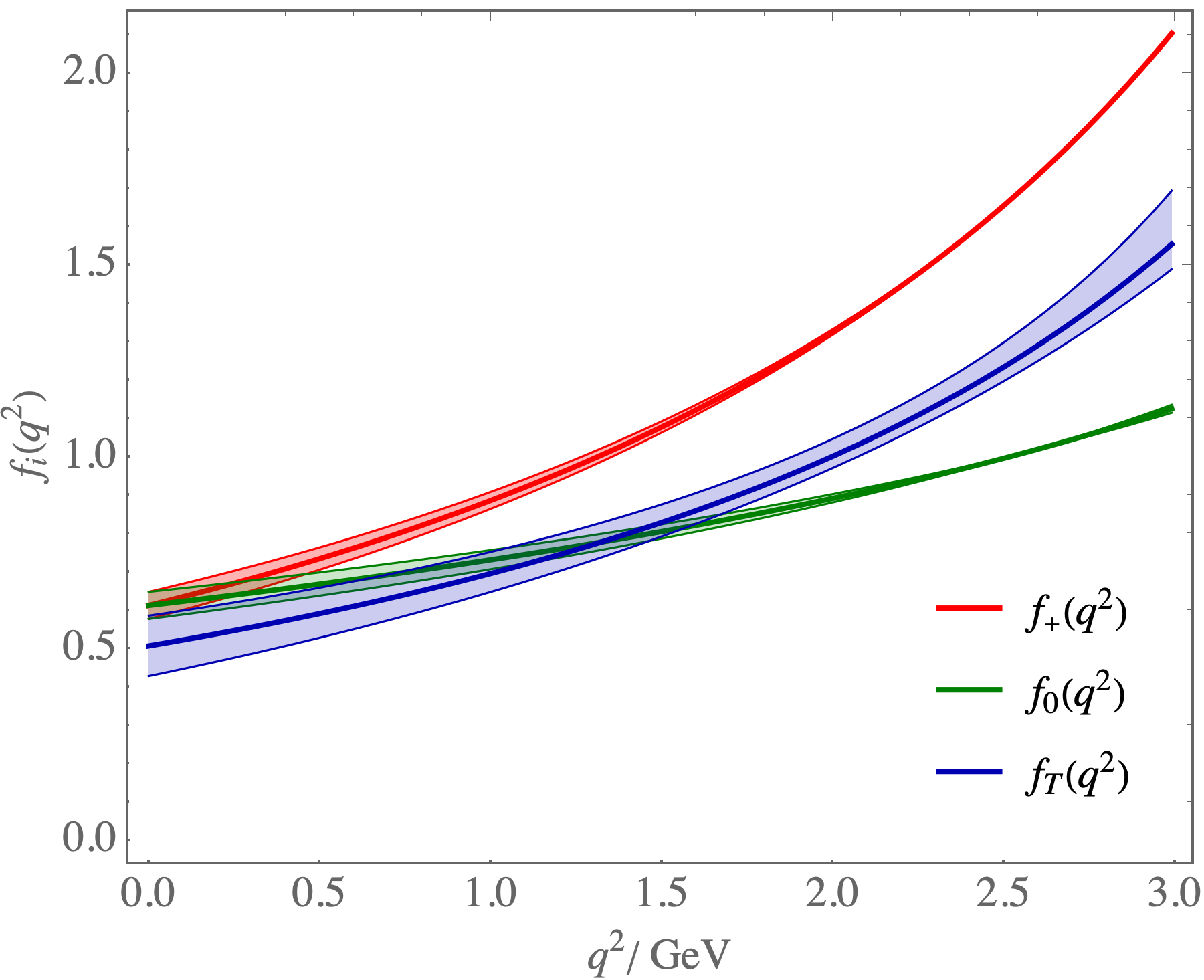}
				\caption{Form factors for $D\to\pi$, $f_+(q^2)$ and $f_0(q^2)$ from Ref.~\cite{Lubicz:2017syv} and $f_T(q^2)$ from Ref.~\cite{Lubicz:2018rfs}}
				\label{Fig:FF}
			\end{figure}

	We adopt recent Lattice QCD results for the vector and scalar $f_+$ and $f_0$ form factors from Ref.~\cite{Lubicz:2017syv} and for the the tensor form factor $f_T$ from Ref.~\cite{Lubicz:2018rfs}. These results are provided in terms of coefficients parameterising an expansion in the variable $z(t,t_0)$ where
			\beq
				z(t,t_0) = \frac{\sqrt{t_+ - t} - \sqrt{t_+ -t_0}}{\sqrt{t_+ - t} + \sqrt{t_+ - t_0}},\quad \mbox{and}\quad z_0 \equiv z(0,t_0)
			\eeq
for $t_+= (m_D +m_\pi)^2$ and $t_0=(m_D+m_\pi)(\sqrt{m_D}-\sqrt{m_\pi})^2$.
The parameterisation is then written as
			\begin{align}
\nonumber		\hspace{-1cm}		f_+(q^2) =\,&\frac{1}{1-P_V q^2}\left[f(0) + c_+ (z(q^2,t_0)-z_0)\left(1+\frac{z(q^2,t_0)+z_0}{2}\right)\right],\\
\nonumber				f_0(q^2) =\,&\frac{1}{1-P_S\,q^2}\left[f(0) \,+ c_0\, (z(q^2,t_0)-z_0)\left(1+\frac{z(q^2,t_0)+z_0}{2}\right)\right],\\
				f_T(q^2) =\,& \frac{1}{1-P_T q^2}\left[f_T(0) + c_T (z(q^2,t_0)-z_0)\left(1+\frac{z(q^2,t_0)+z_0}{2}\right)\right].
			\end{align}
 The value of the parameters $f(0)$, $f_T(0)$, $c_+$, $c_0$, $c_T$, $P_V$, $P_S$ and  $P_T  $ are given in Tab.~\ref{Tab:Num_inputs}. The distributions for the three form-factors are shown in Fig.~\ref{Fig:FF}. They are found to be in good agreement with the ones obtained by LCSR calculation in Ref. \cite{Khodjamirian:2009ys}.

		\subsection{Definition and parametrisation of the \boldmath \texorpdfstring{$\pi$}{pi} and \texorpdfstring{$D$}{D} meson LCDAs}\label{App:LCDA}

	The LCDA of the $\pi$ meson, $ \phi_\pi(u,\mu^2)$, at the scale $\mu$ and with fractional momentum of the quark $u$ can be defined by \cite{Braun:1988qv}:
		\beq
			\langle \pi(k)|\bar u(0)\gamma_\mu \gamma_5 d(x) | 0\rangle = -i f_\pi k_\mu \int_0^1 du~e^{i \bar u k \cdot x} \phi_\pi(u,\mu^2).
			\label{Eq:LCDA_pi}
		\eeq
			The standard parametrisation for $ \phi_\pi(u,\mu^2)$ is via a series in Gegenbauer polynomials $C_n^\alpha (z)$:
		    \beq
		    	\phi_\pi(u, \mu^2) = 6u(1-u)\sum_{n=0}^{\infty} a_n(\mu^2)C_n^{3/2}(2u-1),
		   	\eeq
		   where $a_n(\mu^2)$ are the Gegenbauer moments, $a_0=1$ from the normalisation condition 
		     
		    and the odd moments are equal to zero for the case of pions. 
		    The values of these moments at the scale $\mu=1~\GeV$ is taken from \cite{Bharucha:2012wy}:
		    \beq
		    	\quad a_2(1\text{ GeV}) =0.17 \pm 0.08, \quad a_4(1\text{ GeV}) = 0.06\pm0.10.
		    \eeq
		   	The moments are required at the scale $\mu_c$, obtained via \cite{Ball:2004ye}
		    \beq
		   		c_i(\mu) = L^{\gamma_{c_i}/\beta_0} c_i(1~\GeV),
		   	\eeq
		   	where $L=\alpha_s(\mu)/\alpha_s(1~\GeV)$ and $\gamma_{c_i}$ is given by
		   	\beq
		   		\gamma_{c_i} = C_F \left(1-\frac{2}{(n+1)(n+2)} - \sum_{m=2}^{n+1} \frac{1}{m}\right).
		   	\eeq

	The LCDA of the $D$ meson is defined by (see e.g.~Ref.~\cite{Beneke:2004dp})
		\beq
			\langle 0|\bar{d}_\beta(z) P(z,0)c_\alpha(0)|\bar D(p)\rangle = -\frac{i f_D m_D}{4} \left[ \frac{1+\cancel{v}}{2} \left( 2\phi_D^+(t) + \frac{\phi_D^-(t) - \phi_D^+(t) }{t} \cancel{z}\right)\right]_{\alpha \beta},
			\label{Eq:LCDA_D}
		\eeq
		where $v$ is the velocity of the $D$ meson.
		    We choose to parametrise the $D$ meson LCDA using a simple exponential model~\cite{Grozin:1996pq}:
		    \beq
		    	\phi_D^-(\omega)=\frac{1}{\omega_0}e^{-\omega/\omega_0}, \qquad  \phi_D^+(\omega)=\frac{1}{\omega_0^2}e^{-\omega/\omega_0},
		    	\label{Eq:Dmeson_LCDA}
		    \eeq
			where $\omega_0$ is the sole input parameter.

    \section{Calculation of the Wilson coefficients} \label{App:WC}

Here we recall the two-step running of the Wilson coefficients for the \cull transition~\cite{Boer_WC}. These coefficients are computed at the electroweak scale $\mu_W \sim M_W$ and then run down to the typical mass scale of the decay under consideration, here $\mu_c \sim m_c$. This running involves an intermediate scale $\mu_b \sim m_b$, where the bottom quark is ``integrated out''. The matching coefficients and anomalous dimension matrices are taken to the required order by generalising and extending results from $b\to d/s$ transitions~\cite{BFS2001}. The Wilson coefficients for the $c\to u\tl$ transition at NNLL accuracy were calculated for the first time in Ref.~\cite{Boer_WC}.

		The full procedure to calculate the Wilson coefficients can be broken down in the following steps:
		\begin{itemize}

			\item We determine the Wilson coefficients via matching at $\mu \sim \mu_W$ to second order in $a_s(\mu_W)$, of which only $C_{1/2}$ receive non-zero contributions~\cite{Boer_WC}.

			\item Secondly, $C_1$ and $C_2$ are run down to the scale $\mu_b$ following Ref.~\cite{BFS2001}.
			The full $8\times8$ anomalous dimension matrices, $\gamma^i$ where $i=0$--$2$, required for the RGE running can be broken down in the following way
			\beq
				\gamma^{(i)} \equiv
				\begin{pmatrix}
				    Q_1^{(i)} & Q_2^{(i)} \\
				    Q_3^{(i)} & Q_4^{(i)}
				\end{pmatrix},
			\eeq
			where $Q_1^{(i)}$ are the $6\times6$ three-loop anomalous dimension matrices describing the mixing of the four quark operators $C_{1-6}$, they are taken from \cite{Gorbahn:2004my}.  
			 
			$Q_2^{(i)}$ are the $6\times2$ matrices describing the mixing of the four quark operators and the dipole operators $C_{7/8}$, taken from Ref.~\cite{Boer_WC}. 
			 
			$Q_3^{(i)}=0$ at any order. Finally, the $2\times2$ matrix from self-mixing in the dipole operator sector, $Q_4^{(i)}$, is extracted from \cite{Gorbahn:2005sa}.
			  
			\item We then perform the matching from the five-quark ($n_f=5$) to the four-quark ($n_f=4$) effective field theory at the scale $\mu_b$ following Ref.~\cite{Boer_WC}. The matching matrix $R$ is different from the unit matrix because the operators $\res_{1/2}^b$ are absent below the $b$-quark threshold. $C_{3/9}$ receive non-zero contributions only from the matching of the five-flavor effective  theory above the scale $\mu_b$ to the four-flavor EFT below that scale and from the mixing of $\res_{1/2}$ into $\res_{3/9}$. The $R$ matrix is given by:
			\beq
				R= \delta_{ij} + a_s(m_b) R^{(1)}_{ij}+...,
			\eeq 
			where the non-zero elements of $R^{(1)}_{ij}$ can be found in Ref.~\cite{Boer_WC}.
			\item Finally, $C_{1-8}$ are run down to the scale $\mu_c$ via the  $8\times8$ evolution matrix $U_2(\mu_c, \mu_b)$, again calculated following Ref.~\cite{BFS2001}.
			\item 		For $C_9$, the running procedure follows Ref.~\cite{Boer_WC} and in turn Ref.~\cite{BFS2001}, where the six-dimensional list of 4-quark operators at the weak scale, $C'(\mu_W)$, is evolved down to the scale $\mu_b$ using the six-dimensional anomalous dimension matrices $Q_1^{(i)}$, and matched using the six-dimensional $R_6$. 
			\item Below $\mu_b$ the contribution of these operators to $\res_9$ are calculated via the
		$1\times6$ matrix $W^{n_f=4}(\mu_c,\mu_b)$,
		\beq
			W^{n_f=4}(\mu_c, \mu_b ) = -\frac{1}{2}\int_{a_s(\mu_b)}^{a_s(\mu_c)} da_s \frac{\kappa(a_s)}{\beta(a_s)}U^{n_f=4}(\mu_c, \mu_b).
			\label{Eq:W}
		\eeq
		Here $U_6^{n_f=4}(\mu_c, \mu_b)$ and $R_6$ are $6\times6$ sub-matrices of  $U^{n_f=4}(\mu_c, \mu_b)$ and $R$ are $6\times6$ mentioned earlier for the $C_1$--$C_8$ case. The solution of Eq.~(\ref{Eq:W}) can be found in appendix C of \cite{BFS2001}. The leading order contribution to the initial condition for $C_9$ at the scale $m_b$  is given by
			\beq
			C_9(\mu_b) = -\frac{8}{27}\left[C_1(\mu_b)+ \frac{3}{4}C_2(\mu_b)\right].
		\eeq
		\end{itemize}
		The full procedure  for $C_1$--$C_8$ can be summarised by the following equation:
		\beq
			C(\mu) = U^{(n_f=4)}(\mu, \mu_b) ~R~ U^{(n_f=5)}(\mu_b, \mu_W) ~C(\mu_W)
			\label{Eq:FullEqC18}
		\eeq
		and for $C_9$ by 
		\beq
			C_9(\mu) = C_9(\mu_b) + W^{n_f=4}(\mu, \mu_b) \,R_6\,  U_6^{n_f=5}(\mu_b, \mu_W)~ C'(\mu_W).
		\eeq

		Note that $C_{10}$ does not mix under renormalisation and thus is zero at all scales to leading order in the $1/M_W$ expansion. Moreover, the framework introduced above does not compute the coefficients $C_{7/8}$ but the renormalisation-scheme independent effective ones, defined as:

		\beq
			C_{7/8}^{\rm eff} = C_{7/8} + \sum_{i=1}^6 y_i^{(7/8)} C_i,
		\eeq

		with $y^{(7)} = Q_d \left( 0, 0, 1, \frac{4}{3}, 20, \frac{80}{3} \right)$, $Q_d=-1/3$ and  $y^{(8)} = \left( 0, 0, 1, -\frac{1}{6}, 20, -\frac{10}{3} \right)$.

\bibliographystyle{JHEP-2-2}
\bibliography{Biblio/All.bib}

\providecommand{\href}[2]{#2}\begingroup\raggedright\begin{thebibliography}{10}

\bibitem{Crivellin:2016vjc}
A.~Crivellin, G.~D'Ambrosio, M.~Hoferichter and L.~C. Tunstall, ``{Violation of
  lepton flavor and lepton flavor universality in rare kaon
  decays},''\href{http://dx.doi.org/10.1103/PhysRevD.93.074038}{\emph{Phys.
  Rev. D} {\bf 93} (2016) 074038},
  [\href{https://arxiv.org/abs/1601.00970}{{\tt 1601.00970}}].

\bibitem{Aaij:2013mga}
{\scshape LHCb} collaboration, R.~Aaij et~al., ``{Prompt charm production in pp
  collisions at sqrt(s)=7
  TeV},''\href{http://dx.doi.org/10.1016/j.nuclphysb.2013.02.010}{\emph{Nucl.
  Phys. B} {\bf 871} (2013) 1--20},
  [\href{https://arxiv.org/abs/1302.2864}{{\tt 1302.2864}}].

\bibitem{Aaij:2010gn}
{\scshape LHCb} collaboration, R.~Aaij et~al., ``{Measurement of $\sigma(pp \to
  b \bar{b} X)$ at $\sqrt{s}=7~\rm{TeV}$ in the forward
  region},''\href{http://dx.doi.org/10.1016/j.physletb.2010.10.010}{\emph{Phys.
  Lett. B} {\bf 694} (2010) 209--216},
  [\href{https://arxiv.org/abs/1009.2731}{{\tt 1009.2731}}].

\bibitem{dBH15}
S.~de~Boer and G.~Hiller, ``{Flavor and new physics opportunities with rare
  charm decays into
  leptons},''\href{http://dx.doi.org/10.1103/PhysRevD.93.074001}{\emph{Phys.
  Rev.} {\bf D93} (2016) 074001}, [\href{https://arxiv.org/abs/1510.00311}{{\tt
  1510.00311}}].

\bibitem{FK2015}
S.~Fajfer and N.~Košnik, ``{Prospects of discovering new physics in rare charm
  decays},''\href{http://dx.doi.org/10.1140/epjc/s10052-015-3801-2}{\emph{Eur.
  Phys. J.} {\bf C75} (2015) 567},
  [\href{https://arxiv.org/abs/1510.00965}{{\tt 1510.00965}}].

\bibitem{FMS2017}
T.~Feldmann, B.~Müller and D.~Seidel, ``{$D \to \rho \,\ell^+\ell^-$ decays in
  the QCD factorization
  approach},''\href{http://dx.doi.org/10.1007/JHEP08(2017)105}{\emph{JHEP} {\bf
  08} (2017) 105}, [\href{https://arxiv.org/abs/1705.05891}{{\tt 1705.05891}}].

\bibitem{LZ14}
J.~Lyon and R.~Zwicky, ``{Resonances gone topsy turvy - the charm of QCD or new
  physics in $b \to s \ell^+ \ell^-$?},''
  \href{https://arxiv.org/abs/1406.0566}{{\tt 1406.0566}}.

\bibitem{Kruger:1996cv}
F.~Kruger and L.~M. Sehgal, ``{Lepton polarization in the decays b
  ---\ensuremath{>} X(s) mu+ mu- and B ---\ensuremath{>} X(s) tau+
  tau-},''\href{http://dx.doi.org/10.1016/0370-2693(96)00413-3}{\emph{Phys.
  Lett. B} {\bf 380} (1996) 199--204},
  [\href{https://arxiv.org/abs/hep-ph/9603237}{{\tt hep-ph/9603237}}].

\bibitem{Fajfer:2001sa}
S.~Fajfer, S.~Prelovsek and P.~Singer, ``{Rare charm meson decays $D \to P
  \ell^+ \ell^-$ and $c\to u \ell^+ \ell^-$ in SM and
  MSSM},''\href{http://dx.doi.org/10.1103/PhysRevD.64.114009}{\emph{Phys. Rev.}
  {\bf D64} (2001) 114009}, [\href{https://arxiv.org/abs/hep-ph/0106333}{{\tt
  hep-ph/0106333}}].

\bibitem{Fajfer:2005ke}
S.~Fajfer and S.~Prelovsek, ``{Effects of littlest Higgs model in rare D meson
  decays},''\href{http://dx.doi.org/10.1103/PhysRevD.73.054026}{\emph{Phys.
  Rev.} {\bf D73} (2006) 054026},
  [\href{https://arxiv.org/abs/hep-ph/0511048}{{\tt hep-ph/0511048}}].

\bibitem{Fajfer:2007dy}
S.~Fajfer, N.~Kosnik and S.~Prelovsek, ``{Updated constraints on new physics in
  rare charm
  decays},''\href{http://dx.doi.org/10.1103/PhysRevD.76.074010}{\emph{Phys.
  Rev.} {\bf D76} (2007) 074010}, [\href{https://arxiv.org/abs/0706.1133}{{\tt
  0706.1133}}].

\bibitem{Bause:2019vpr}
R.~Bause, M.~Golz, G.~Hiller and A.~Tayduganov, ``{The new physics reach of
  null tests with $D \rightarrow \pi \ell \ell $ and $D_s \rightarrow K \ell
  \ell $
  decays},''\href{http://dx.doi.org/10.1140/epjc/s10052-020-7621-7}{\emph{Eur.
  Phys. J. C} {\bf 80} (2020) 65},
  [\href{https://arxiv.org/abs/1909.11108}{{\tt 1909.11108}}].

\bibitem{Bause:2020obd}
R.~Bause, H.~Gisbert, M.~Golz and G.~Hiller, ``{Exploiting $CP$-asymmetries in
  rare charm
  decays},''\href{http://dx.doi.org/10.1103/PhysRevD.101.115006}{\emph{Phys.
  Rev. D} {\bf 101} (2020) 115006},
  [\href{https://arxiv.org/abs/2004.01206}{{\tt 2004.01206}}].

\bibitem{Beylich:2011aq}
M.~Beylich, G.~Buchalla and T.~Feldmann, ``{Theory of $B \to K^{(*)}\ell^+
  \ell^-$ decays at high $q^2$: OPE and quark-hadron
  duality},''\href{http://dx.doi.org/10.1140/epjc/s10052-011-1635-0}{\emph{Eur.
  Phys. J. C} {\bf 71} (2011) 1635},
  [\href{https://arxiv.org/abs/1101.5118}{{\tt 1101.5118}}].

\bibitem{Adolph:2020ema}
N.~Adolph, J.~Brod and G.~Hiller, ``{Radiative three-body D-meson decays in and
  beyond the standard model},'' \href{https://arxiv.org/abs/2009.14212}{{\tt
  2009.14212}}.

\bibitem{deBoer:2017que}
S.~de~Boer and G.~Hiller, ``{Rare radiative charm decays within the standard
  model and
  beyond},''\href{http://dx.doi.org/10.1007/JHEP08(2017)091}{\emph{JHEP} {\bf
  08} (2017) 091}, [\href{https://arxiv.org/abs/1701.06392}{{\tt 1701.06392}}].

\bibitem{Lubicz:2017syv}
{\scshape ETM} collaboration, V.~Lubicz, L.~Riggio, G.~Salerno, S.~Simula and
  C.~Tarantino, ``{Scalar and vector form factors of $D \to \pi(K) \ell \nu$
  decays with $N_f=2+1+1$ twisted
  fermions},''\href{http://dx.doi.org/10.1103/PhysRevD.96.054514}{\emph{Phys.
  Rev.} {\bf D96} (2017) 054514}, [\href{https://arxiv.org/abs/1706.03017}{{\tt
  1706.03017}}].

\bibitem{Lubicz:2018rfs}
{\scshape ETM} collaboration, V.~Lubicz, L.~Riggio, G.~Salerno, S.~Simula and
  C.~Tarantino, ``{Tensor form factor of $D \to \pi(K) \ell \nu$ and $D \to
  \pi(K) \ell \ell$ decays with $N_f=2+1+1$ twisted-mass
  fermions},''\href{http://dx.doi.org/10.1103/PhysRevD.98.014516}{\emph{Phys.
  Rev.} {\bf D98} (2018) 014516}, [\href{https://arxiv.org/abs/1803.04807}{{\tt
  1803.04807}}].

\bibitem{Boer_WC}
S.~de~Boer, B.~Müller and D.~Seidel, ``{Higher-order Wilson coefficients for
  $c \to u$ transitions in the standard
  model},''\href{http://dx.doi.org/10.1007/JHEP08(2016)091}{\emph{JHEP} {\bf
  08} (2016) 091}, [\href{https://arxiv.org/abs/1606.05521}{{\tt 1606.05521}}].

\bibitem{BFS2001}
M.~Beneke, T.~Feldmann and D.~Seidel, ``{Systematic approach to exclusive $B
  \to V l^+ l^-$, $V \gamma$
  decays},''\href{http://dx.doi.org/10.1016/S0550-3213(01)00366-2}{\emph{Nucl.
  Phys.} {\bf B612} (2001) 25--58},
  [\href{https://arxiv.org/abs/hep-ph/0106067}{{\tt hep-ph/0106067}}].

\bibitem{Beneke:2004dp}
M.~Beneke, T.~Feldmann and D.~Seidel, ``{Exclusive radiative and electroweak $b
  \to d$ and $b \to s$ penguin decays at
  NLO},''\href{http://dx.doi.org/10.1140/epjc/s2005-02181-5}{\emph{Eur. Phys.
  J.} {\bf C41} (2005) 173--188},
  [\href{https://arxiv.org/abs/hep-ph/0412400}{{\tt hep-ph/0412400}}].

\bibitem{Lyon:2012fk}
J.~Lyon and R.~Zwicky, ``{Anomalously large ${\cal O}_8$ and long-distance
  chirality from $A_{\rm CP}[D^0 \to (\rho^0,\omega) \gamma](t)$},''
  \href{https://arxiv.org/abs/1210.6546}{{\tt 1210.6546}}.

\bibitem{Lyon:2013gba}
J.~Lyon and R.~Zwicky, ``{Isospin asymmetries in $B\to(K^*,\rho)\gamma/l^+l^-$
  and $B\to Kl^+l^-$ in and beyond the standard
  model},''\href{http://dx.doi.org/10.1103/PhysRevD.88.094004}{\emph{Phys. Rev.
  D} {\bf 88} (2013) 094004}, [\href{https://arxiv.org/abs/1305.4797}{{\tt
  1305.4797}}].

\bibitem{Grinstein:2004vb}
B.~Grinstein and D.~Pirjol, ``{Exclusive rare $B \to K^*\ell^+\ell^-$ decays at
  low recoil: Controlling the long-distance
  effects},''\href{http://dx.doi.org/10.1103/PhysRevD.70.114005}{\emph{Phys.
  Rev. D} {\bf 70} (2004) 114005},
  [\href{https://arxiv.org/abs/hep-ph/0404250}{{\tt hep-ph/0404250}}].

\bibitem{Shifman:2000jv}
M.~A. Shifman, ``{Quark hadron duality},'' in \emph{{At the frontier of
  particle physics. Handbook of QCD. Vol. 1-3}}, (Singapore), pp.~1447--1494,
  World Scientific, World Scientific, 2001.
\newblock \href{https://arxiv.org/abs/hep-ph/0009131}{{\tt hep-ph/0009131}}.
\newblock \href{http://dx.doi.org/10.1142/9789812810458_0032}{DOI}.

\bibitem{Boito:2018yvl}
D.~Boito, M.~Golterman, A.~Keshavarzi, K.~Maltman, D.~Nomura, S.~Peris et~al.,
  ``{Strong coupling from $e^+e^-\to$ hadrons below
  charm},''\href{http://dx.doi.org/10.1103/PhysRevD.98.074030}{\emph{Phys.\
  Rev.\ D} {\bf 98} (2018) 074030},
  [\href{https://arxiv.org/abs/1805.08176}{{\tt 1805.08176}}].

\bibitem{Daub:2015xja}
J.~Daub, C.~Hanhart and B.~Kubis, ``{A model-independent analysis of
  final-state interactions in $ {\overline{B}}_{d/s}^0\to J/\psi \pi \pi
  $},''\href{http://dx.doi.org/10.1007/JHEP02(2016)009}{\emph{JHEP} {\bf 02}
  (2016) 009}, [\href{https://arxiv.org/abs/1508.06841}{{\tt 1508.06841}}].

\bibitem{Davier:2013sfa}
M.~Davier, A.~Höcker, B.~Malaescu, C.-Z. Yuan and Z.~Zhang, ``{Update of the
  ALEPH non-strange spectral functions from hadronic $\tau$
  decays},''\href{http://dx.doi.org/10.1140/epjc/s10052-014-2803-9}{\emph{Eur.\
  Phys.\ J.\ C} {\bf 74} (2014) 2803},
  [\href{https://arxiv.org/abs/1312.1501}{{\tt 1312.1501}}].

\bibitem{Ackerstaff:1998yj}
{\scshape OPAL} collaboration, K.~Ackerstaff et~al., ``{Measurement of the
  strong coupling constant alpha(s) and the vector and axial vector spectral
  functions in hadronic tau
  decays},''\href{http://dx.doi.org/10.1007/s100529901061}{\emph{Eur.\ Phys.\
  J.\ C} {\bf 7} (1999) 571--593},
  [\href{https://arxiv.org/abs/hep-ex/9808019}{{\tt hep-ex/9808019}}].

\bibitem{Boito:2012cr}
D.~Boito, M.~Golterman, M.~Jamin, A.~Mahdavi, K.~Maltman, J.~Osborne et~al.,
  ``{An Updated determination of $\alpha_s$ from $\tau$
  decays},''\href{http://dx.doi.org/10.1103/PhysRevD.85.093015}{\emph{Phys.\
  Rev.\ D} {\bf 85} (2012) 093015},
  [\href{https://arxiv.org/abs/1203.3146}{{\tt 1203.3146}}].

\bibitem{Boito:2014sta}
D.~Boito, M.~Golterman, K.~Maltman, J.~Osborne and S.~Peris, ``{Strong coupling
  from the revised ALEPH data for hadronic $\tau$
  decays},''\href{http://dx.doi.org/10.1103/PhysRevD.91.034003}{\emph{Phys.\
  Rev.\ D} {\bf 91} (2015) 034003},
  [\href{https://arxiv.org/abs/1410.3528}{{\tt 1410.3528}}].

\bibitem{PDG2018}
{\scshape Particle Data Group} collaboration, M.~Tanabashi et~al., ``{Review of
  Particle
  Physics},''\href{http://dx.doi.org/10.1103/PhysRevD.98.030001}{\emph{Phys.
  Rev.} {\bf D98} (2018) 030001}.

\bibitem{Ablikim:2009ad}
{\scshape BES} collaboration, M.~Ablikim et~al., ``{R value measurements for e+
  e- annihilation at 2.60-GeV, 3.07-GeV and
  3.65-GeV},''\href{http://dx.doi.org/10.1016/j.physletb.2009.05.055}{\emph{Phys.\
  Lett.\ B} {\bf 677} (2009) 239--245},
  [\href{https://arxiv.org/abs/0903.0900}{{\tt 0903.0900}}].

\bibitem{Anashin:2015woa}
V.~Anashin et~al., ``{Measurement of $R_{\text{uds}}$ and $R$ between 3.12 and
  3.72 GeV at the KEDR
  detector},''\href{http://dx.doi.org/10.1016/j.physletb.2015.12.059}{\emph{Phys.\
  Lett.\ B} {\bf 753} (2016) 533--541},
  [\href{https://arxiv.org/abs/1510.02667}{{\tt 1510.02667}}].

\bibitem{Anashin:2016hmv}
V.~V. Anashin et~al., ``{Measurement of $R$ between 1.84 and 3.05 GeV at the
  KEDR
  detector},''\href{http://dx.doi.org/10.1016/j.physletb.2017.04.073}{\emph{Phys.
  Lett.} {\bf B770} (2017) 174--181},
  [\href{https://arxiv.org/abs/1610.02827}{{\tt 1610.02827}}].

\bibitem{Abazov:2007aj}
{\scshape D0} collaboration, V.~Abazov et~al., ``{Search for
  flavor-changing-neutral-current $D$ meson
  decays},''\href{http://dx.doi.org/10.1103/PhysRevLett.100.101801}{\emph{Phys.
  Rev. Lett.} {\bf 100} (2008) 101801},
  [\href{https://arxiv.org/abs/0708.2094}{{\tt 0708.2094}}].

\bibitem{Zyla:2020zbs}
{\scshape Particle Data Group} collaboration, P.~Zyla et~al., ``{Review of
  Particle Physics},''\href{http://dx.doi.org/10.1093/ptep/ptaa104}{\emph{PTEP}
  {\bf 2020} (2020) 083C01}.

\bibitem{Aaij:2013sua}
{\scshape LHCb} collaboration, R.~Aaij et~al., ``{Search for $D^+_{(s)} \to
  \pi^+ \mu^+ \mu^-$ and $D^+_{(s)} \to \pi^- \mu^+ \mu^+$
  decays},''\href{http://dx.doi.org/10.1016/j.physletb.2013.06.010}{\emph{Phys.
  Lett.} {\bf B724} (2013) 203--212},
  [\href{https://arxiv.org/abs/1304.6365}{{\tt 1304.6365}}].

\bibitem{BHP2007}
C.~Bobeth, G.~Hiller and G.~Piranishvili, ``{Angular distributions of $\bar{B}
  \to \bar{K} \ell^+\ell^-$
  decays},''\href{http://dx.doi.org/10.1088/1126-6708/2007/12/040}{\emph{JHEP}
  {\bf 12} (2007) 040}, [\href{https://arxiv.org/abs/0709.4174}{{\tt
  0709.4174}}].

\bibitem{Braun:2015axa}
V.~Braun, S.~Collins, M.~G\"ockeler, P.~P\'erez-Rubio, A.~Sch\"afer, R.~Schiel
  et~al., ``{Second Moment of the Pion Light-cone Distribution Amplitude from
  Lattice
  QCD},''\href{http://dx.doi.org/10.1103/PhysRevD.92.014504}{\emph{Phys. Rev.
  D} {\bf 92} (2015) 014504}, [\href{https://arxiv.org/abs/1503.03656}{{\tt
  1503.03656}}].

\bibitem{Bali:2019dqc}
G.~S. Bali, V.~M. Braun, S.~B\"urger, M.~G\"ockeler, M.~Gruber, F.~Hutzler
  et~al., ``{Light-cone distribution amplitudes of pseudoscalar mesons from
  lattice QCD},''\href{http://dx.doi.org/10.1007/JHEP08(2019)065}{\emph{JHEP}
  {\bf 08} (2019) 065}, [\href{https://arxiv.org/abs/1903.08038}{{\tt
  1903.08038}}].

\bibitem{Khodjamirian:2011ub}
A.~Khodjamirian, T.~Mannel, N.~Offen and Y.-M. Wang, ``{$B \to \pi \ell \nu_l$
  Width and $|V_{ub}|$ from QCD Light-Cone Sum
  Rules},''\href{http://dx.doi.org/10.1103/PhysRevD.83.094031}{\emph{Phys. Rev.
  D} {\bf 83} (2011) 094031}, [\href{https://arxiv.org/abs/1103.2655}{{\tt
  1103.2655}}].

\bibitem{Bijnens:2002mg}
J.~Bijnens and A.~Khodjamirian, ``{Exploring light cone sum rules for pion and
  kaon
  form-factors},''\href{http://dx.doi.org/10.1140/epjc/s2002-01042-1}{\emph{Eur.
  Phys. J. C} {\bf 26} (2002) 67--79},
  [\href{https://arxiv.org/abs/hep-ph/0206252}{{\tt hep-ph/0206252}}].

\bibitem{Huber:2008id}
{\scshape Jefferson Lab} collaboration, G.~Huber et~al., ``{Charged pion
  form-factor between $Q^2 = 0.60$ GeV$^2$ and 2.45 GeV$^2$. II. Determination
  of, and results for, the pion
  form-factor},''\href{http://dx.doi.org/10.1103/PhysRevC.78.045203}{\emph{Phys.
  Rev. C} {\bf 78} (2008) 045203}, [\href{https://arxiv.org/abs/0809.3052}{{\tt
  0809.3052}}].

\bibitem{Cheng:2020vwr}
S.~Cheng, A.~Khodjamirian and A.~V. Rusov, ``{The pion light-cone distribution
  amplitude from the pion electromagnetic form
  factor},''\href{http://dx.doi.org/10.1103/PhysRevD.102.074022}{\emph{Phys.
  Rev. D} {\bf 102} (2020) 7}, [\href{https://arxiv.org/abs/2007.05550}{{\tt
  2007.05550}}].

\bibitem{Khodjamirian:1998vk}
A.~Khodjamirian, R.~Ruckl and C.~Winhart, ``{The Scalar B ---\ensuremath{>} pi
  and D ---\ensuremath{>} pi form-factors in
  QCD},''\href{http://dx.doi.org/10.1103/PhysRevD.58.054013}{\emph{Phys. Rev.
  D} {\bf 58} (1998) 054013}, [\href{https://arxiv.org/abs/hep-ph/9802412}{{\tt
  hep-ph/9802412}}].

\bibitem{Alekhin:2012py}
S.~Alekhin, A.~Djouadi and S.~Moch, ``{The top quark and Higgs boson masses and
  the stability of the electroweak
  vacuum},''\href{http://dx.doi.org/10.1016/j.physletb.2012.08.024}{\emph{Phys.
  Lett. B} {\bf 716} (2012) 214--219},
  [\href{https://arxiv.org/abs/1207.0980}{{\tt 1207.0980}}].

\bibitem{PDG2016}
{\scshape Particle Data Group} collaboration, C.~Patrignani et~al., ``{Review
  of Particle
  Physics},''\href{http://dx.doi.org/10.1088/1674-1137/40/10/100001}{\emph{Chin.
  Phys.} {\bf C40} (2016) 100001}.

\bibitem{deBoer:2016dcg}
S.~de~Boer, B.~Müller and D.~Seidel, ``{Higher-order Wilson coefficients for
  $c \to u$ transitions in the standard
  model},''\href{http://dx.doi.org/10.1007/JHEP08(2016)091}{\emph{JHEP} {\bf
  08} (2016) 091}, [\href{https://arxiv.org/abs/1606.05521}{{\tt 1606.05521}}].

\bibitem{Fuentes-Martin:2020lea}
J.~Fuentes-Martin, A.~Greljo, J.~Martin~Camalich and J.~D. Ruiz-Alvarez,
  ``{Charm physics confronts high-p$_{T}$ lepton
  tails},''\href{http://dx.doi.org/10.1007/JHEP11(2020)080}{\emph{JHEP} {\bf
  11} (2020) 080}, [\href{https://arxiv.org/abs/2003.12421}{{\tt 2003.12421}}].

\bibitem{Burdman:2001tf}
G.~Burdman, E.~Golowich, J.~L. Hewett and S.~Pakvasa, ``{Rare charm decays in
  the standard model and
  beyond},''\href{http://dx.doi.org/10.1103/PhysRevD.66.014009}{\emph{Phys.
  Rev. D} {\bf 66} (2002) 014009},
  [\href{https://arxiv.org/abs/hep-ph/0112235}{{\tt hep-ph/0112235}}].

\bibitem{Aaij:2013cza}
{\scshape LHCb} collaboration, R.~Aaij et~al., ``{Search for the rare decay
  $D^0 \to \mu^+
  \mu^-$},''\href{http://dx.doi.org/10.1016/j.physletb.2013.06.037}{\emph{Phys.
  Lett.} {\bf B725} (2013) 15--24},
  [\href{https://arxiv.org/abs/1305.5059}{{\tt 1305.5059}}].

\bibitem{Wang:2014uiz}
R.-M. Wang, J.-H. Sheng, J.~Zhu, Y.-Y. Fan and Y.-G. Xu, ``{Decays
  $D^+_{(s)}\to \pi(K)^{+}\ell^+\ell^-$ and $D^0\to \ell^+\ell^-$ in the MSSM
  with and without
  R-parity},''\href{http://dx.doi.org/10.1142/S0217751X15500633}{\emph{Int. J.
  Mod. Phys.} {\bf A30} (2015) 1550063},
  [\href{https://arxiv.org/abs/1409.0181}{{\tt 1409.0181}}].

\bibitem{Paul:2011ar}
A.~Paul, I.~I. Bigi and S.~Recksiegel, ``{On $D\to X_u l^+ l^-$ within the
  Standard Model and Frameworks like the Littlest Higgs Model with T
  Parity},''\href{http://dx.doi.org/10.1103/PhysRevD.83.114006}{\emph{Phys.
  Rev.} {\bf D83} (2011) 114006}, [\href{https://arxiv.org/abs/1101.6053}{{\tt
  1101.6053}}].

\bibitem{Crivellin:2017zlb}
A.~Crivellin, D.~M\"uller and T.~Ota, ``{Simultaneous explanation of
  R($D^{(*)}$) and $b\rightarrow s\mu^{+}\mu^{-}$: the last scalar leptoquarks
  standing},''\href{http://dx.doi.org/10.1007/JHEP09(2017)040}{\emph{JHEP} {\bf
  09} (2017) 040}, [\href{https://arxiv.org/abs/1703.09226}{{\tt 1703.09226}}].

\bibitem{Becirevic:2018afm}
D.~Be\v{c}irevi\'c, I.~Dor\v{s}ner, S.~Fajfer, N.~Ko\v{s}nik, D.~A. Faroughy
  and O.~Sumensari, ``{Scalar leptoquarks from grand unified theories to
  accommodate the $B$-physics
  anomalies},''\href{http://dx.doi.org/10.1103/PhysRevD.98.055003}{\emph{Phys.
  Rev. D} {\bf 98} (2018) 055003},
  [\href{https://arxiv.org/abs/1806.05689}{{\tt 1806.05689}}].

\bibitem{Angelescu:2018tyl}
A.~Angelescu, D.~Be\v{c}irevi\'c, D.~Faroughy and O.~Sumensari, ``{Closing the
  window on single leptoquark solutions to the $B$-physics
  anomalies},''\href{http://dx.doi.org/10.1007/JHEP10(2018)183}{\emph{JHEP}
  {\bf 10} (2018) 183}, [\href{https://arxiv.org/abs/1808.08179}{{\tt
  1808.08179}}].

\bibitem{Crivellin:2019dwb}
A.~Crivellin, D.~Müller and F.~Saturnino, ``{Flavor Phenomenology of the
  Leptoquark Singlet-Triplet Model},''
  \href{https://arxiv.org/abs/1912.04224}{{\tt 1912.04224}}.

\bibitem{Lees:2011hb}
{\scshape BaBar} collaboration, J.~P. Lees et~al., ``{Searches for Rare or
  Forbidden Semileptonic Charm
  Decays},''\href{http://dx.doi.org/10.1103/PhysRevD.84.072006}{\emph{Phys.
  Rev.} {\bf D84} (2011) 072006}, [\href{https://arxiv.org/abs/1107.4465}{{\tt
  1107.4465}}].

\bibitem{Ablikim:2018gro}
{\scshape BESIII} collaboration, M.~Ablikim et~al., ``{Search for the rare
  decays $D\to
  h(h')e^+e^-$},''\href{http://dx.doi.org/10.1103/PhysRevD.97.072015}{\emph{Phys.
  Rev. D} {\bf 97} (2018) 072015},
  [\href{https://arxiv.org/abs/1802.09752}{{\tt 1802.09752}}].

\bibitem{Prospects19}
A.~Contu, ``{Rare Charm Decays and asymmetries}.'' Talk at
  \href{https://cds.cern.ch/record/2670156/files/AContu_RareCharm_TUPFPhop2019.pdf}{Towards
  the Ultimate Precision in Flavour Physics}, Durham, UK, April, 2019.

\bibitem{Prospects20}
D.~Mitzel, ``{LOI Rare Charm decays}.'' Talk at
  \href{https://indico.fnal.gov/event/45713/contributions/198320/attachments/135507/168086/Snowmass_RareCharm.pdf}{RPF
  Town Hall Meeting}, October, 2020.

\bibitem{Khodjamirian:2009ys}
A.~Khodjamirian, C.~Klein, T.~Mannel and N.~Offen, ``{Semileptonic charm decays
  $D \to \pi\ell\nu(\ell)$ and $D \to \pi\ell\nu(\ell)$ from QCD Light-Cone Sum
  Rules},''\href{http://dx.doi.org/10.1103/PhysRevD.80.114005}{\emph{Phys.
  Rev.} {\bf D80} (2009) 114005}, [\href{https://arxiv.org/abs/0907.2842}{{\tt
  0907.2842}}].

\bibitem{Braun:1988qv}
V.~M. Braun and I.~E. Filyanov, ``{QCD Sum Rules in Exclusive Kinematics and
  Pion Wave Function},''\href{http://dx.doi.org/10.1007/BF01548594}{\emph{Z.
  Phys.} {\bf C44} (1989) 157}.

\bibitem{Bharucha:2012wy}
A.~Bharucha, ``{Two-loop Corrections to the $B \to \pi$ Form Factor from QCD
  Sum Rules on the Light-Cone and
  $|V_{ub}|$},''\href{http://dx.doi.org/10.1007/JHEP05(2012)092}{\emph{JHEP}
  {\bf 05} (2012) 092}, [\href{https://arxiv.org/abs/1203.1359}{{\tt
  1203.1359}}].

\bibitem{Ball:2004ye}
P.~Ball and R.~Zwicky, ``{New results on $B \to \pi, K, \eta$ decay formfactors
  from light-cone sum
  rules},''\href{http://dx.doi.org/10.1103/PhysRevD.71.014015}{\emph{Phys.
  Rev.} {\bf D71} (2005) 014015},
  [\href{https://arxiv.org/abs/hep-ph/0406232}{{\tt hep-ph/0406232}}].

\bibitem{Grozin:1996pq}
A.~G. Grozin and M.~Neubert, ``{Asymptotics of heavy meson
  form-factors},''\href{http://dx.doi.org/10.1103/PhysRevD.55.272}{\emph{Phys.
  Rev.} {\bf D55} (1997) 272--290},
  [\href{https://arxiv.org/abs/hep-ph/9607366}{{\tt hep-ph/9607366}}].

\bibitem{Gorbahn:2004my}
M.~Gorbahn and U.~Haisch, ``{Effective Hamiltonian for non-leptonic $|\Delta F|
  = 1$ decays at NNLO in
  QCD},''\href{http://dx.doi.org/10.1016/j.nuclphysb.2005.01.047}{\emph{Nucl.
  Phys.} {\bf B713} (2005) 291--332},
  [\href{https://arxiv.org/abs/hep-ph/0411071}{{\tt hep-ph/0411071}}].

\bibitem{Gorbahn:2005sa}
M.~Gorbahn, U.~Haisch and M.~Misiak, ``{Three-loop mixing of dipole
  operators},''\href{http://dx.doi.org/10.1103/PhysRevLett.95.102004}{\emph{Phys.
  Rev. Lett.} {\bf 95} (2005) 102004},
  [\href{https://arxiv.org/abs/hep-ph/0504194}{{\tt hep-ph/0504194}}].

\end{thebibliography}\endgroup
\end{document}